\documentclass[aps,longbibliography,superscriptaddress,eqsecnum,twocolumn,footinbib]{revtex4-1}
\usepackage{graphicx}
\usepackage{stmaryrd}
\usepackage{amsmath, amsfonts}
\usepackage{mathrsfs}
\usepackage{braket}
\usepackage{color}
\usepackage{xspace}
\usepackage{amscd}
\usepackage{comment}
\usepackage{bm}
\definecolor{lightblue}{rgb}{0.13, 0.26, 0.99}

\usepackage[
colorlinks=true,
urlcolor=blue,
citecolor=blue,
linkcolor=blue,
hyperfootnotes=false]{hyperref}

\newcommand{\bacd}{\mathrm{BaCdVO(PO_4)_2{}}}
\newcommand{\amo}{A\mathrm{MoOPO_4Cl}{}}

\allowdisplaybreaks

\begin{document}

\title{Control of superexchange interactions with DC electric fields}
\author{Shunsuke C. Furuya}
\affiliation{Department of Physics, Ibaraki University, Mito, Ibaraki 310-8512, Japan}
\author{Kazuaki Takasan}
\affiliation{Department of Physics, University of California, Berkeley, California 94720, USA}
\affiliation{Materials Sciences Division, Lawrence Berkeley National Laboratory, Berkeley, California 94720, USA}
\author{Masahiro Sato}
\affiliation{Department of Physics, Ibaraki University, Mito, Ibaraki 310-8512, Japan}

\begin{abstract}
    We discuss DC electric-field controls of superexchange interactions.
    We first present generic results about antiferromagnetic and ferromagnetic superexchange interactions valid in a broad class of Mott insulators, where we also estimate typical field strength to observe DC electric-field effects: $\sim 1~\mathrm{MV/cm}$ for inorganic Mott insulators such as transition-metal oxides and $\sim 0.1~\mathrm{MV/cm}$ for organic ones.
    Next, we apply these results to geometrically frustrated quantum spin systems.
    Our theory widely applies to (quasi-)two-dimensional and thin-film systems and one-dimensional quantum spin systems on various lattices such as square, honeycomb, triangular, and kagome ones.
    In this paper, we give our attention to those on
    the square lattice and on the chain.
    For the square lattice, we show that DC electric fields can control a ratio of the nearest-neighbor and next-nearest-neighbor exchange interactions.
    In some realistic cases, DC electric fields make the two next-nearest-neighbor interactions nonequivalent and eventually turns the square-lattice quantum spin system into a deformed triangular-lattice one.
    For the chain, DC electric fields can induce singlet-dimer and Haldane-dimer orders.
    We show that the DC electric-field-induced spin gap $\propto |\bm E|^{2/3}$ in the Heisenberg antiferromagnetic chain will reach $\sim 10~\%$ of the dominant superexchange interaction in the case of a spin-chain compound $\mathrm{KCuMoO_4(OH)}$ when the DC electric field of $\sim 1~\mathrm{MV/cm}$ is applied.
\end{abstract}
\date{\today}
\maketitle

\section{Introduction}

Controlling quantum states of matter has been a long-standing subject of condensed-matter physics and other related fields.
Historically, the condensed-matter-physics community has made much effort for the challenging task of searching for novel quantum states of matter such as, in quantum magnetism, spin liquids~\cite{anderson_spinliquid, savary_review_2016, zhou_review_spinliquid,knolle_spinliquid_review,jiang_qsl_sq, eggert_j1j2} and spin-nematic states~\cite{chubukov_nematic_91,shannon_sq_nematic, lauchli_blbq, hikihara_nematic_2008, sudan_J1J2_chain,nematic_review}.
This task includes a search for experimental realizations of theoretical models that can host such quantum states.
However, the synthesis of a compound that faithfully realizes the theoretical model does not mean realizing the quantum phase of our interest, which depends on the parameters of the compound.
In general, it is even more challenging to obtain a model compound with a parameter set suitable for realizing the desired phase.

Microscopic controls of quantum states by external forces then become important.
External forces can induce a quantum phase transition into the phase of our interest and can bring us the microscopic information of the quantum state through the response to the external forces.
An external DC (i.e., static) magnetic field is a typical example.
The DC magnetic field adds the Zeeman interaction to the Hamiltonian and induces quantum phase transitions such as Bose-Einstein condensation of magnetic excitations~\cite{giamarchi_bec}.
The pressure is another interesting example.
The pressure can change the microscopic parameters of the Hamiltonian without introducing additional interactions~\cite{zayed_scbo,sakurai_scbo,zvyagin_cs2cucl4}.

Recently, AC-field controls of quantum states of matter have become a vigorous research subject~\cite{oka_photovolatic,kitagawa_floquet, wang_floquet-bloch, jotzu_floquet_haldane}.
The Floquet theory~\cite{Shirley_floquet, sambe_floquet} provides us with a fundamental framework of such AC-field controls, which are often referred to as the Floquet engineering~\cite{bukov_floquet,eckardt_rmp,oka_kitamura_floquet, sato_floquet_book}.
The Floquet engineering has also been discussed in quantum magnetism~\cite{oka_kitamura_floquet,takayoshi_laser_2014a,takayoshi_laser_2014b,sato_laser_dm, mentink_ultrafast, kitamura_mott, takasan_kondo, claassen_mott, chaudhary_orbital_floquet, sato_floquet_majorana, higashikawa_floquet}.

\begin{figure}[t!]
    \centering
    \includegraphics[viewport = 0 0 1500 1100, width=\linewidth]{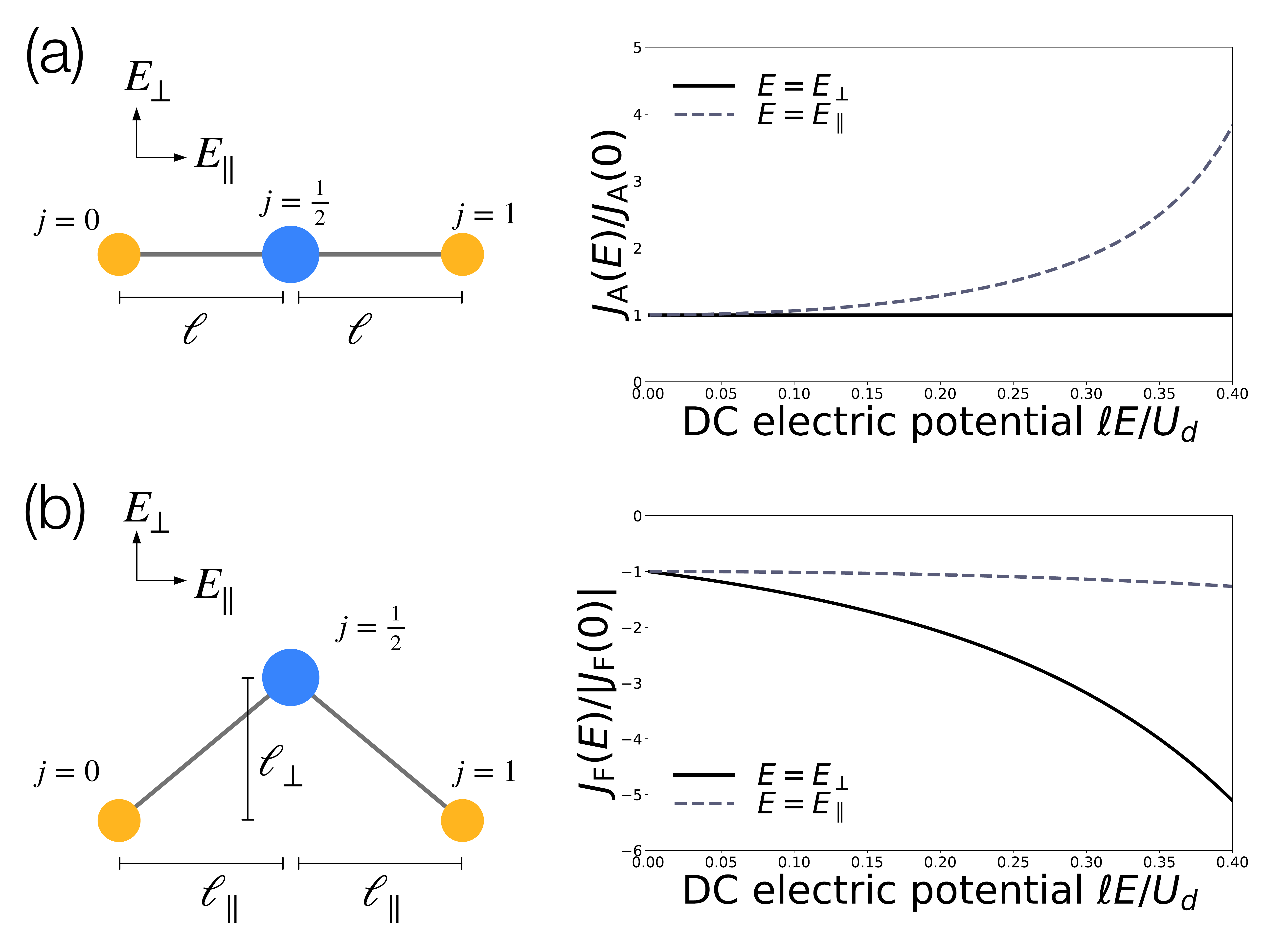}
    \caption{(a) The antiferromagnetic superexchange interaction \eqref{JA} between two magnetic ions (the orange balls) is plotted as a function of $\ell E/U_d$, where $\ell$ is the distance between the $j=0$ ($j=1$) and $j=1/2$ sites, and $U_d$ is the on-site Coulomb repulsion at the $d$-orbital sites.
    We assume that the two magnetic-ion sites and the ligand (the blue ball) site are linearly aligned along the unit vector $\bm e_{0,1}$ that connects the $j=0,1$ sites.
    The DC electric field is applied either parallel ($\bm E_\parallel$) or perpendicular ($\bm E_\perp$) to a unit vector $\bm e_{0,1}$.
    (b) When the ligand site is located to form the right angle as shown in the left panel, the superexchange interaction between the magnetic ions can become ferromagnetic.
    The ferromagnetic interaction \eqref{JF} is plotted in the right panel. 
    When the DC electric field is applied parallel (perpendicular) to $\bm e_{0,1}$, we take $\ell = \ell_\parallel$ ($\ell=\ell_\perp$, respectively).
    For (a) and (b), we used typical values, $U_d=5$~eV, $U_p=1$~eV, and $\Delta_{dp}=2$~eV.
    Also, we took $J_{\rm H}=U_p = 1$~eV for (b).
    The hoppings $t_0$ and $t_1$ can be arbitrary when we consider ratios $J_{\rm A}(\bm E)/J_{\rm A}(\bm 0)$ and $J_{\rm F}(\bm E)/|J_{\rm F}(\bm 0)|$ [see Eqs.~\eqref{JA} and \eqref{JF}].
    }
    \label{fig:JA_JF}
\end{figure}

\begin{figure}[t!]
    \centering
    \includegraphics[viewport = 0 0 1600 900, width=\linewidth]{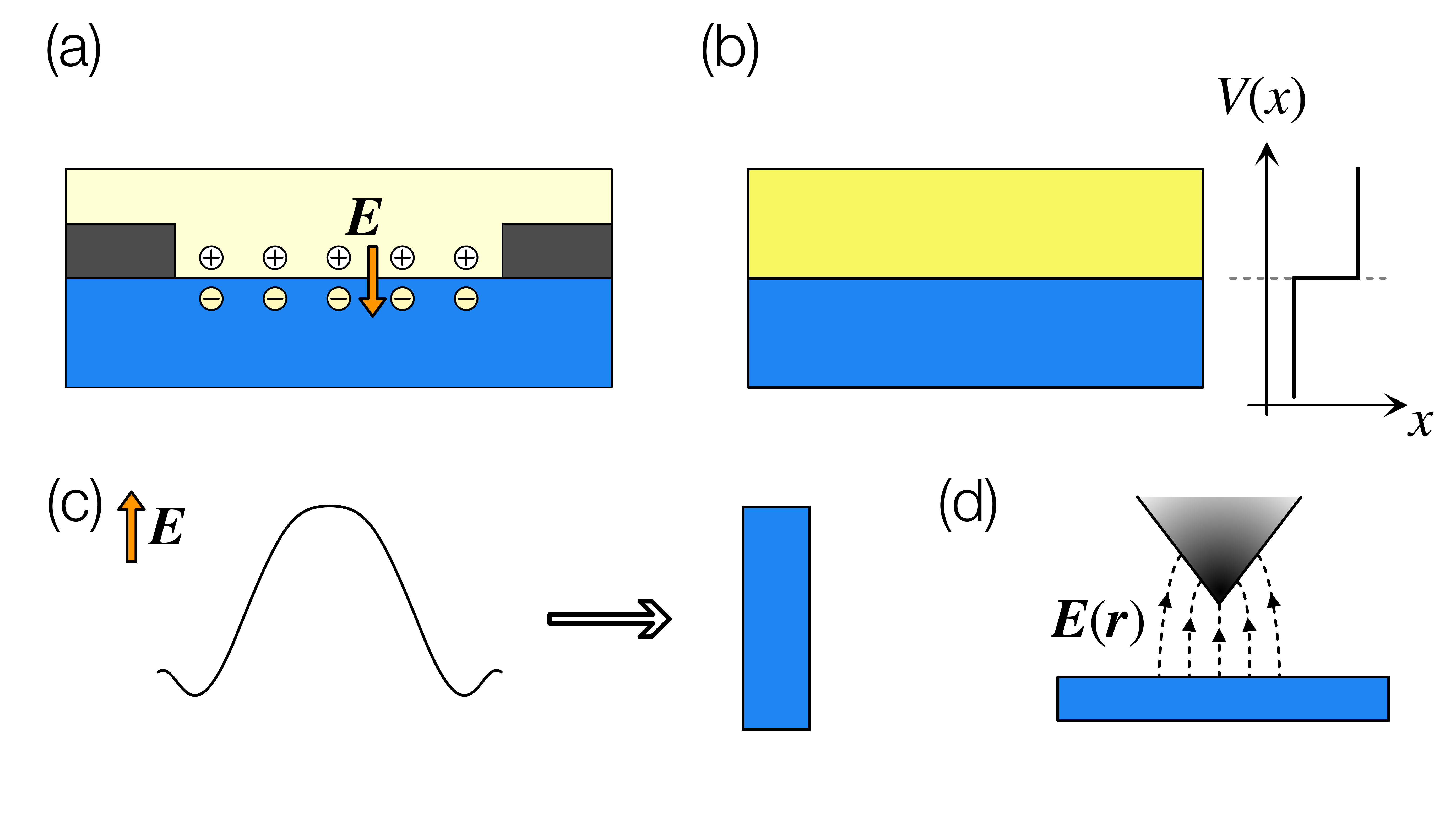}
    \caption{Three exemplary methods to impose a site-dependent potential to a sample. (a) We can employ a field-effect transistor to generate the DC electric field indicated by an orange arrow~\cite{ueno_fet, bisri_fet}. (b) A surface potential at a boundary of two different materials can be used as a source of the gradient of the potential $V(x)$ due to the explicit inversion-symmetry breaking at the surface. (c) A single-cycle terahertz laser pulse can also be a DC electric-field source within a time scale sufficiently shorter than the pulse width (see Sec.~\ref{sec:thz}.).
    (d) A needle-like electrode device such as scanning tunneling microscopes can yield a spatially local DC electric field~\cite{Romming2013,Hsu2016}.
    The dashed curves depict electric force lines.
    }
    \label{fig:fet_surface_pulse}
\end{figure}

Despite these recent developments, microscopic DC \emph{electric}-field controls, which should be theoretically simpler than AC ones, are less considered in quantum spin systems partly because these systems are usually realized in Mott insulators where the charge degree of freedom is frozen.
The DC electric field actually affects the Hamiltonian of quantum spin systems, as shown in Fig.~\ref{fig:JA_JF}, because the exchange interaction has an electronic origin.
The superexchange interaction of spins comes from hoppings of electrons carrying the spin.
It also motivates us to study DC electric-field effects that the DC field is free from heating effects that the AC one inevitably induces.

Previously, some of the authors investigated DC electric-field controls of ``direct superexchange'' interactions~\footnote{In this paper, we define superexchange interactions as exchange interactions originating from hoppings between magnetic ions and intermediate nonmagnetic ions.
We refer to the exchange interactions that originate from direct hoppings between magnetic ions as ``direct superexchange'' interactions to distinguish it from the above-mentioned superexchange and the direct exchange interaction~\cite{koch_exchange}.}, originating from direct hoppings between magnetic ions, by starting from a fundamental electron model, the Hubbard model~\cite{takasan_dc}.
Reference~\cite{takasan_dc} dealt with the DC electric potential as a site-dependent on-site potential.
This treatment of the DC electric field applies to general site-dependent potentials and is convenient to discuss several related phenomena on an equal footing, as shown in Fig.~\ref{fig:fet_surface_pulse}.
Note that strong DC electric fields are often required to induce sizable effects on magnetism, where the surface geometry [Fig.~\ref{fig:fet_surface_pulse}~(b)]~\cite{matsukura_nat_nanotech_2015, matsukura_GaMnAs} and the single-cycle terahertz (THz) laser pulse [Fig.~\ref{fig:fet_surface_pulse}~(c)]~\cite{hirori_laser_2011,mukai_laser_2016,nicoletti_laser_2016} are helpful for that purpose.
DC electric field generated by these methods can be deemed spatially uniform.
Note that we can also apply a local DC electric field to the material, for example, by using a needle-like device [Fig.~\ref{fig:fet_surface_pulse}~(d)]~\cite{Romming2013,Hsu2016}~\footnote{Generic results in Sec.~\ref{sec:results_gen} will hold also for the spatially nonuniform DC electric fields whereas we will mainly focus on the spatially uniform one throughout the paper.}

As of this writing, DC electric-field controls of superexchange interactions mediated by nonmagnetic ions are largely unexplored despite their importance to a broad class of Mott-insulating materials such as transition-metal oxides.
Reference~\cite{takasan_dc} discussed DC electric-field controls of exchange interactions, which is focused on controlling the spatial anisotropy by the DC electric field.
It was not yet discussed how to control exchange interactions by keeping the spatial anisotropy intact.

This paper develops a theory of DC electric-field controls of superexchange interactions in geometrically frustrated quantum spin systems, starting from simple electron models.
We are mainly focused on controlling microscopic Hamiltonian without affecting the dimensionality of the sample, in contrast to our previous work~\cite{takasan_dc}.
Our theory is widely applicable to (quasi-)two-dimensional thin-film and one-dimensional quantum spin systems on various lattices such as square, honeycomb, triangular, and kagome ones.
We discuss basic exemplary applications of our theory to frustrated quantum spin systems on the square lattice and those on the chain.

This paper is organized as follows.
In Sec.~\ref{sec:models}, we define two models that give a firm foundation for DC electric-field controls of superexchange interactions in specific cases.
In Sec.~\ref{sec:results_gen}, we perform fourth-order perturbation expansions on those two models and derive superexchange interactions in generic forms.
Also, we estimate typical values of DC electric-field strength from our results.
We apply generic results of Sec.~\ref{sec:results_gen} to specific situations in Secs.~\ref{sec:2d} and \ref{sec:1d}.
Section~\ref{sec:2d} is devoted to geometrically frustrated quantum spin systems on the square lattice, where the nearest-neighbor $J_1$ and next-nearest-neighbor $J_2$ exchange interactions compete with each other (see Sec.~\ref{sec:sq}).
First, we deal with a simple toy model in Sec.~\ref{sec:toy} to demonstrate controlling a ratio of $J_1/J_2$ by a DC electric field without affecting the spatial anisotropy.
Next, we investigate a more realistic situation corresponding to a compound $\bacd$~\cite{nath_bacdvopo4_2008} (Sec.~\ref{sec:bacd}).
To get insight into the essential effects of DC electric fields, we employ an electron model that significantly simplifies the structures of those compounds.
The simplified model that emulates the experimental reality is studied in Secs.~\ref{sec:toy2} and \ref{sec:toy2_spin}.
Section~\ref{sec:1d} discusses another application of our generic results to frustrated quantum spin chains formed on a CuO$_2$ chain.
We show that the DC electric field induces an alternation of nearest-neighbor exchange interactions, called bond alternation.
We discuss experimentally observable consequences of the DC electric-field-induced bond alternation, which turns out to differ from the DC magnetic-field-induced one.
Section~\ref{sec:other_eff} discusses other major DC electric-field effects not incorporated in our analysis.
We summarize the paper in Sec.~\ref{sec:summary}.

\section{Generic models}\label{sec:models}

\begin{figure}[t!]
    \centering
    \includegraphics[viewport = 0 0 1000 450, width=\linewidth]{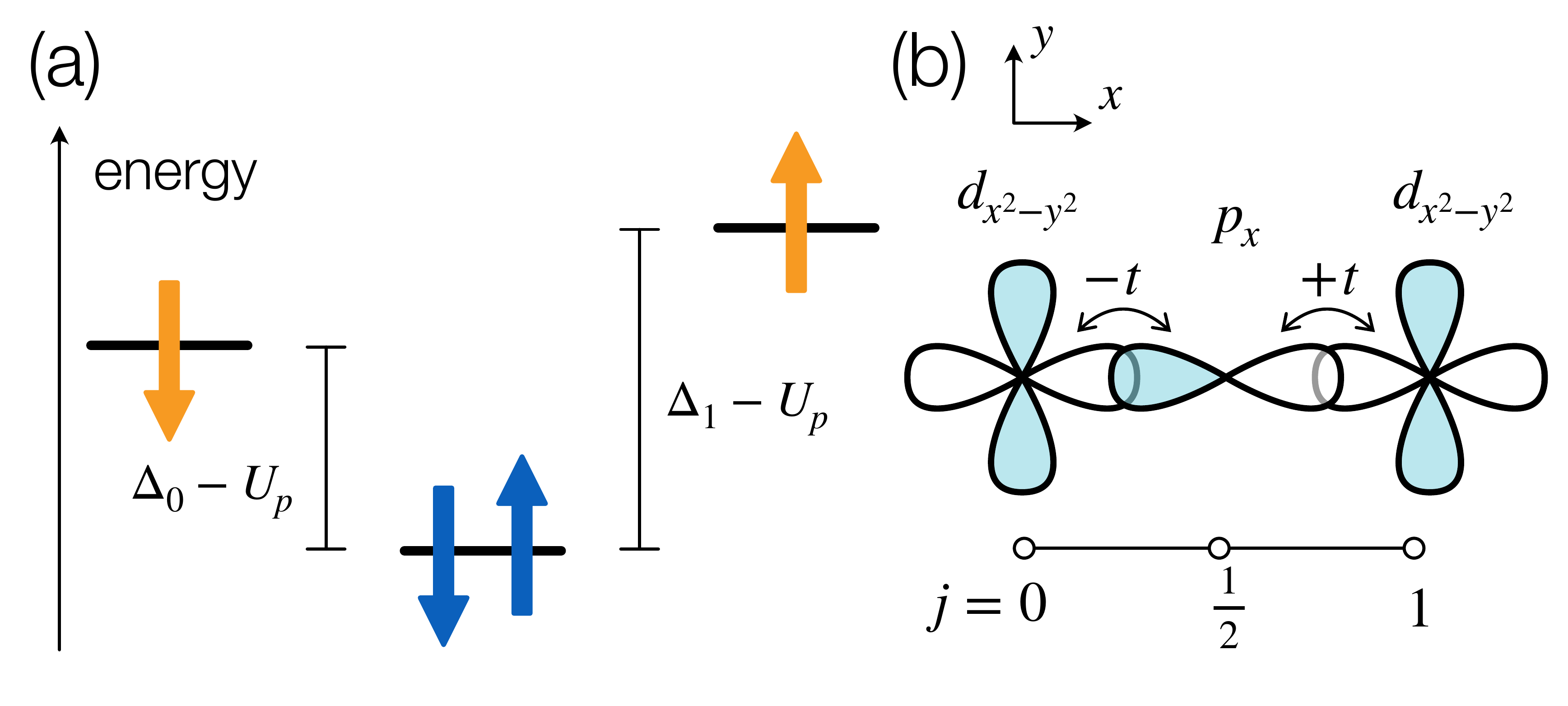}
    \caption{(a) An electron configuration of an unperturbed ground state of the model \eqref{HA} is schematically drawn. Bars represent the unperturbed energy levels of the $j=0, \frac 12$, and $1$ site from left to right, whose energy differences are depicted as those among heights.
    Each bar corresponds to a single orbital.
    Namely, each bar can have two electrons with the opposite spins.
    Since the $p$ orbital at the $j=1/2$ site is fully occupied, the eigenenergy is raised by the repulsive potential energy $U_p$.
    (b) Example of the $d$ and $p$ orbitals for the model \eqref{HA}, where the $j=0$ and $1$ sites have the $d_{x^2-y^2}$ orbital and the $j=1/2$ site has the $p_x$ orbital.
    The hopping amplitudes satisfy $-t_0=t_1=t$.
    }
    \label{fig:HA_levels}
\end{figure}
\begin{figure}[t!]
    \centering
    \includegraphics[viewport = 0 0 1000 1200, width=\linewidth]{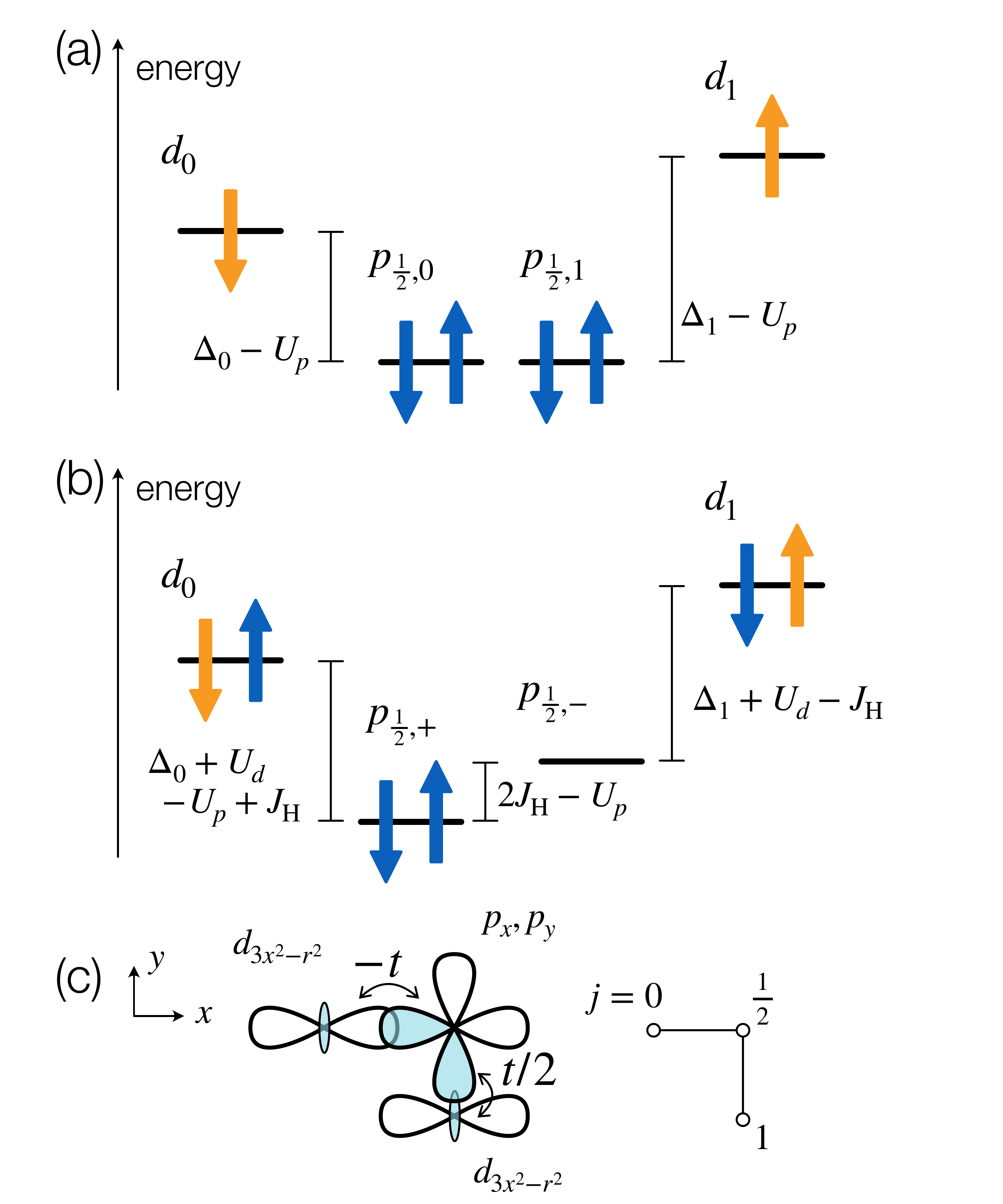}
    \caption{(a) Unperturbed ground state of the model \eqref{HF}. The model has doubly degenerate $p$ orbitals labeled by $p_{\frac 12,0}$ and $p_{\frac 12, 1}$ at the $j=1/2$ site unless both of them are half occupied.
    (b) Unperturbed excited state of the model \eqref{HF}. Since the $p$ orbitals are half occupied, the Coulomb exchange interaction acts on the $p_{\frac 12, \mu}$ orbitals ($\mu=0,1$) and lift their degeneracy by reconstructing them.
    The reconstructed $p$ orbitals at the $j=1/2$ site are $p_{\frac 12, \pm } = p_{\frac 12, 0} \pm p_{\frac 12, 1}$. The eigenenergies of the two $p$ orbitals differ by $2J_{\rm H}-U_p$ when they are half occupied, that is, when the lower-energy level $p_{\frac 12, +}$ is fully occupied and the higher-energy one $p_{\frac 12, -}$ is empty.
    (c) Example of the $d$ and $p$ orbitals. Here, we took $d_0 = d_1 = d_{3x^2-r^2}$, $p_{\frac 12, 0} = p_x$, and $p_{\frac 12, 1} = p_y$. There are two hoppings, $t_0 = -t$ between the $d_{3x^2-r^2}$ and $p_x$ orbitals and $t_1 = \frac 12 t$ between the $d_{3x^2-r^2}$ and $p_y$ orbitals.
    }
    \label{fig:HF_levels}
\end{figure}

Our argument is based on a degenerate perturbation theory of single-band or multi-band Hubbard models with a site-dependent potential, whose Hamiltonian $\mathcal H$ is given by the following generic form:
\begin{align}
    \mathcal H &= \mathcal H_U + \mathcal H_t.
    \label{H_gen}
\end{align}
Here, $\mathcal H_U$ represents potential terms such as the on-site Coulomb repulsion and site-dependent potentials, and $\mathcal H_t$ represents hopping terms of electrons.
Note that we include the DC electric field in the site-dependent potential.
Throughout the paper, $\mathcal H_t$ is regarded as a perturbation to $\mathcal H_U$.

We take two simple and generic models.
Both models are made of three sites, two of which have $d$ orbitals, and the other site in the middle has $p$ orbitals.
The number of sites is minimized but large enough to yield the superexchange interaction.
To discuss the superexchange interaction between $d$ electrons, we assume that each $d$ orbital is half occupied, and the $p$ orbitals are fully occupied in the subspace of the degenerate unperturbed ground states.
The difference of two models lies in a degeneracy of $p$ orbitals at the middle site (Figs.~\ref{fig:HA_levels} and \ref{fig:HF_levels}).

The first model (Fig.~\ref{fig:HA_levels}) is the three-site single-band Hubbard model with the following Hamiltonian:
\begin{align}
    \mathcal H_{\rm A}
    &= \mathcal H_{U;\mathrm{A}} + \mathcal H_{t;\mathrm{A}},
    \label{HA} \\
    \mathcal H_{U;\mathrm{A}}
    &= U_d\sum_{j=0,1} n_{j,\uparrow} n_{j,\downarrow} + U_p  n_{\frac 12, \uparrow} n_{\frac 12, \downarrow}
    \notag \\
    &\quad + \sum_{j=0,\frac 12, 1}\sum_\sigma V_j n_{j,\sigma} - \frac h2 \sum_{j=0, \frac 12, 1} (n_{j,\uparrow}-n_{j,\downarrow}),
    \label{HA_U} \\
    \mathcal H_{t;\mathrm{A}}
    &= -  \sum_{\sigma}\{ (t_0d_{0,\sigma}^\dag + t_1d_{1,\sigma}^\dag) p_{\frac 12, \sigma} + \mathrm{H.c.} \},
    \label{HA_t}
\end{align}
where the $j$th site has a $d$ orbital for $j=0,1$ and a $p$ orbital for $j=1/2$.
$d_{j,\sigma}$ and $p_{j,\sigma}$ are annihilation operators of $d$ and $p$ electrons with the spin $\sigma = \uparrow, \downarrow$.
Hopping amplitudes, $t_0, t_1 \in \mathbb C$ are supposed to be complex to incorporate $\mathrm{U(1)}$ fluxes, if necessary.
$V_j$ represents the site-dependent and spin-independent potential of electrons, into which the DC electric potential enters.
Generally, $p$ and $d$ orbitals have different eigenenergies.
These atomic orbital eigenenergies are incorporated into our model through the on-site potential term $V_j$.
The last term of Eq.~\eqref{HA_U} represents the Zeeman energy, where $h$ is the DC magnetic field.
Though $h=g\mu_B B$ with $B=\mu_0 H$ is the Zeeman splitting rather than the DC magnetic field $H$, we roughly identify $h$ and $H$, and call them simply the DC magnetic field in this paper.
Both the external magnetic field is spatially uniform. If nonuniform, the exchange interactions would depend on the magnetic field.

We assume that the on-site repulsions $U_d > U_p > 0$ are much larger than $|V_j|$ and $|t_j|$ to validate a degenerate perturbation expansion about $|t_j|/U_d$ and $|t_j|/U_p$.
The condition of small $|V_j|$ is required to ensure that the low-energy physics involves magnetic excitations only.
The same parameter conditions apply to the other model given below.

The second model (Fig.~\ref{fig:HF_levels}) has two $p$ orbtals at the $j=1/2$ site:
\begin{align}
    \mathcal H_{\rm F}
    &= \mathcal H_{U;\mathrm{F}} + \mathcal H_{t;\mathrm{F}},
    \label{HF} \\
    \mathcal H_{U;\mathrm{F}}
    &= U_d\sum_{j=0,1} n_{j,\uparrow} n_{j,\downarrow} + U_p \sum_{\mu=0,1}  n_{\frac 12, \uparrow, \mu} n_{\frac 12, \downarrow, \mu}
    \notag \\
    &\quad + \sum_{j=0, 1}\sum_\sigma V_j  n_{j,\sigma} + \sum_{\sigma}\sum_{\mu=0,1} V_{\frac 12} n_{j, \sigma, \mu}
    \notag \\
    &\quad - \frac h2 \biggl[\sum_{j=0,1} (n_{j,\uparrow} - n_{j,\downarrow}) + \sum_{\mu=0,1} (n_{\frac 12,\mu,\uparrow}  - n_{\frac 12, \mu,\downarrow}) \biggr]
    \notag \\
    &\quad - J_{\rm H} \bm s_0 \cdot \bm s_1,
    \label{HF_U} \\
    \mathcal H_{t;\mathrm{F}}
    &= -  \sum_{\sigma} ( t_0d_{0,\sigma}^\dag p_{\frac 12, \sigma, 0} + t_1 d_{1,\sigma}^\dag p_{\frac 12, \sigma, 1} + \mathrm{H.c.}).
    \label{HF_t}
\end{align}
The $j=1/2$ site has two $p$ orbitals labeled by the index $\mu = 0,1$.
The operator $\bm s_\mu=\frac \hbar 2 \sum_{s,s'=\uparrow,\downarrow} p_{\frac 12,s,\mu}^\dag \bm \sigma^{ss'} p_{\frac 12,s',\mu}$ denotes the spin-$1/2$ hosted by the $p$ orbital $\mu=0,1$, where $\bm \sigma = (\sigma^x, \, \sigma^y, \sigma^z)$ is a set of the Pauli matrices.
The operator $\bm s_\mu$ makes sense only when the $p$ orbitals $\mu=0,1$ are half occupied.
We include the interaction, $-J_{\rm H}\bm s_0 \cdot \bm s_1$, only when both the two $p$ orbitals are half occupied and otherwise may drop it from Eq.~\eqref{HF_U}.
The coupling $-J_{\rm H}<0$, which is the ferromagnetic direct exchange between the $p$ orbitals and related to the Hund's rule, affects the $p$ orbitals only when they are half occupied.
Though the inter-band Coulomb potentials are omitted in this model for simplicity, it is straightforward to take them into account.

For simplicity, we focus on a situation where the eigenenergy, $\mathcal E_p$, of the empty $p$ orbital is equal to or lower than the eigenenergy, $\mathcal E_d$, of the empty $d$ orbital in both models, that is, $\mathcal E_p\le \mathcal E_d$.
These eigenenergies are encoded in the on-site potential $V_j$.
Here, we introduce a parameter $\Delta_j(\bm E)$ for $j=0,1$,
\begin{align}
    \Delta_j(\bm E) &= V_j - V_{\frac 12}.
    \label{Delta_j_def}
\end{align}
$\bm E$ denotes the DC electric field, and
$\Delta_j(\bm 0)=\mathcal E_d-\mathcal E_p$ represents the eigenenergy difference of the $d$ orbital at the $j$ site from the $p$ orbital aside from the Coulomb repulsion energy.
For example, when we apply $\bm E=\bm E_\parallel$ in Fig.~\ref{fig:JA_JF}~(a), we obtain $\Delta_0(\bm E_\parallel) = \mathcal E_d-\mathcal E_p -|e|\ell E_\parallel$ and $\Delta_1(\bm E_\parallel)= \mathcal E_d-\mathcal E_p+|e|\ell E_\parallel$, where $-e<0$ is the electron charge.

Though we use parameters that meet the condition,
\begin{align}
    \Delta_j(\bm 0) \ge 0 \quad (j=0,1),
    \label{cond_Delta_0}
\end{align}
for simplicity throughout this paper, nothing forbids real materials from violating the inequality \eqref{cond_Delta_0}.
We emphasize that our results also hold for $\Delta_j(\bm 0)<0$.
The eigenenergy differences can be shifted by the amount of the Coulomb repulsive energy when the orbitals are partially or fully occupied [Figs.~\ref{fig:HA_levels}~(a) and \ref{fig:HF_levels}~(a,b)].

We conclude this section by mentioning the orbital degeneracy.
We assumed above the single $d$ orbital at each magnetic-ion site, namely, the absence of the $d$-orbital degeneracy.
This assumption can be easily relaxed.
Let us consider situations where only one $d$ orbital at each magnetic-ion site can participate in hoppings between the intermediate nonmagnetic-ion site because of symmetries.
For example, in the setup of Fig.~\ref{fig:HF_levels}~(c), the $p_x$ orbital at the $j=0$ has a nonzero hopping amplitude to the $d_{3x^2-r^2}$ orbital at the $j=1/2$ site and has vanishing hopping amplitude to the other four $d$ orbitals, $d_{xy},\, d_{yz}, \, d_{zx},$ and $d_{y^2-z^2}$.
The low-energy physics in these multi-$d$-orbital cases are also effectively described by the models \eqref{HA} and \eqref{HF} unless symmetry-breaking structural distortion occurs.

\section{Superexchange interactions}\label{sec:results_gen}

This section describes the degenerate perturbation theory of the two models \eqref{HA} and \eqref{HF} and shows that the former (the latter) model gives rise to a Heisenberg exchange interaction with the antiferromagnetic (ferromagnetic, respectively) coupling.

\subsection{Definition of the effective Hamiltonian}

The unperturbed ground state is $2^N$-fold degenerate for the two models \eqref{HA} and \eqref{HF},  where $N=2$ is the number of $d$-orbital sites.
Let us define a projection operator $P$ onto the subspace of the Hilbert space spanned by the degenerate ground state.
$Q=1-P$ is a projection operator onto its complementary space spanned by the unperturbed excited states.

We carry out the perturbation expansion as follows.
First, we perform the Schrieffer-Wolf canonical transformation~\cite{schriefferwolf_transf} on the full Hamiltonian \eqref{H_gen}, $\mathcal H  \to e^{\eta} \mathcal H e^{-\eta}$, with an anti-Hermitian operator $\eta$.
The effective Hamiltonian $\mathcal H_{\rm eff}$ that describes the low-energy physics is then defined as
\begin{align}
    \mathcal H_{\rm eff} &= P e^{\eta}  \mathcal H e^{-\eta}P.
    \label{H_eff_def}
\end{align}
The unitary operator $e^{\eta}$ keeps the excitation spectrum of the model unchanged but can simplify the effective Hamiltonian \eqref{H_eff_def}.
$\eta$ is determined so that the $e^{\eta}\mathcal H e^{-\eta}$ is commutative with $P$~\cite{slagle_deg_pert}.
Generic forms of the perturbation expansion of $\mathcal H_{\rm eff}$ up to the sixth order are listed in the appendix of Ref.~\cite{slagle_deg_pert}.

In our case, a relation, $P \mathcal H_t P = 0$, simplifies the perturbation significantly.
Up to the fourth-order perturbation expansion, the effective Hamiltonian \eqref{H_eff_def} is expanded as
\begin{align}
    \mathcal H_{\rm eff}
    &= \mathcal H_0^{\rm eff} + \mathcal H_2^{\rm eff} + \mathcal H_4^{\rm eff},
    \label{H_eff_expanded} \\
    \mathcal H_0^{\rm eff}
    &= P \mathcal H_U P,
    \label{H_eff_0} \\
    \mathcal H_2^{\rm eff}
    &=P \mathcal H_t  \frac{1}{E_g-\mathcal H_U} Q \mathcal H_t P,
    \label{H_eff_2} \\
    \mathcal H_4^{\rm eff}
    &=P \mathcal H_t \biggl(\frac{1}{E_g-\mathcal H_U} Q \mathcal H_t \biggr)^3P,
    \label{H_eff_4}
\end{align}
where $E_g$ is the unperturbed ground-state energy and  $(\frac{1}{E_g-\mathcal H_U} Q\mathcal H_t)^3$ is an abbreviation of $\frac{1}{E_g-\mathcal H_U} Q\mathcal H_t\frac{1}{E_g-\mathcal H_U} Q\mathcal H_t\frac{1}{E_g-\mathcal H_U} Q\mathcal H_t$.
The first- and third-order terms vanish trivially in our models.

The zeroth-order term \eqref{H_eff_0} is mostly constant but contains one important term, the Zeeman energy:
\begin{align}
    \mathcal H_0^{\rm eff} &= -h (S_0^z + S_1^z) + \mathrm{const}.
    \label{H_0_zeeman}
\end{align}
$\bm S_j = \frac \hbar 2 \sum_{s,s'=\uparrow, \downarrow} d_{j,s}^\dag \bm \sigma^{ss'} d_{j,s'}$ is the $S=1/2$ spin operator at the $j=0,1$ site.
Note that the spins of the $p$ orbitals do not appear in Eq.~\eqref{H_0_zeeman} since they are fully occupied in the unperturbed ground-state subspace.
The DC magnetic field $h$ appears only in the zeroth-order term \eqref{H_0_zeeman} because it is spatially uniform and our model Hamiltonians do not have spin-orbit couplings.
The second-order term \eqref{H_eff_2} is a constant that has no impact on low-energy physics and thus discarded hereafter.

\subsection{Antiferromagnetic superexchange}

Performing the fourth-order perturbation expansion on the model \eqref{HA} (see the  Appendix Sec.~\ref{app:JA}), we obtain the following effective Hamiltonian:
\begin{align}
    \mathcal H_{\rm A}^{\rm eff}
    &= J_{\rm A}(\bm E) \bm S_0 \cdot \bm S_1 -h(S_0^z+S_1^z),
\end{align}
where $J_{\rm A}$ is the antiferromagnetic superexchange interaction,
\begin{widetext}
\begin{align}
    J_{\rm A}(\bm E)
    &= 2|t_0t_1|^2 \biggl( \frac{1}{\Delta_0(\bm E) + U_d-U_p} + \frac{1}{\Delta_1(\bm E) + U_d - U_p} \biggr)^2 \frac{1}{\Delta_0(\bm E) + \Delta_1 (\bm E) +2U_d -U_p}
    \notag \\
    &\quad +
    2|t_0t_1|^2 \biggl( \frac{1}{(\Delta_0(\bm E) + U_d-U_p)^2} \frac{1}{\Delta_0(\bm E) - \Delta_1(\bm E) + U_d} + \frac{1}{(\Delta_1(\bm E) +U_d-U_p)^2} \frac{1}{\Delta_1 (\bm E) - \Delta_0(\bm E) +U_d} \biggr),
    \label{JA}
\end{align}
\end{widetext}
where the DC electric field $\bm E$ enters into Eq.~\eqref{JA} through the energy difference $\Delta_j(\bm E)$.
When $\bm E=\bm 0$, the Heisenberg superexchange coupling \eqref{JA} is antiferromagnetic.
In fact,
\begin{align}
    J_{\rm A}(\bm 0)
    &= \frac{4|t_0t_1|^2}{(\Delta_0(\bm 0) + U_d-U_p)^2}  \biggl(\frac{2}{2\Delta_0(\bm 0) + 2U_d - U_p }+  \frac{1}{U_d} \biggr)
\end{align}
is positive when $\Delta_j(\bm 0)\ge 0$ and $U_d>U_p>0$.
The latter inequalities are the case.
The former condition \eqref{cond_Delta_0} is likely to be the case but can be violated, as we mentioned below Eq.~\eqref{Delta_j_def}.
Small DC electric field modifies the strength of the antiferromagnetic exchange coupling \eqref{JA} with keeping its sign.

\subsection{Ferromagnetic superexchange}

When applied to the model \eqref{HF}, the fourth-order perturbation expansion (appendix~\ref{app:JF}) leads to the following effective Hamiltonian of the model \eqref{HF}:
\begin{align}
    \mathcal H_{\rm F}^{\rm eff} &= J_{\rm F}(\bm E) \bm S_0 \cdot \bm S_1 - h(S_0^z+S_1^z),
\end{align}
where $J_{\rm F}(\bm E)$ is the DC-electric-field dependent ferromagnetic superexchange interaction,
\begin{widetext}
\begin{align}
    J_{\rm F}(\bm E)
    &=-2|t_0t_1|^2 \biggl( \frac{1}{\Delta_0(\bm E) + U_d - U_p} + \frac{1}{\Delta_1(\bm E) + U_d - U_p} \biggr)^2 \frac{J_{\rm H}}{[\Delta_0(\bm E) + \Delta_1(\bm E) + 2(U_d-U_p)]^2 - {J_{\rm H}}^2}.
    \label{JF}
\end{align}
\end{widetext}
The sign of $J_{\rm F}(\bm E)$ is determined by that of $J_{\rm H}$.
The coupling $J_{\rm H}$ must be positive since it represents the direct Coulomb exchange interaction~\cite{koch_exchange}.
The additional condition $\Delta_0 (\bm E)+\Delta_1 (\bm E) + 2(U_d-U_p) - J_{\rm H}>0$ is also required to guarantee that the right hand side of Eq.~\eqref{JF} is negative.
The inequality is usually satisfied because $J_d/U_d=O(1)$, $U_d \gg U_p$, and $\Delta_j(\bm E)\ge 0$ [Eq.~\eqref{cond_Delta_0}].
When the inequality \eqref{cond_Delta_0} is violated, the superexchange coupling \eqref{JF} can become antiferromagnetic.
Therefore, the superexchange interaction $J_{\rm F}(\bm E)$ is ferromagnetic.

\subsection{Estimates in typical situations}
\label{sec:est}

Here, we estimate the DC electric-field dependence of the 
antiferromagnetic \eqref{JA} and ferromagnetic \eqref{JF} superexchange interaction parameters for typical situations.
As Fig.~\ref{fig:JA_JF} shows, the DC electric field affects the antiferromagnetic (ferromagnetic) superexchange interaction drastically when the DC electric field is applied parallel to $\bm e_\parallel$  ($\bm e_\perp$), where the unit vector $\bm e_\parallel$ ($\bm e_\perp$) is parallel (perpendicular, respectively) to the vector that connects the $j=0$ site and the $j=1$ site.

Hence, to estimate the typical field strength, we give our attention to the following situations:
\begin{align}
    \bm E = E_\parallel \bm e_\parallel,
\end{align}
for the antiferromagnetic case (Fig.~\ref{fig:HA_levels}) and
\begin{align}
    \bm E = E_\perp \bm e_{\perp}
\end{align}
for the ferromagnetic case (Fig.~\ref{fig:HF_levels}).
Note that the unit vector $\bm e_\perp$, perpendicular to $\bm e_\parallel$, is on the plane where $j=0, \frac 12$, and $1$ sites are put.
$J_{\rm A}(\bm E)$ of Eq.~\eqref{JA} and $J_{\rm F}(\bm E)$ of Eq.~\eqref{JF} are plotted for $\ell E/U_d$ in Fig.~\ref{fig:JA_JF}, where $E$ is either $E_\perp$ or $E_\parallel$, and $\ell$ is a length scale between the $d$-orbital and the $p$-orbital sites.
As shown in Fig.~\ref{fig:JA_JF}~(b), we take $\ell = \ell_\perp$ for $E_\perp$.

Supposing transition-metal oxides, we employ their typical values $U_d = 5$~eV, $U_p=1$~eV, $\Delta_{dp} = 2$~eV, $J_{\rm H} = 1$~eV, and $\ell =5$~\AA{} (e.g., Ref.~\cite{kim_sr2vo4}).
Hopping amplitudes $t_0$ and $t_1$ can be arbitrary, though they should be perturbative.

$J_{\rm A}(\bm E)$ and $J_{\rm F}(\bm E)$ depend on the direction of the DC electric field.
$J_{\rm A}(\bm E)$ is sensitive to $E_\parallel$ but independent of $E_\perp$, which is obvious because the latter case only shifts $V_j$ uniformly for $j=0, \frac 12, 1$.
By contrast, $J_{\rm F}(\bm E)$ is more sensitive to $E_\perp$ than $E_\parallel$.
To increase $J_{\rm A}(\bm E)$ and $J_{\rm F}(\bm E)$ by 1~\% of their original values at $\bm E=\bm 0$, we need
$\ell |E_\parallel| /U_d \approx 0.05$ for the former and $\ell_\perp |E_\perp|/U_d \approx 0.003$ for the latter case, namely, 
\begin{align}
    |E_\parallel| &\approx 5~\mathrm{MV/cm},
\end{align}
for the antiferromagnetic superexchange and
\begin{align}
    |E_\perp| &\approx 0.3~\mathrm{MV/cm},
\end{align}
for the ferromagnetic superexchange.
These values are large but feasible with current experimental techniques. 
As summarized in Ref.~\cite{ueno_fet}, the DC electric field in the conventional field-effect transistor setup [Fig.~\ref{fig:fet_surface_pulse}~(a)] reaches the order of 1~$\mathrm{MV/cm}$. 
Also, recently developing techniques, such as the electric double layer transistors, are capable of inducing the electric field stronger than 10~$\mathrm{MV/cm}$~\cite{ueno_fet,bisri_fet}.

Despite these technical developments, realization of the DC electric field with $O(1)~\mathrm{MV/cm}$ is still challenging.
It is thus important to look into other approaches that facilitate the achievement of the large DC electric field. 
There are basically two options.
One is to use an alternative sample with more suitable parameters.
The other is to use an alternative method to induce the DC electric field.

The simplest way in the first option is to reduce the thickness of the sample.
If we can use a thin-film sample, we will be able to use larger DC electric field though the electric field is then applied only in a direction perpendicular to the thin film [Fig.~\ref{fig:fet_surface_pulse}~(a)].
Reducing the sample thickness meets our purpose of controlling microscopic parameters without changing the composition of the compound.
However, for comparison, it is now worth considering the use of a completely different sample to enhance the DC electric-field effects.
Recall that the superexchange interactions \eqref{JA} and \eqref{JF} are the functions of $\ell E/U_d$.
Instead of increasing $|E|$, we may increase $\ell/U_d$.
Organic Mott insulators will be suitable to this purpose for their long $\ell$ and weak $U_d$~\footnote{Even when the spins $\bm S_0$ and $\bm S_1$ come from $p$ orbitals, our conclusion will not change since we hardly rely on the fact that the magnetic moment is attributed to the $d$ orbital.
We will come back to this point later in Sec.~\ref{sec:toy2_spin}.
}.
Let us assume, for example, $U_d=1$~eV, $U_p=0.5$~eV, $\Delta_{dp}=0.8$~eV, $\ell=10$~\AA{} (e.g., Refs.~\cite{nakamura_k-et, shimizu_organic_triangular}).
Then, $J_{\rm A}(\bm E)/J_{\rm A}(\bm 0)$ and $|J_{\rm F}(\bm E)/J_{\rm F}(\bm 0)|$ are increased by 1~\% when 
\begin{align}
    E_\parallel &= 0.5~\mathrm{MV/cm},
\end{align}
for the antiferromagnetic superexchange and
\begin{align}
    E_\perp &= 0.04~\mathrm{MV/cm}
\end{align}
for the ferromagnetic superexchange.

The last option we consider here is to employ an alternative DC electric-field source, the THz laser pulse (Fig.~\ref{fig:fet_surface_pulse}), instead of using modifying the sample~\footnote{The THz laser pulse typically has the $O(1)$~ps pulse width, which is much longer than the typical time scale of hoppings. The hopping amplitude is $O(10^{-1}U_d)=O(10^{-1})$~eV. Accordingly, the time scale of hoppings is $O(10^1)$~fs $=O(10^{-2})$~ps. In other words, electrons can hop $O(10^2)$ times during the single-cycle THz laser pulse is applied to the system.
Under such circumstances, we may regard the THz laser pulse as an effective DC electric field. 
}.
The THz laser can induce a larger electric field than the DC one~\cite{fulop_terahertz}.
We can regard the THz laser pulse effectively as a DC electric field in a shorter time than the temporal width of the pulse. 
To adopt the THz laser pulse for the DC electric-field control of quantum magnetic states, we need to use a quantum magnet with fast enough spin dynamics (see Sec.~\ref{sec:thz}).

\section{Frustrated ferromagnets on square lattice}\label{sec:2d}

\begin{figure}[b!]
    \centering
    \includegraphics[bb = 0 0 842 595, width=\linewidth]{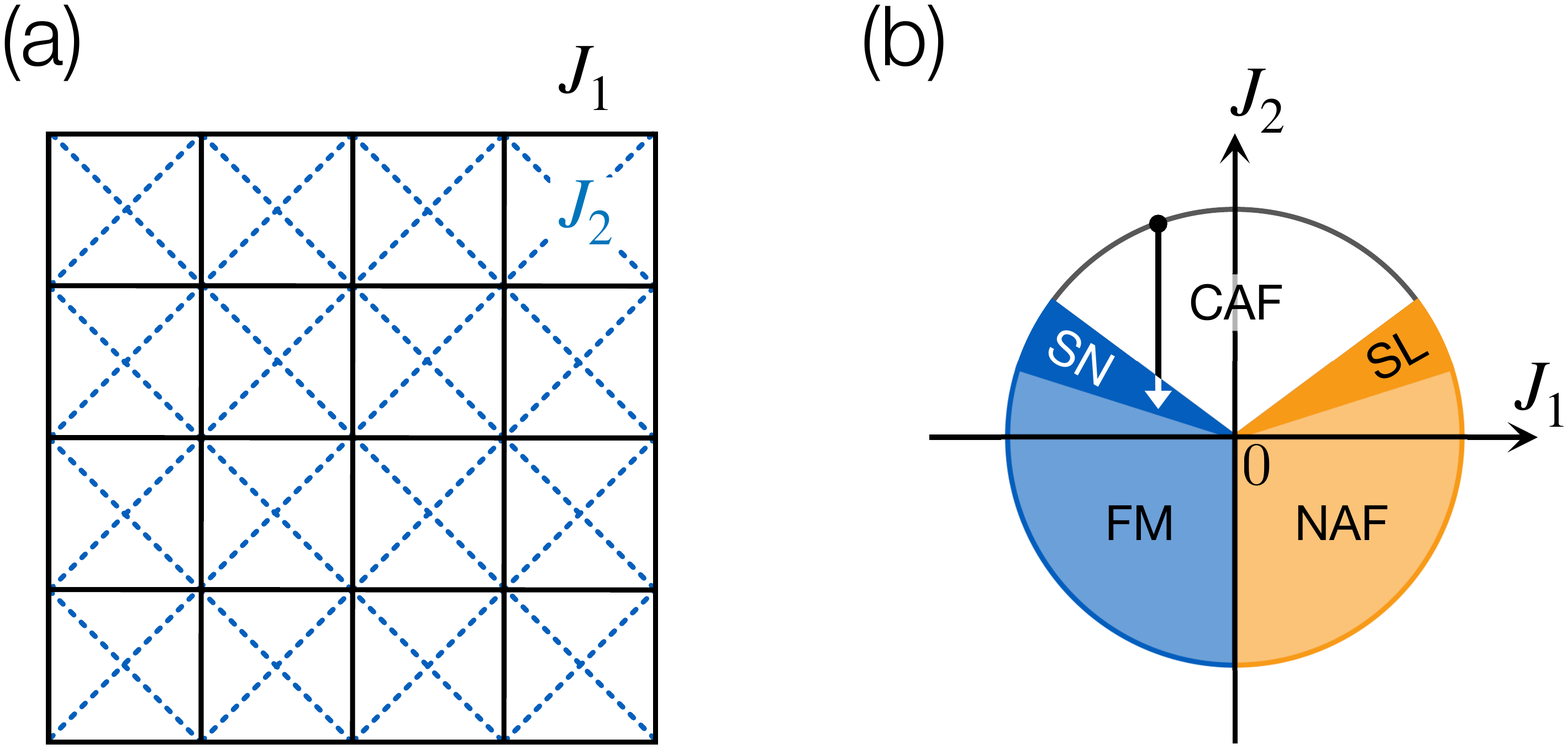}
    \caption{(a) Lattice structure of the spin-$1/2$ $J_1$--$J_2$ Heisenberg model on the square lattice \eqref{H_sq}. The spin is put on every crossing of the square and interact with each other via the nearest-neighbor $J_1$ interaction (the solid line segment) and the next-nearest-neighbor $J_2$ interaction (the dashed line segment). (b) The ground-state phase diagram of the $J_1$--$J_2$ model \eqref{H_sq}~\cite{schmidt_square,shannon_sq_nematic,jiang_qsl_sq}. The angle $\theta$ around the origin corresponds to the ratio $J_1/J_2$: $\tan \theta = J_2/J_1$. The labels CAF, SN, FM, NAF, and SL are shorthands for the canted antiferromagnetic, spin-nematic, ferromagnetic, N\'eel antiferromagnetic, and spin-liquid phases, respectively.}
    \label{fig:sq}
\end{figure}

Sections~\ref{sec:2d} and \ref{sec:1d} are devoted to applications of the generic results of Sec.~\ref{sec:results_gen} to simple frustrated quantum spin systems.

\subsection{Model}\label{sec:sq}

This section deals with a frustrated quantum spin system on the square lattice.
We give our attention to a frustrated spin-$1/2$ $J_1$--$J_2$ model~\cite{chandra_sq_afm_1988,dagotto_sq_afm_1989,read_sq_afm_1991,nomura_zigzag_1994,singh_sq_afm_1999,capriotti_sq_afm_2000,sirker_sq_afm_2006,ueda_j1j2,jiang_qsl_sq,hu_sq_afm_2013,wang_sq_afm_2016,wang_sq_afm_2018,shannon_sq_nematic,shindou_sq_fm_2011,iqbal_sq_fm_2016,morisaku_nematic_laser} with the following Hamiltonian,
\begin{align}
    \mathcal H_{\rm sq}
    &= J_1 \sum_{\braket{\bm r,\bm r'}_1} \bm S_{\bm r} \cdot \bm S_{\bm r'} +J_2 \sum_{\braket{\bm r,\bm r'}_2} \bm S_{\bm r} \cdot \bm S_{\bm r'},
    \label{H_sq}
\end{align}
where $J_1$ and $J_2$ represent the nearest-neighbor and next-nearest-neighbor interactions, respectively.
$\braket{\bm r,\bm r'}_1$ ($\braket{\bm r,\bm r'}_2$) denotes a pair of a nearest-neighbor (next-nearest-neighbor, respectively) bond connecting two sites $\bm r= (r_x, \, r_y)$ and $\bm r' = (r'_x,\, r'_y)$ of the square lattice [Fig.~\ref{fig:sq}~(a)].
The model \eqref{H_sq} is realized in various compounds~\cite{melzi_J1J2_sq_2001, Rosner_J1J2_sq_2003,Kaul_J1J2_sq_2004,bombardi_j1j2_2004,bombardi_j1j2_2005,kageyama_j1j2_2005,oba_j1j2_2006,nath_bacdvopo4_2008,tsirlin_j1j2_2008,tsirlin_J1J2_sq_2009,ishikawa}

We can apply the DC magnetic field $h$ to the model \eqref{H_sq} through the zeroth-order term \eqref{H_0_zeeman}, but in what follows, we focus on the $h=0$ case.
The model \eqref{H_sq} shows a rich ground-state phase diagram~\cite{schmidt_square, shannon_sq_nematic} depending on the ratio of $J_1$ and $J_2$ [Fig.~\ref{fig:sq}~(b)], which contains the spin-liquid phase and the spin-nematic phase.
Hereafter, we employ the unit system $\hbar = e = a_0 = 1$ for simplicity, where $-e<0$ is the electron charge and $a_0$ is the lattice spacing.

\subsection{Toy model}\label{sec:toy}

\begin{figure}[b!]
    \centering
    \includegraphics[viewport = 0 0 2500 900, width=\linewidth]{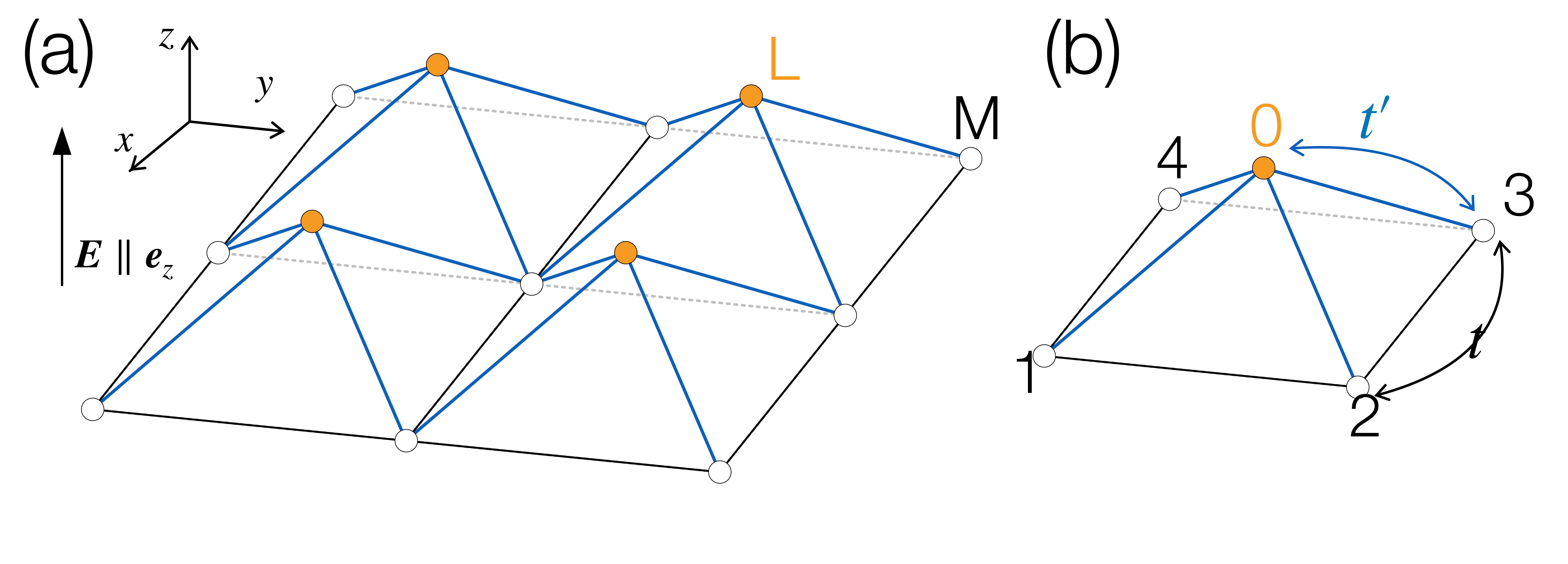}
    \caption{(a) Toy model to demonstrate the DC electric-field control of the ratio $J_1/J_2$. The $d$ orbitals are put on the square lattice (the white balls) and the $p$ orbitals are put on top of the pyramids (the orange balls). (b) Five-site (single-pyramid) model \eqref{H_py} with two kinds of hoppings, $t$ for the $d$-$d$ hoppings and $t'$ for the $d$-$p$ hoppings.}
    \label{fig:pyramids}
\end{figure}

Before addressing an experimentally feasible case, we first consider a simple toy model on a lattice of Fig.~\ref{fig:pyramids}~(a) to get insight into the DC electric-field effects on $J_1/J_2$.
The lattice contains edge-sharing pyramids, where the empty balls form a square lattice.
Every crossing of the square lattice has one magnetic-ion site with a single $d$ orbital that hosts an $S=1/2$ spin.
The top of the pyramid has one ligand site, which is depicted as a filled ball in Fig.~\ref{fig:pyramids}~(a).
We assume that this ligand site hosts two $p$ orbitals.
Similarly to the model \eqref{HF}, these $p$ orbitals are fully occupied and degenerate in the unperturbed ground-state subspace.
When the system is excited to a state with the half-filled $p$ orbitals, the Coulomb exchange $J_{\rm H}$ lifts the orbital degeneracy.
For simplicity, we consider only two hoppings: $t$ on the square edge and $t'$ that climbs up the pyramid to the top [Fig.~\ref{fig:pyramids}~(b)].

As we did in Sec.~\ref{sec:results_gen}, we minimize the number of sites down to 5 [Fig.~\ref{fig:pyramids}~(b)] and discuss nearest-neighbor and next-nearest-neighbor exchange interactions of spins.
The five-site model on the single pyramid has the following Hamiltonian:
\begin{align}
    \mathcal H_{\rm py}
    &= \mathcal H_{U;\mathrm{py}} + \mathcal H_{t;\mathrm{py}},
    \label{H_py} \\
    \mathcal H_{U;\mathrm{py}}
    &= U_d \sum_{j=1}^4 n_{j,\uparrow} n_{j,\downarrow} + U_p \sum_{\mu= 0,1} n_{0,\uparrow,\mu} n_{0,\downarrow,\mu}
    \notag \\
    &\quad + \sum_{j=0}^4\sum_\sigma V_j n_{j,\sigma},
    \label{HU_py} \\
    \mathcal H_{t;\mathrm{py}} &= -\sum_{j=1}^4\sum_{\sigma}  (t d_{j,\sigma}^\dag d_{j+1,\sigma} +\mathrm{H.c.}) 
    \notag \\
    &\quad
    -\sum_{\sigma} [t' (d_{1,\sigma}^\dag +d_{3,\sigma}^\dag) p_{0,\sigma,1} 
    \notag \\
    &\qquad + t' (d_{2,\sigma}^\dag +d_{4,\sigma}^\dag) p_{0,\sigma, 0} + \mathrm{H.c.}]
    \notag \\
    &\qquad - J_{\rm H} \bm s_0\cdot \bm s_1,
    \label{Ht_py}
\end{align}
where $d_{j,\sigma}^\dag$ for $j=1,2,3$, and $4$ and $p_{0,\sigma,\mu}^\dag$ for $\mu=0,1$ are creation operators of the $d$-orbital electron and the $p$-orbital one, respectively.
The index $\mu=0,1$ distinguishes two $p$ orbitals.
Note that we employed the periodic boundary condition in Eq.~\eqref{H_py}, $d_{j+4,\sigma}^\dag = d_{j,\sigma}^\dag$.

Here, we apply the DC electric field to this system so that the electric field points perpendicular to the square lattice.
Namely, we assume the following relations among on-site potentials, $V_j$:
\begin{align}
    V_0 &=  V_1- \Delta_{dp} + \ell E, \\
    V_1 &= V_2 = V_3 = V_4,
\end{align}
where $\Delta_{dp}=\Delta_0(\bm 0) =\mathcal E_d-\mathcal E_p\ge 0$ is the eigenenergy difference of the $d$ orbital and the $p$ orbital and
$\ell>0$ denotes the height of the $p$-orbital site from the square-lattice plane.
The electric field $\bm E$ is applied so that
\begin{align}
    \bm E = E \bm e_z,
    \label{E_z}
\end{align}
where $\bm e_z$ is perpendicular to the basal square lattice of the pyramids (Fig.~\ref{fig:pyramids}).
The situation \eqref{E_z} is in contrast to Ref.~\cite{takasan_dc} where the DC electric field is applied within the square-lattice plane.

We designed the Hamiltonian \eqref{H_py} so that the superexchange interaction along the path $1\to 0 \to 2$ is ferromagnetic and that along the path $1 \to 0 \to 3$ is antiferromagnetic.
The other superexchange intearctions are determined similarly.
Applying the generic results for $\mathcal H_{\rm A}$ [Eq.~\eqref{HA}] and $\mathcal H_{\rm F}$ [Eq.~\eqref{HF}] to this pyramid, we obtain an effective spin model with the Hamiltonian,
\begin{align}
    \mathcal H_{\rm py}^{\rm eff}
    &= \sum_{j=1}^4[J_1(\bm E)  \bm S_j \cdot \bm S_{j+1} + J_2 (\bm E) \bm S_j \cdot \bm S_{j+2}],
\end{align}
with coupling constants,
\begin{align}
    J_1(\bm E)
    &= \frac{4|t|^2}{U_d} -8|t'|^4 \biggl( \frac{1}{U_d-U_p  + \Delta_{dp}-
    \ell E} \biggr)^2
    \notag \\
    &\qquad \times \frac{J_{\rm H}}{4[U_d-U_p+\Delta_{dp} -\ell E]^2 - {J_{\rm H}}^2}, 
    \label{J1_py} \\
    J_2(\bm E)
    &=  4|t'|^4 \biggl( \frac{1}{U_d-U_p + \Delta_{dp} - \ell E}\biggr)^2 
    \notag \\
    &\qquad \times \biggl(\frac{2}{2U_d-U_p + 2\Delta_{dp}-2\ell E} + \frac{1}{U_d} \biggr),
    \label{J2_py}
\end{align}
within the fourth-order perturbation expansion about the hoppings.
Note that we implicitly assumed that the second-order direct superexchange and the fourth-order superexchange interactions are comparable with each other.
In other words, we assumed a relation $O(|t|)=O(|t' |^2)$ so that the leading terms of the ``direct superexchange'' and superexchange interactions are of the same order.

\begin{figure}[b!]
    \centering
    \includegraphics[viewport= 0 0 864 504, width=\linewidth]{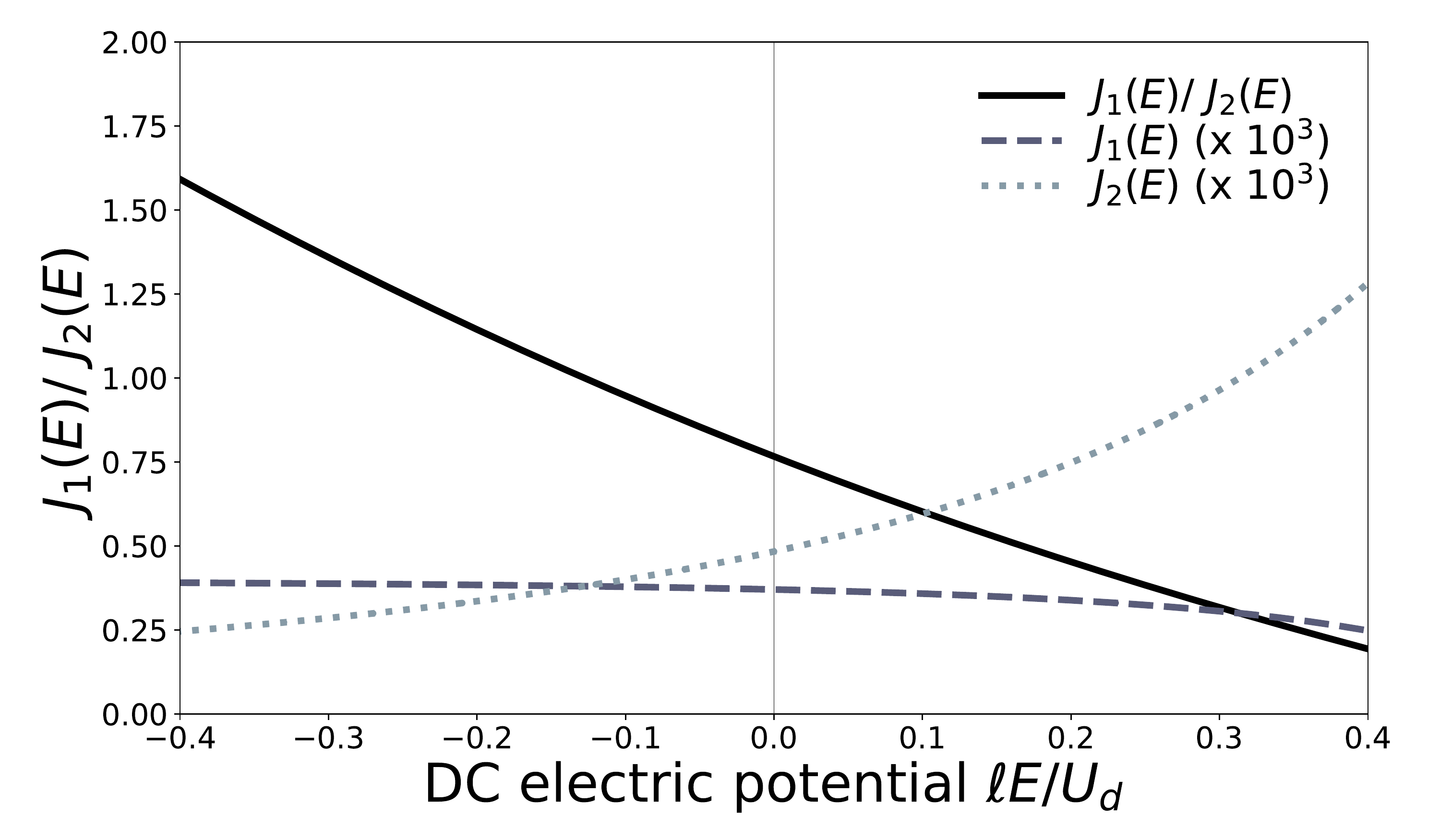}
    \caption{The ratio $J_1(\bm E)/J_2(\bm E)$ is plotted against the DC electric potential $\ell E/U_d$. We used parameters: $U_d=1$, $U_p=J_{\rm H}=0.3$, $\Delta_{dp}=0.5$, $t=0.01$, and $t'=0.1$.}
    \label{fig:ratio_pyramid}
\end{figure}

The first term of $J_1(\bm E)$ is the direct superexchange interaction, and the second term is the ferromagnetic superexchange interaction.
$J_2(\bm E)$ is the antiferromagnetic superexchange interaction.
We can strengthen ($E>0$) or weaken ($E<0$) the exchange interactions thanks to the explicit breaking of the $z\to -z$ inversion symmetry.
Since the electric field \eqref{E_z} is antisymmetric in this inversion, the superexchange interactions \eqref{J1_py} and \eqref{J2_py} can also contain antisymmetric terms under the $z\to -z$ inversion, which is in contrast to the results of Ref.~\cite{takasan_dc}.
Figure~\ref{fig:ratio_pyramid} shows how $E$ controls the ratio $J_1(\bm E)/J_2(\bm E)$.

\subsection{Motivation from experiments}\label{sec:bacd}

We demonstrated that the DC electric field indeed controls the ratio of the nearest-neighbor and next-nearest-neighbor interactions of the toy model \eqref{H_py}.
Since both $J_1$ and $J_2$ are antiferromagnetic in this model, the DC electric field does not drive the system into the spin-nematic phase [Fig.~\ref{fig:sq}~(b)].

Many experiments were reported about $J_1-J_2$ square-lattice quantum magnets with $J_1<0<J_2$, for example,
$\bacd$~\cite{nath_bacdvopo4_2008,kohama_bacdvopo4_2019,skoulatos_bacdvopo4_2019} and $\amo$ ($A=\mathrm{K}, \mathrm{Rb}$)~\cite{ishikawa}.
The former compound has $(J_1, J_2) \approx (-3.6~\mathrm{K}, 3.2~\mathrm{K})$~\cite{nath_bacdvopo4_2008}, which is in the vicinity of the phase boundary between the canted antiferromagnetic phase and the spin-nematic phase.
The latter compound has $(J_1, J_2) \approx (-2~\mathrm{K}, 19~\mathrm{K})$ for $A=\mathrm{K}$ and $(0~\mathrm{K}, 29~\mathrm{K})$ for $A=\mathrm{Rb}$~\cite{ishikawa}, both of which are supposed to be deep in the canted antiferromagnetic phase.

As we show below, these compounds have characteristic crystal structures that allow for the DC electric-field control of $J_1/J_2$ in a different mechanism from that for the pyramid model \eqref{H_py}.
In what follows, we describe the essential characteristics of the crystal structure relevant to our purpose and build a model that simplifies the crystal structure without interfering with the essence.

\begin{figure}[t!]
    \centering
    \includegraphics[viewport = 0 0 1000 1200, width=0.8\linewidth]{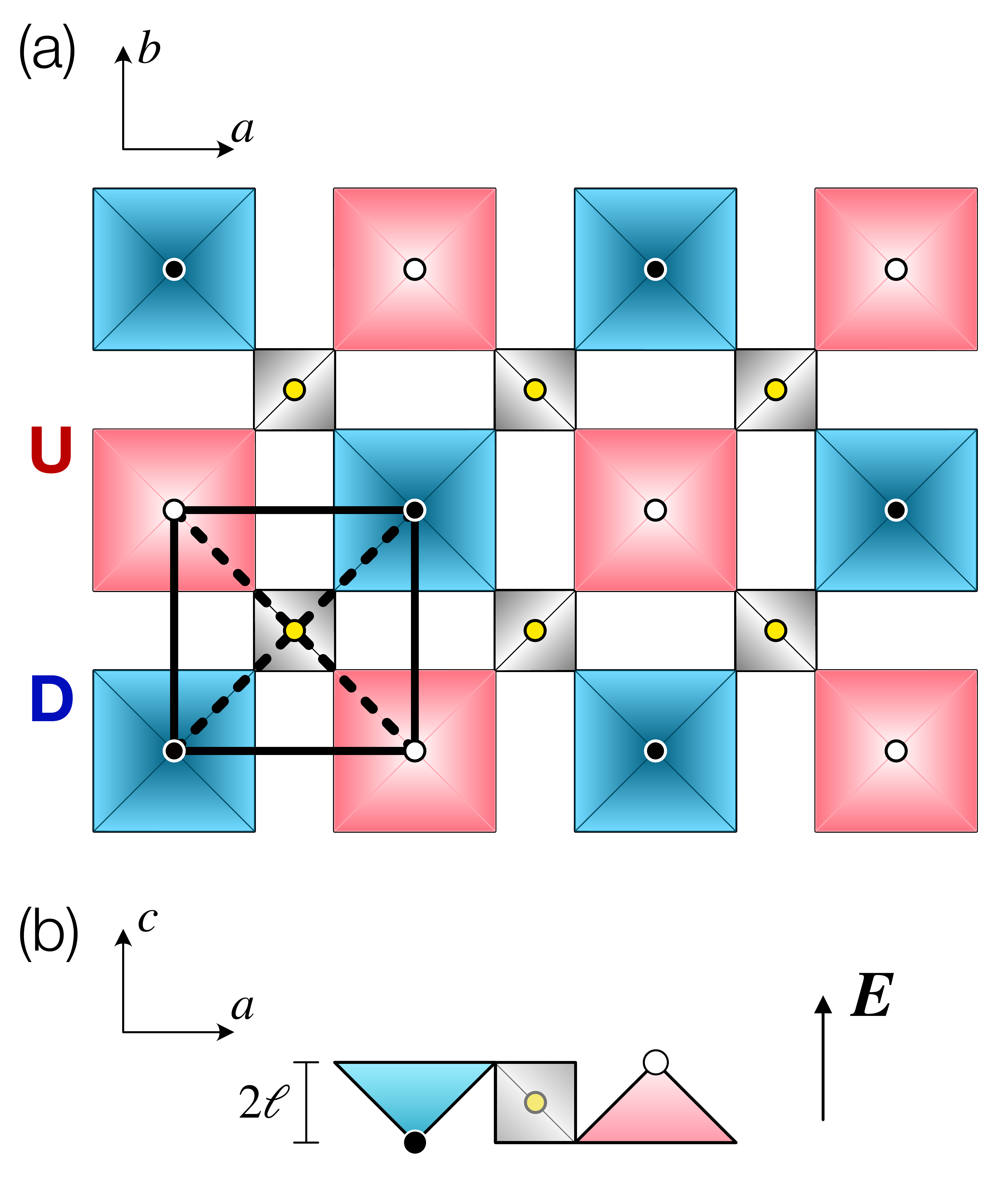}
    \caption{(a) Schematic drawing of the crystal structure of $\bacd$ projected onto the $ab$ plane~\cite{nath_bacdvopo4_2008}.
    Large blue and red squares represent $\mathrm{VO_4}$ pyramids. Small gray squares represent $\mathrm{PO_4}$ tetrahedra. Note that oxygen atoms are located on each corner of the squares, which are omitted. We depicted the atoms important for the later discussion of Sec.~\ref{sec:toy2}.
    The black and white balls represent the magnetic $\mathrm{V}$ ions and the yellow ones represent the nonmagnetic $\mathrm P$ ions.
    The $V$ ions are located at the top of the pyramid, and the $P$ ions are at the center of the tetrahedron.
    (b) When the upward and downward pyramids are viewed along the $b$ axis, we find that their magnetic ions are shifted by $2\ell$ ($>0$) along the $c$ axis. This height difference of the magnetic ions plays a crucial role in response to the DC electric field \eqref{E_uni}.
    }
    \label{fig:crystal_ab}
\end{figure}

Figure~\ref{fig:crystal_ab}~(a) shows a schematic picture of the crystal structure of $\bacd$ viewed along the $c$ axis.
The single layer that hosts the $J_1$--$J_2$ model is projected onto the $ab$ plane in this figure.
Large squares and small gray squares represent $\mathrm{VO_4}$ pyramids and $\mathrm{PO_4}$ tetrahedra, respectively [Fig.~\ref{fig:crystal_ab}~(b)].
Interestingly, there are two kinds of $\mathrm{VO_4}$ pyramids, distinguished by the label U and D in Fig.~\ref{fig:crystal_ab}~(a).

The single layer of the $J_1$--$J_2$ Heisenberg model \eqref{H_sq} is composed of U and D pyramids connected by the $\mathrm{PO_4}$ tetrahedra~\cite{nath_bacdvopo4_2008}.
It is crucial in the following argument that the $c$ coordinate of the magnetic ion depends on whether it belongs to the pyramid U or D [Fig.~\ref{fig:crystal_ab}~(b)].

To change the ratio $J_1/J_2$, we apply a uniform DC electric field, 
\begin{align}
    \bm E =  E \bm e_c,
    \label{E_uni}
\end{align}
to the system, where $\bm e_c$ is the unit vector along the $c$ axis.
If we define the origin of the electric potential at the height of the $P$ ions, the V ion of the upward pyramid feels the electric potential $-E\ell$ and that of the downward pyramid feels $+E\ell$ with $2\ell>0$ being the height of the pyramid.
Though the external field is uniform, the magnetic ion at $\bm r= (r_x,r_y)$ feels a staggered electric potential $(-1)^{r_x+r_y}E\ell$ due to the staggered structure of the pyramids [Fig.~\ref{fig:crystal_ab}~(a)].

\subsection{Simplified electron model}\label{sec:toy2}

\begin{figure}[t!]
    \centering
    \includegraphics[viewport = 0 0 1000 800, width=0.8\linewidth]{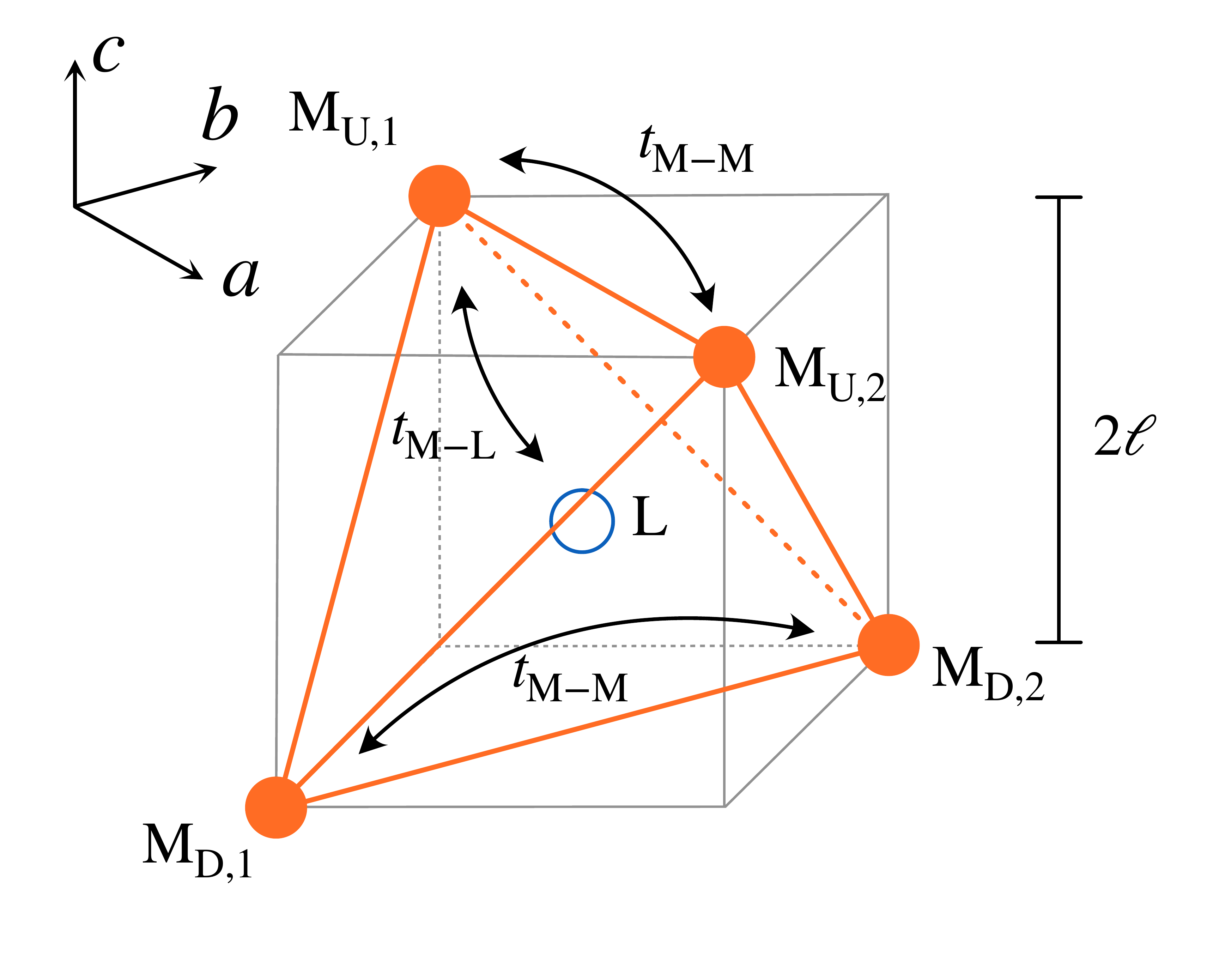}
    \caption{Hopping amplitudes inside a tetrahedron that connects the upper and the lower layers. Magnetic ions (filled balls) on the upward (downward) pyramid are M$_{\rm I,1}$ and M$_{\rm I,2}$ (M$_{\rm II,1}$ and M$_{\rm II,2}$). There is a direct hopping $t_{\rm M-M}$ between two magnetic ions that belong to the same kind of pyramid.
    To hop between two magnetic ions on the two different kinds of pyramids, the electron first hops to the ligand ion (the empty ball) at the center of the tetrahedron and next to the other magnetic ion.
    The hopping amplitude between the magnetic ion and the ligand is given by $t_{\rm M-L}$.
    }
    \label{fig:tetrahedron}
\end{figure}

To investigate DC electric-field effects on $\bacd$, we develop a model that emulates $\bacd$ with a much simpler structure.
The compound $\bacd$ has complicated exchange paths between neighboring magnetic V ions.
Starting from a V ion site, the path goes through an O ion, a P ion, and another O ion before reaching the other V ion.
The miscellaneous O, Ba, and Cd atoms are omitted in Fig.~\ref{fig:crystal_ab}.
In our simplified model, the miscellaneous atoms of $\bacd$ are removed.
Namely, we consider more direct hoppings between the magnetic ions along paths such as $\mathrm{V \to V}$ or $\mathrm{V \to P \to V}$ as shown in Fig.~\ref{fig:tetrahedron} instead of $\mathrm{V\to O\to P \to O \to V}$.
The U and D pyramids are replaced by the magnetic ions with the same labels.
Figure~\ref{fig:tetrahedron} represents the U (D) magnetic ions as $\mathrm M_{\rm U,1}$ and $\mathrm M_{\rm U,2}$ ($\mathrm M_{\rm D,1}$ and $\mathrm M_{\rm D,2}$, respectively).
The ligand site that corresponds to the P ion is also simplified in our model.
The ligand site of our model hosts two $p$ orbitals.
The Coulomb exchange interaction lifts the orbital degeneracy of the $p$ orbitals when they are half occupied.
One of the $p$ orbitals admits hoppings of electrons from/to the $d$ orbitals of the magnetic ions, $M_{\rm U,1}$ and $M_{\rm U,2}$, and the other admits hoppings from/to the $d$ orbitals of the magnetic ions, $M_{\rm D,1}$ and $M_{\rm D,2}$.

Our simplified model inherits the essential structure of $\bacd$.
First, we keep the ligand ion at the center of a tetrahedron formed by the each of four magnetic-ion sites to yield superexchange interactions (Fig.~\ref{fig:tetrahedron}).
Second, our model has the alternating structure that the magnetic-ion site $\bm r=(r_x,r_y)$ is located above (below) the ligand site for $r_x+r_y$ is even (odd) [Fig.~\ref{fig:crystal_ab}~(a)].

\begin{figure}[t!]
    \centering
    \includegraphics[viewport = 0 0 1000 400, width=\linewidth]{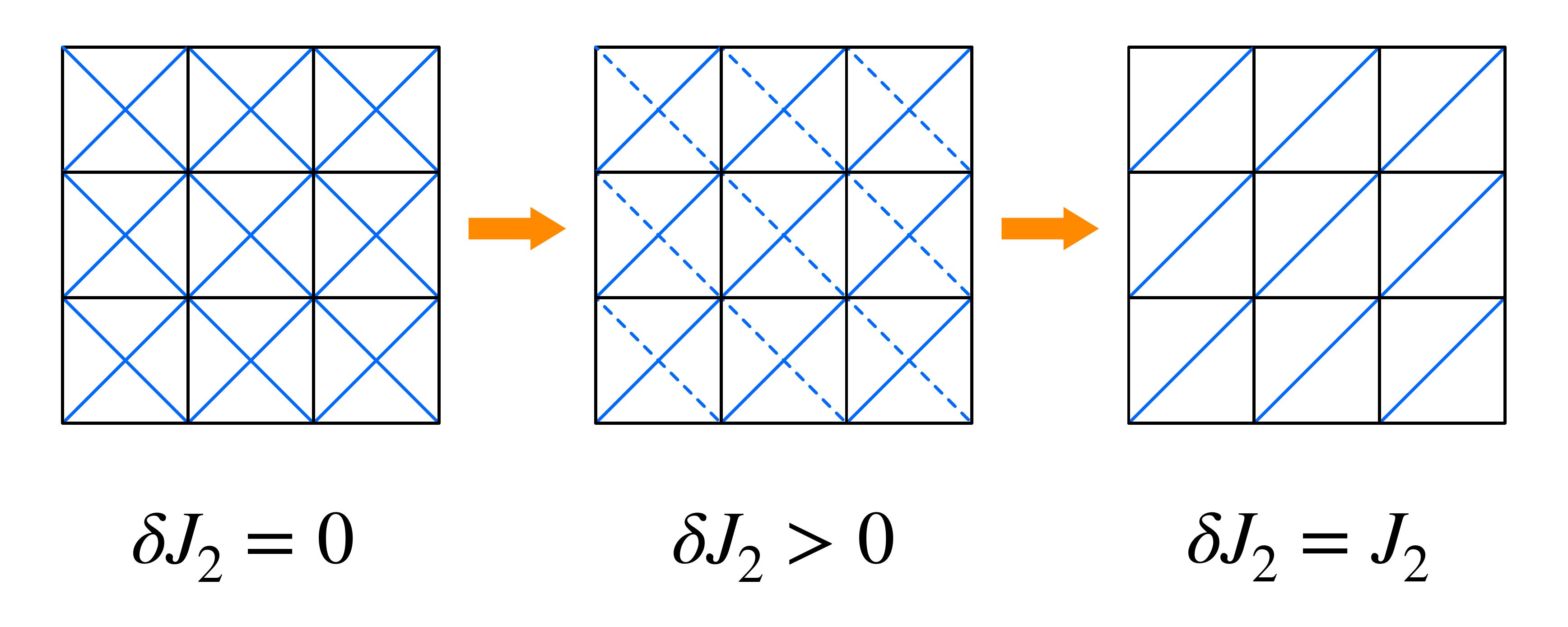}
    \caption{Lattice deformation by the DC electric field. The $J_1$--$J_2$ square-lattice model acquires $\delta J_2(\bm E)>0$ from the DC electric field and eventually turns into the deformed triangular-lattice antiferromagnet when $\delta J_2$ reaches $J_2$.}
    \label{fig:sq2tri}
\end{figure}
\begin{figure}[t!]
    \centering
    \includegraphics[viewport= 0 0 1000 1100, width=\linewidth]{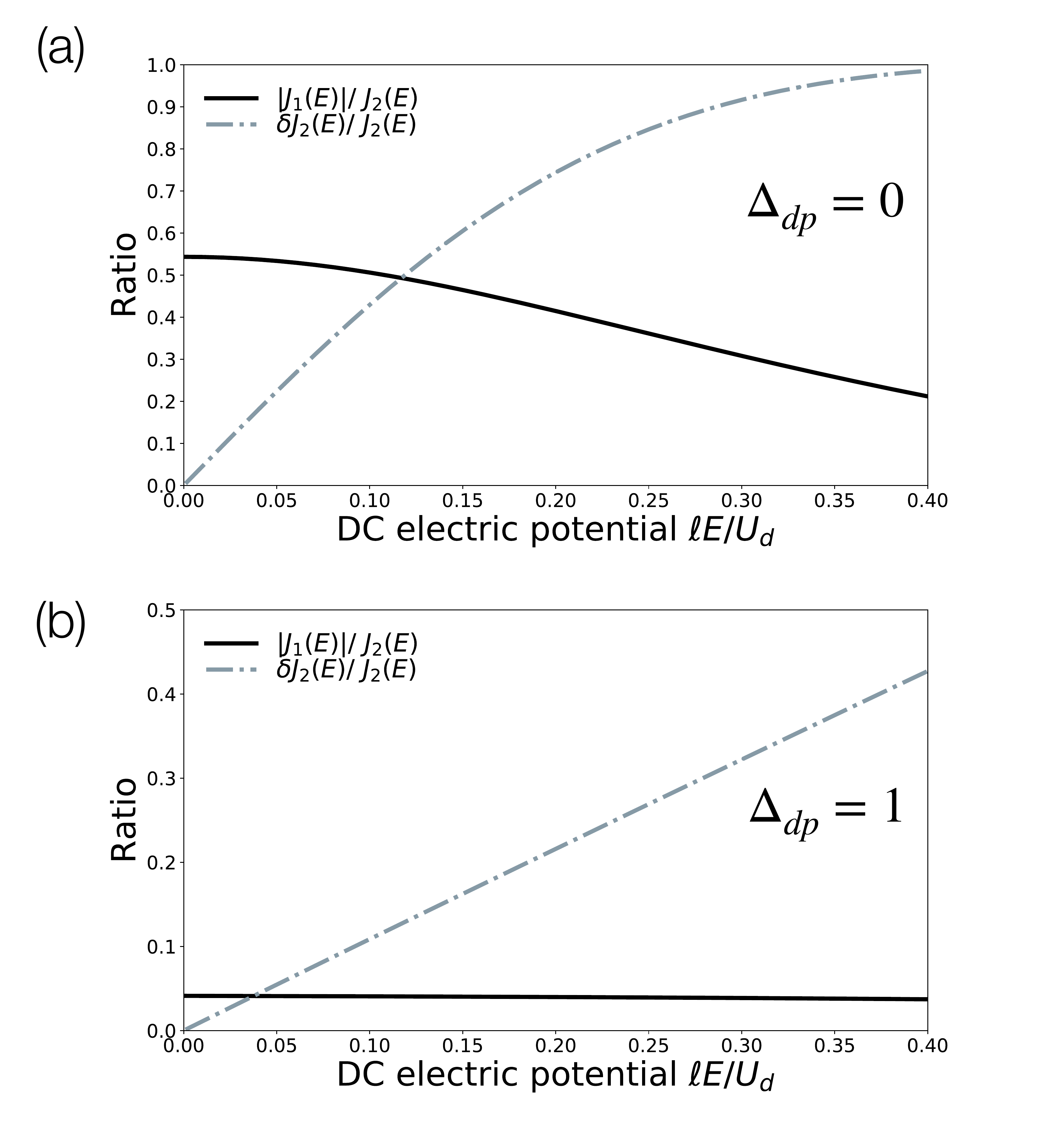}
    \caption{The ratio $J_1(\bm E)/J_2(\bm E)$ of Eqs.~\eqref{J1} and \eqref{J2} are plotted.
    In panels (a) and (b), we take $\Delta_{dp}=0$ and $\Delta_{dp}=1$, respectively.
    The other parameters are taken in common as $U_d=1$, $U_p=0.5$, $J_H=0.5$, $t=0.01$, and $t'=0.12$.
    The ratio $\delta J_2(\bm E)/J_2(\bm E)$ grows with an increase of $\ell E/U_d$ for the two cases, where the square lattice becomes the anisotropic triangular lattice.
    In particular, panel (b) shows that the DC electric field changes only $\delta J_1(\bm E)/J_2(\bm E)$.
    }
    \label{fig:ratio_bacd}
\end{figure}

Following the generic argument of Sec.~\ref{sec:results_gen},
we minimize the number of sites and consider a five-site electron model with a Hamiltonian,
\begin{align}
    \mathcal H_{\rm tetra}
    &= \mathcal H_{U;\mathrm{tetra}} + \mathcal H_{t;\mathrm{tetra}},
    \label{H_tetra} \\
    \mathcal H_{U;\mathrm{tetra}}
    &= U_d \sum_{j=1,2}\sum_{\alpha = \mathrm{U,D}} n_{j,\alpha, \uparrow} n_{j,\alpha, \downarrow}
    \notag \\
    &\qquad + U_p \sum_{\alpha = \mathrm{U,D}} n_{\rm L, \alpha,\uparrow} n_{\rm L,\alpha, \downarrow}
    \notag \\
    &\qquad + \sum_{j=1,2}\sum_{\alpha=\mathrm{U,D}} \sum_{\sigma} V_{\alpha} n_{j,\alpha, \sigma},
    \label{HU_tetra} \\
    \mathcal H_{t;\mathrm{tetra}}
    &= - \sum_{\sigma}\sum_{\alpha=\mathrm{U,D}} \{(t_{\rm M-M} d_{1,\alpha,\sigma}^\dag d_{2,\alpha,\sigma} +\mathrm{H.c.})
    \notag \\
    &\quad  +[ t_{\rm M-L} (d_{1,\alpha,\sigma}^\dag + d_{2,\alpha,\sigma}^\dag) p_{\mathrm{L},\alpha, \sigma}+ \mathrm{H.c.}]
    \}
    \notag \\
    &\qquad -J_{\rm H}\bm s_{\rm U}\cdot \bm s_{\rm D}.
    \label{Ht_tetra}
\end{align}
The model has a $d$ orbital at four magnetic-ion sites at the vertices of the tetrahedron and two orbitals at the ligand site at the center.
$d_{j,\alpha, \sigma}^\dag$ represents the annihilation operator of electron at the magnetic-ion site $j$ with the spin $\sigma$. The index $\alpha = \mathrm{U,D}$ indicates the upward and downward magnetic ions, respectively.
Accordingly, the on-site potential is given by $V_{\rm U} = \bar V - \ell E$ and $V_{\rm D} = \bar V +\ell E$ with an $\bm E$-independent constant $\bar V$.
The $p_{\mathrm{L},\alpha, \sigma}$ operator annihilates the $p$ electron at the ligand site.
As the hopping term \eqref{Ht_tetra} shows, the $p$ orbitals labeled by the index $\alpha=\mathrm{U,D}$ have a hopping term from/to the $d$ orbital with the same $\alpha$ index.
$-J_{\rm H}<0$ is the Coulomb exchange interaction between $\bm s_{\alpha}=\frac 12 \sum_{s,s'} p_{\mathrm{L},\alpha,s}^\dag \bm \sigma^{s,s'} p_{\mathrm{L},\alpha,s'}$ for $\alpha=\mathrm{U,D}$ at the ligand site.

\subsection{Effective spin model}\label{sec:toy2_spin}

The low-energy effective Hamiltonian of the model \eqref{H_tetra} is given by
\begin{align}
    \mathcal H_{\rm tetra}^{\rm eff}
    &= J_1(\bm E) (\bm S_{\rm U,1} +\bm S_{\rm U,2}) \cdot (\bm S_{\rm D,1} + \bm S_{\rm D,2})
    \notag \\
    &\quad + J_2(\bm E) (\bm S_{\rm U,1} \cdot \bm S_{\rm U,2} + \bm S_{\rm D,1} \cdot \bm S_{\rm D,2}) 
    \notag \\
    &\quad +  \delta J_2(\bm E) (\bm S_{\rm U,1} \cdot \bm S_{\rm U,2}  - \bm S_{\rm D,1} \cdot \bm S_{\rm D,2}).
    \label{H_tetra_eff}
\end{align}
We dealt with the hoppings by $t_{\rm M-M}$ and $t_{\rm M-L}$ as perturbations to the remainder of the Hamiltonian.

Let us embed the effective model \eqref{H_tetra_eff} on the tetrahedron into the square lattice of Fig.~\ref{fig:sq}~(a).
$J_1(\bm E)$ and $J_2(\bm E)$ are the nearest-neighbor and the next-nearest-neighbor interactions, respectively.
The perturbation expansion gives a side effect that the effective Hamiltonian \eqref{H_tetra_eff} contains the third interaction of $\delta J_2(\bm E)$, which makes the two diagonal bonds nonequivalent (the middle panel of Fig.~\ref{fig:sq2tri}).
The appearance of nonzero $\delta J_2(\bm E)$ is due to the explicit violation of an inversion symmetry by $\bm E$.
When $\bm E=\bm 0$, the bond connecting $\mathrm{M_{U,1}}$ and $\mathrm{M_{U,2}}$ is symmetrically equivalent to that connecting $\mathrm{M_{D,1}}$ and $\mathrm{M_{D,2}}$.
The DC electric field along the $c$ axis makes these bonds nonequivalent, that is, $\delta J_2(\bm E)\not=0$.
Figure~\ref{fig:ratio_bacd} shows the DC electric-field dependence of the exchange interaction for some parameter sets.

The nearest-neighbor interaction is a superexchange interaction, and the next-nearest-neighbor interaction is mainly a direct superexchange interaction though its subleading terms are superexchange ones as shown below.
Up to the fourth order of the perturbation expansion,  $J_1(\bm E)$, 
\begin{widetext}
$J_2(\bm E)$, and $\delta J_2(\bm E)$ are given by
\begin{align}
    J_1(\bm E) &=  -2|t_{\rm M-L}|^4 \biggl( \frac{1}{U_d-U_p + \Delta_{dp}- \ell E} + \frac{1}{U_d-U_p + \Delta_{dp} +\ell E} \biggr)^2
    \frac{J_{\rm H}}{[2(U_d-U_p + \Delta_{dp})]^2 -{J_{\rm H}}^2},
    \label{J1} \\
    J_2(\bm E) &= \frac{4|t_{\rm M-M}|^2}{U_d}
    +  |t_{\rm M-L}|^4\biggl[ \biggl(\frac{2}{U_d-U_p + \Delta_{dp}-\ell E}\biggr)^2 \frac{1}{2U_d-U_p+2\Delta_{dp}-2\ell E}
    \notag \\
    &\qquad 
    +\biggl(\frac{2}{U_d-U_p+\Delta_{dp}+\ell E}\biggr)^2 \frac{1}{2U_d-U_p+2\Delta_{dp}+2\ell E}
    \notag \\
    &\qquad + \frac{2}{U_d} \biggl\{ \biggl(\frac{1}{U_d-U_p+\Delta_{dp}-\ell E}\biggr)^2 + \biggl(\frac{1}{U_d-U_p+\Delta_{dp}+\ell E}\biggr)^2 \biggr\}
    \biggr],
    \label{J2} \\
    \delta J_2(\bm E)
    &= |t_{\rm M-L}|^4\biggl[ \biggl(\frac{2}{U_d-U_p+\Delta_{dp}-\ell E}\biggr)^2 \frac{1}{2U_d-U_p+2\Delta_{dp}-2\ell E}-\biggl(\frac{2}{U_d-U_p+\Delta_{dp}+\ell E}\biggr)^2 \frac{1}{2U_d-U_p+\Delta_{dp}+2\ell E}
    \notag \\
    &\qquad + \frac{2}{U_d} \biggl\{\biggl(\frac{1}{U_d-U_p+\Delta_{dp}-\ell E}\biggr)^2 - \biggl(\frac{1}{U_d-U_p+\Delta_{dp}+\ell E}\biggr)^2 \biggr\} 
    \biggr].
    \label{dJ2}
\end{align}
\end{widetext}
$J_1(\bm E)$ and $J_2(\bm E)$ are even functions of $E$ as they should be from the symmetry point of view.
By contrast, $\delta J_2(\bm E)$ is an odd function of $E$.
Just like we did in Eqs.~\eqref{J1_py} and \eqref{J2_py}, we assumed that $O(|t_{\rm M-M}|) = O(|t_{\rm M-L}|^2)$ to make the leading terms of the exchange and superexchange interactions comparable.

The side effect of nonzero $\delta J_2(\bm E)$ is an interesting phenomenon by itself.
In the extreme situation of $\delta J_2 = J_2$, the spin model turns into a deformed triangular-lattice Heisenberg antiferromagnet (Fig.~\ref{fig:sq2tri})~\cite{coldea_aniso_tri,starykh_aniso_tri,bishop_aniso_tri}.
Each triangle unit has one $J_2$ bond and two $J_1$ bonds.
We can thus expect that the DC electric field changes the square-lattice antiferromagnet eventually into the triangular-lattice one.
The deformed triangular-lattice antiferromagnet also exhibits a rich phase diagram under magnetic fields~\cite{starykh_aniso_tri}.
It will be interesting to apply the DC electric and magnetic fields simultaneously to the model \eqref{H_tetra}, searching for $\bm E$-induced quantum phase transitions.

We emphasize that the argument in Secs.~\ref{sec:toy2} and \ref{sec:toy2_spin} also applies to $\amo$.
There are minor differences between $\bacd$ and $\amo$ from the DC electric-field viewpoint.
$\amo$ has the Mo ion whose $4f$ orbital contributes to quantum magnetism, however, which can be dealt with on equal footing with the model \eqref{H_tetra}.
The orbital hosting the $S=1/2$ spin does not need to have the $d$-orbital symmetry though we used the notation of the $d$ orbital symbolically.
Our model here as well as the generic ones of Sec.~\ref{sec:models} assume that the orbital at the magnetic-ion site is nondegenerate and has a large on-site repulsion, $U_d>U_p$.
The $4f$ orbital fulfills this assumption.

$\bacd$ and $\amo$ have the same characteristics in response to the DC electric field for the following reasons.
First, $\amo$ also has two kinds of magnetic-ion sites with two different electric potential energies.
Instead of the upward and downward pyramids, $\amo$ has two kinds of octahedra shifted in the opposite direction along the $c$ axis that results in the same staggered electric-field potential $(-1)^{r_x+r_y} \ell E$ as $\bacd$.
Second, the same simplification of the hopping paths to Fig.~\ref{fig:tetrahedron} applies to $\amo$ as well as $\bacd$.
We can expect that the simplification works even better in $\amo$ thanks to the outspread probability distribution of the $4f$ orbital.
We thus conclude that the results \eqref{J1}, \eqref{J2}, and \eqref{dJ2} will thus hold also for $\amo$.

\section{Frustrated ferromagnetic chains}\label{sec:1d}

\subsection{Experimental realizations}

\begin{figure}
    \centering
    \includegraphics[bb = 0 0 1000 600, width=\linewidth]{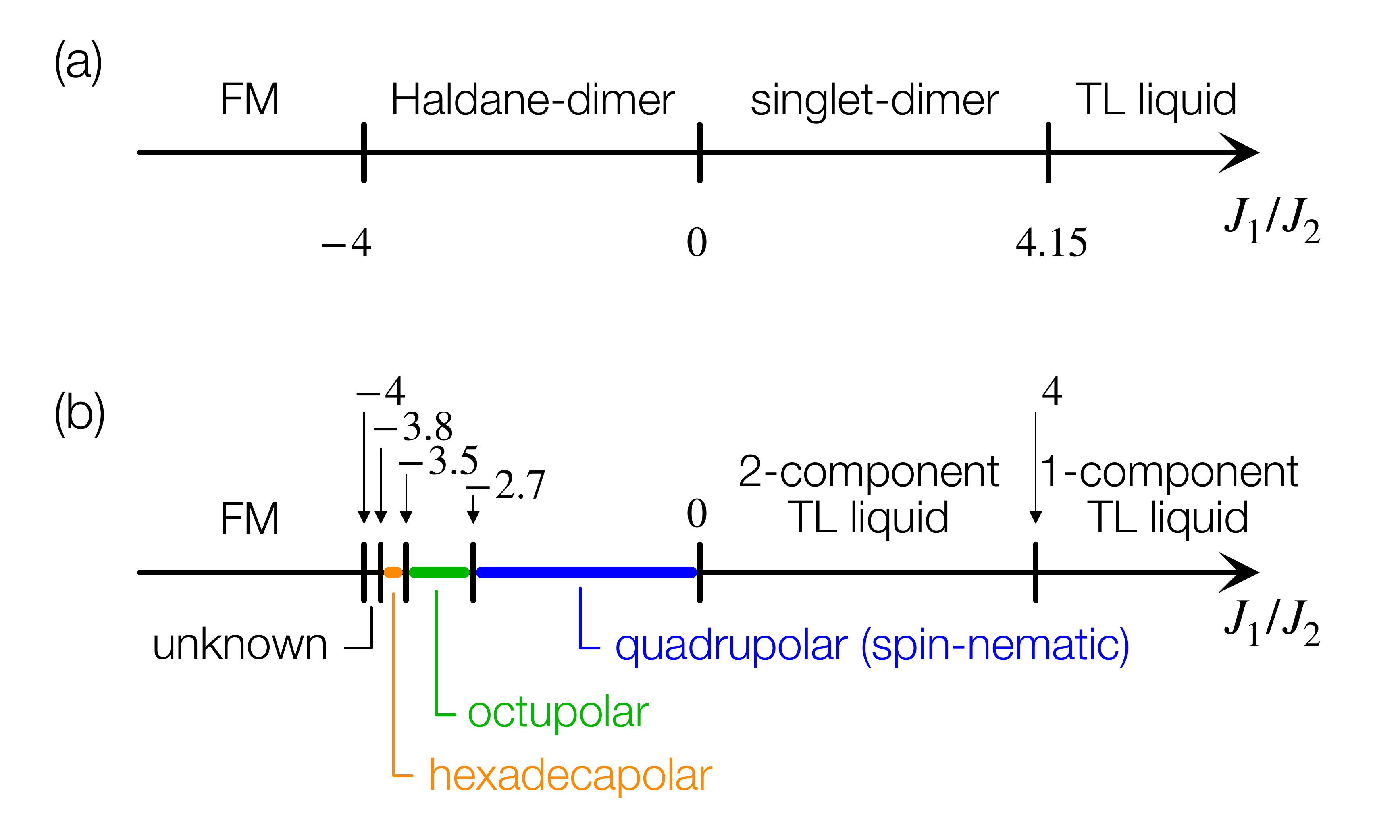}
    \caption{Schematic phase diagrams of the $J_1-J_2$ spin-$1/2$ chain \eqref{H_chain_spin} with $J_2>0$ (a) at zero magnetic field~\cite{furukawa_J1J2_2010b} and (b) near the saturation field~\cite{kecke_multimagnon,hikihara_nematic_2008,sudan_J1J2_chain}.
    (a) The zero-field ground-state phase diagram contains the Tomonaga-Luttinger (TL) liquid, the singlet-dimer, the Haldane-dimer, and the ferromagnetically ordered (FM) phases~\cite{nomura_zigzag_1994, furukawa_J1J2_XXZ}.
    Around the origin, $J_1/J_2=0$, the gapped phases with two different spontaneous dimer orders appear.
    The dimers are formed on nearest-neighbor ($J_1$) bonds in both phases.
    (b) When the magnetic field $h$ is in the vicinity of the saturation field, the phase diagram becomes rich.
    The phase diagram contains quasi-long-range multipolar order phases of spins, the quadrupolar, octapolar, and hexadecapolar phases for $J_1<0$~\cite{kecke_multimagnon, hikihara_nematic_2008}.
    In particular, the quadrupolar phase is known as the (quasi-long-range) spin-nematic phase~\cite{hikihara_nematic_2008, sudan_J1J2_chain, zhitomirsky_licuvo4,sato_nematic_nmr2009, sato_nematic_nmr2011, sato_J1J2_ModPhysLett, sato_nematic_quasi1d,hirobe_nematic_seebeck,morisaku_nematic_laser,furuya_esr_angle}.
    There are one- and two-component TL-liquid phases for $J_1/J_2>4$ and $0<J_1/J_2<4$, respectively~\cite{okunishi_zigzag, hikihara_zigzag}.
    }
    \label{fig:PD_J1J2_chain}
\end{figure}

This section discusses another important frustrated quantum spin system of an $S=1/2$ $J_1$--$J_2$ spin chain described by the Hamiltonian
\begin{align}
    \mathcal H_{\rm chain} &= J_1 \sum_j \bm S_j \cdot \bm S_{j+1} + J_2 \sum_j \bm S_j \cdot \bm S_{j+2} - h \sum_j S_j^z.
    \label{H_chain_spin}
\end{align}
$J_1<0<J_2$ cases~\cite{hikihara_nematic_2008, sudan_J1J2_chain, furukawa_J1J2_2010a, furukawa_J1J2_2010b, sato_J1J2_ModPhysLett,kumar_chain} and  $J_1,J_2>0$ cases~\cite{nomura_zigzag_1994, okunishi_zigzag, hikihara_zigzag} have been intensively studied.  
Figure~\ref{fig:PD_J1J2_chain}~(a) shows the ground-state phase diagram of the model \eqref{H_chain_spin} at $h=0$.
It contains two kinds of dimer phases.
The sign of $J_1$ governs the nature of these dimers.
The dimer is a nonmagnetic singlet type for $J_1>0$ and a magnetic triplet type for $J_1<0$~\cite{furukawa_J1J2_XXZ}.

\begin{figure}[t!]
    \centering
    \includegraphics[viewport = 0 0 1000 1200, width=\linewidth]{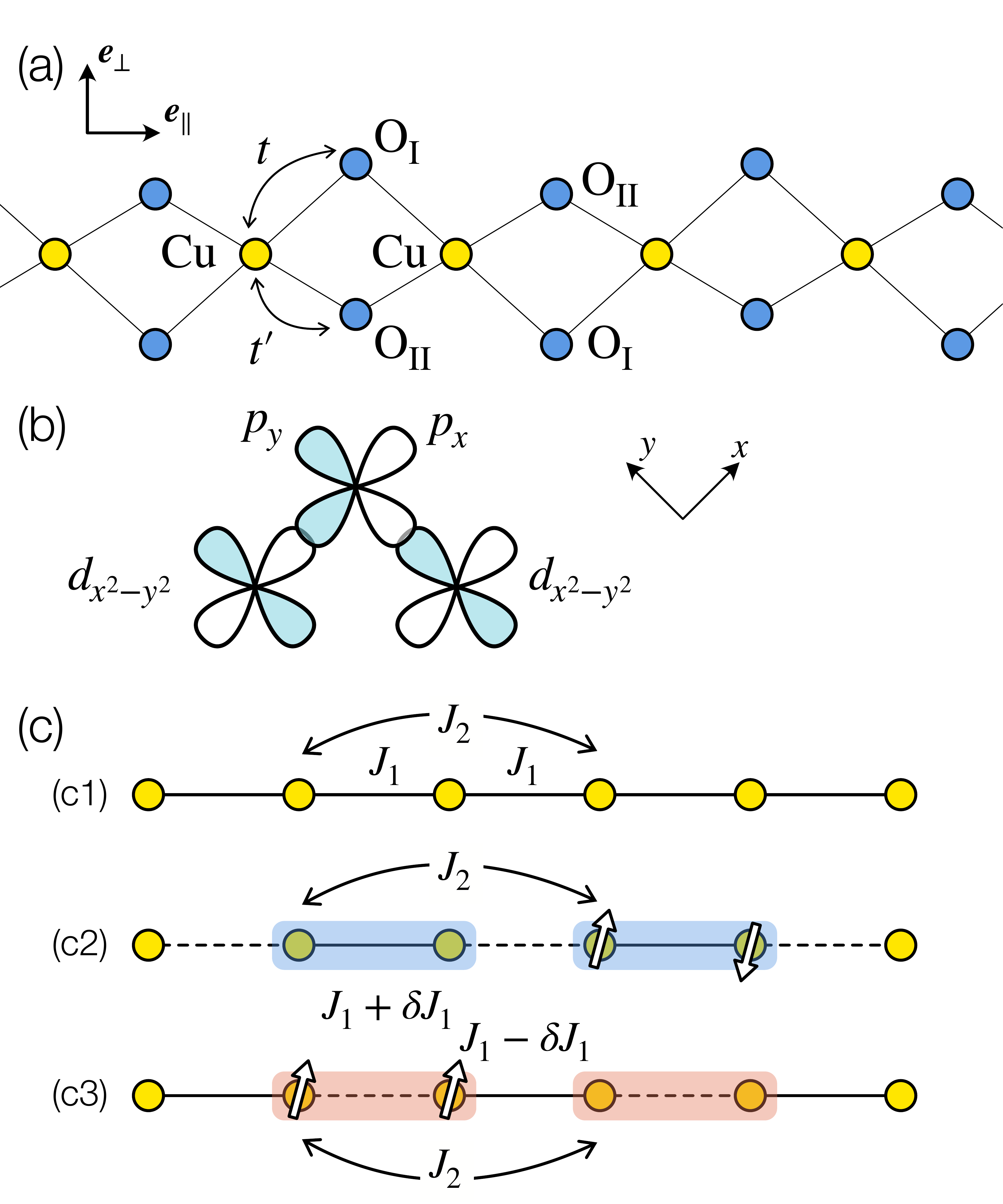}
    \caption{(a) CuO$_2$ chain with two nonequivalent oxygens, O$_{\rm I}$ and $O_{\rm II}$.
    The angles $\angle \mathrm{Cu-O_I-Cu}$ and $\angle \mathrm{Cu-O_{II}-Cu}$ are supposed to be site-independent and differ from each other.
    Accordingly, the CuO$_2$ chain has the twofold screw symmetry in the $\bm e_\parallel$ direction along which the chain extends.
    (b) $3d_{x^2-y^2}$ orbital of Cu ions and  $2p_{x,y}$ orbitals of O ions. When the angle formed by two coppers and one oxygen equals exactly $90^\circ$ as shown here, the electron can hop from the left (right) Cu ion to the $p_x$  ($p_y$, respectively) orbital of the O ion.
    (c) The $J_1$--$J_2$ spin chain and the $\bm E$-induced dimer states are schematically shown. (c1) The $J_1$--$J_2$ chain in the absence of the DC electric field \eqref{E_perp}. (c2) When $J_1>0$, the DC electric field \eqref{E_perp} with $E>0$ yields the bond alternation $\delta J_1>0$, which results in the singlet-dimer ground state. The singlet dimer is formed on the $J_1+\delta J_1$ bond.
    (c3) When $J_1<0$, the DC electric field \eqref{E_perp} with $E>0$ yields the bond alternation $\delta J_1>0$, which results in the Haldane-dimer ground state. The Haldane dimer is formed on the $J_1-\delta J_1$ bond.
    }
    \label{fig:CuO2_chain}
\end{figure}

Some materials with CuO$_2$ chains~\cite{Enderle_2005,Naito_LicuVO4_2007,yasui_cuo2_chain, schapers_cuo2_chain, hase_2004, matsui_2017, nawa_1d_J1J2_InorgChem} are known to be described by 
the $J_1-J_2$ chain with $J_1<0$.
When $-2.7< J_1/J_2 < 0$, this spin chain has the spin nematic phase in the vicinity of the fully polarized phase forced by the strong magnetic field $h$ [Fig.~\ref{fig:PD_J1J2_chain}~(b)].

\subsection{Goodenough-Kanamori rule}

The model \eqref{H_chain_spin} can experimentally be realized, for example, in a CuO$_2$ chain of Fig.~\ref{fig:CuO2_chain}~(a), where the Cu ion carries the $S=1/2$ spin.
Figure~\ref{fig:CuO2_chain}~(a) shows a generic case that two O ions, O$_{\rm I}$ and O$_{\rm II}$, are nonequivalent.
Two O ions can be equivalent but, in what follows, are supposed to be nonequivalent to derive the antiferromagnetic $J_1$ and ferromagnetic $J_1$ from the single model.

The sign $J_1$ is determined by the angle $\angle \mathrm{Cu-O-Cu}$.
The two O ions mediate the nearest-neighbor superexchange interaction.
According to the Goodenough-Kanamori rule~\cite{goodenough,kanamori1,kanamori2}, the superexchange interaction is ferromagnetic when an angle $\angle\mathrm{Cu-O-Cu}$ is $90^\circ$ and antiferromagnetic when the angle is $180^\circ$.
We can easily understand this difference by looking into the $d$ and $p$ orbitals of Fig.~\ref{fig:CuO2_chain}~(b).
When the angle is $90^\circ$, there are no hoppings between the $d_{x^2-y^2}$ orbital of the Cu ion and the $p_x$ orbital of the O ion.
Then, the superexchange interaction is well described by the model \eqref{HF} with the two degenerate $p$ orbitals at the oxygen site.
On the other hand, when the angle is $180^\circ$, the single $p$ orbital allows for hoppings of electrons from the O ion to the two Cu ions on both sides.
Then, the superexchange interaction is modeled by Eq.~\eqref{HA} with the nondegenerate $p$ orbital at the oxygen site [see also Fig.~\ref{fig:HA_levels}~(b) and Fig.~\ref{fig:HF_levels}~(c)].

Generally, the angles are intermediate, when the superexchange process also has an intermediate character of the two models \eqref{HA} and \eqref{HF}.
For simplicity, we consider an ideal situation where the superexchange interaction via the O$_{\mathrm I}$ ion is described by the model \eqref{HF} with $t_0=t_1 = t$ and that via the O$_{\rm II}$ ion is described by the model \eqref{HA} with $t_0=t_1=t'$ [Fig.~\ref{fig:CuO2_chain}~(a)].
It follows from the above modeling that $0<\angle \mathrm{Cu-O_I-Cu} <\angle \mathrm{Cu-O_{II}-Cu} \le 180^\circ$.
We assume that the two oxygens are exactly in the middle of the two nearest-neighbor Cu ions along the chain direction, $\bm e_\parallel$.

\subsection{Twofold screw symmetry}

When the CuO$_2$ plane has two nonequivalent oxygen sites, the CuO$_2$ chain has a twofold screw symmetry along the chain direction, a combination of the discrete translation symmetry in the chain direction and the $\pi$ spatial rotation around that direction.
Let us denote the discrete translation as $T_1$ and the $\pi$ rotation as $R_\pi$.
The CuO$_2$ chain of Fig.~\ref{fig:CuO2_chain}~(a) has neither $T_1$ nor $R_\pi$ symmetries though it has the $T_1R_\pi$ symmetry.
Nevertheless, the low-energy spin-chain model \eqref{H_chain_spin} in the absence of the electric field has both the $T_1$  and  $R_\pi$ symmetries instead when the electric potentials at the two oxygen sites are exactly balanced.

The DC electric field can violate this balance, replacing the emergent high symmetry of the spin chain by the original $T_1R_\pi$ one of the CuO$_2$ chain.
Let us apply the following DC electric field to the system,
\begin{align}
    \bm E = E\bm e_{\perp},
    \label{E_perp}
\end{align}
along the direction perpendicular to the CuO$_2$ chain and on the CuO$_2$ plane [Fig.~\ref{fig:CuO2_chain}~(a)].
We denote the electric potentials at the Cu, O$_{\mathrm I}$, and O$_{\mathrm{II}}$ sites as $0$, $\pm \ell_{\rm I}E$, and $\mp \ell_{\rm II}E$, respectively.
The sign of the latter two potentials is positive if the O ion is below the Cu ion in the $\bm e_\perp$ direction and negative if the O ion is above the Cu ion.
We can assume $\ell_{\rm I} > \ell_{\rm II}$ to be consistent with the assumption of $\angle \mathrm{Cu-O_I-Cu} < \angle \mathrm{Cu-O_{II}-Cu}$.

\begin{figure}[b!]
    \centering
    \includegraphics[viewport = 0 0 864 1008, width=\linewidth]{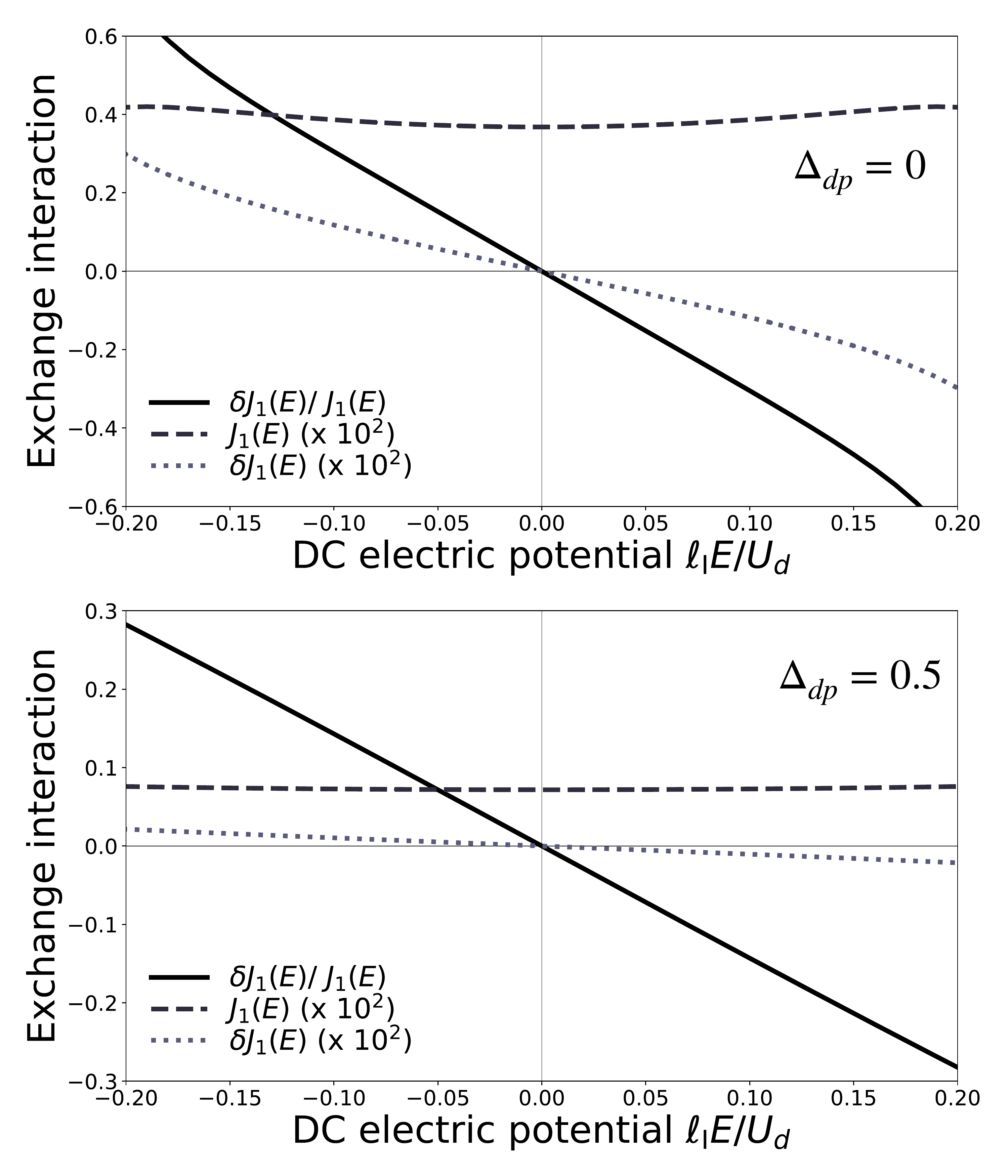}
    \caption{The $\bm E$-induced bond alternation $\delta J_1(\bm E)$ is plotted for the DC electric potential $\ell_{\rm I} E$ for the $J_1, J_2>0$ case with parameters, $U_d=1$, $U_p=0.5$, $J_H = 0.5$, $t=0.04$, $t'=0.1$, and $\ell_{\rm II}/\ell_{\rm I} = 0.60$.
    The ratio $\ell_{\rm II}/\ell_{\rm I}$ is determined from the data for $\mathrm{KCuMoO_4(OH)}$~\cite{nawa_1d_J1J2_InorgChem}.
    We took $\Delta_{dp}=0$ for the upper panel and $\Delta_{dp}=0.5$ for the lower panel.}
    \label{fig:bondalt_chain_AFM}
\end{figure}

\begin{figure}[b!]
    \centering
    \includegraphics[viewport= 0 0 864 1008, width=\linewidth]{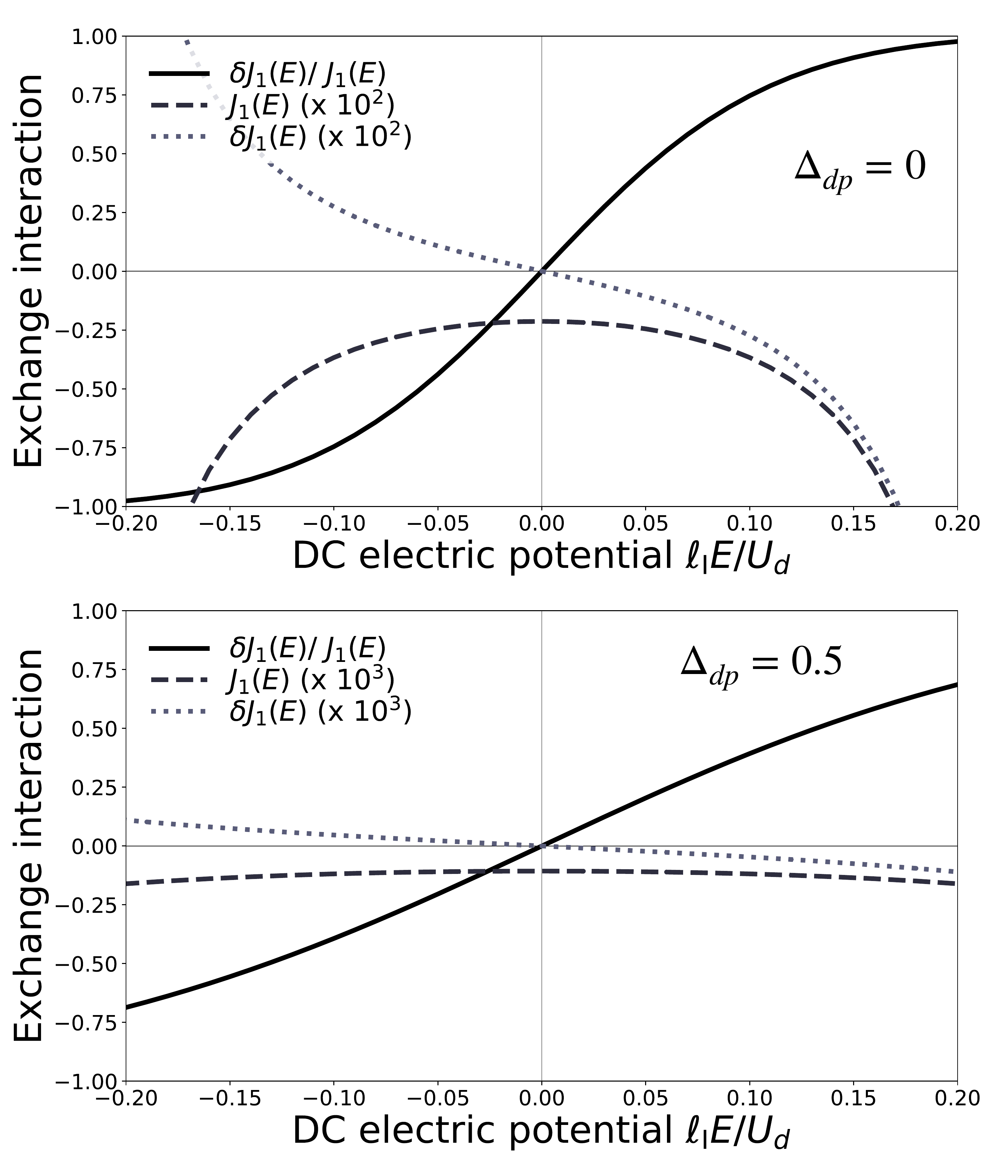}
    \caption{The $\bm E$-induced bond alternation $\delta J_1(\bm E)$ is plotted for the DC electric potential $\ell_{\rm I} E$ for the $J_1<0<J_2$ case with parameters, $U_d=1$, $U_p=0.5$, $J_H = 0.5$, $t=0.1$, $t'=0.01$, and $\ell_{\rm II}/\ell_{\rm I} = 0.81$.
    The ratio $\ell_{\rm II}/\ell_{\rm I}$ is determined from the data for $\mathrm{NaCuMoO_4(OH)}$~\cite{nawa_1d_J1J2_InorgChem}.
    We took $\Delta_{dp}=0$ for the upper panel and $\Delta_{dp}=0.5$ for the lower panel.
    }
    \label{fig:bondalt_chain_FM}
\end{figure}

The effective spin-chain model under the DC electric field \eqref{E_perp} has the following Hamiltonian,
\begin{align}
    \mathcal H_{\rm chain}
    &= J_1 (\bm E) \sum_j \bm S_j \cdot \bm S_{j+1} + J_2(\bm E) \sum_j \bm S_j \cdot \bm S_{j+1}
    \notag \\
    &\qquad - h \sum_j S_j^z + \delta J_1(\bm E) \sum_j (-1)^j \bm S_j \cdot \bm S_{j+1} ,
    \label{H_chain_spin_E}
\end{align}
where the last term, called the bond alternation, is a direct consequence of the imbalance of the electric potentials of O$_{\rm I}$ and O$_{\rm II}$.
The nearest-neighbor interactions, $J_1(\bm E)$ and $\delta J_1(\bm E)$, are given by
\begin{widetext}
\begin{align}
    J_1(\bm E) &= -4|t|^4 \biggl[\biggl(\frac{1}{U_d-U_p+\Delta_{dp} -\ell_{\rm I}E} \biggr)^2 \frac{J_{\rm H}}{4(U_d-U_p+\Delta_{dp} - \ell_{\rm I}E)^2-{J_{\rm H}}^2} 
    \notag \\
    &\qquad + \biggl(\frac{1}{U_d-U_p +\Delta_{dp}+\ell_{\rm I}E} \biggr)^2 \frac{J_{\rm H}}{4(U_d-U_p +\Delta_{dp} + \ell_{\rm I}E)^2-{J_{\rm H}}^2}
    \biggr]
    \notag \\
    &\quad + 2|t'|^4 \biggl[\biggl(\frac{1}{U_d-U_p + +\Delta_{dp}+\ell_{\rm II}E} \biggr)^2 \biggl(\frac{2}{2U_d-U_p +2\Delta_{dp}+ 2\ell_{\rm II}E} + \frac{1}{U_d} \biggr) 
    \notag \\
    &\qquad + \biggl(\frac{1}{U_d-U_p+\Delta_{dp} - \ell_{\rm II}E} \biggr)^2 \biggl(\frac{2}{2U_d-U_p +2\Delta_{dp}-2\ell_{\rm II}E} + \frac{1}{U_d} \biggr)
    \biggr], 
    \label{J1_chain} \\
    \delta J_1(\bm E)
    &=-4|t|^4 \biggl[\biggl(\frac{1}{U_d-U_p +\Delta_{dp}-\ell_{\rm I}E} \biggr)^2 \frac{J_{\rm H}}{4(U_d-U_p +\Delta_{dp}- \ell_{\rm I}E)^2-{J_{\rm H}}^2} 
    \notag \\
    &\qquad - \biggl(\frac{1}{U_d-U_p+\Delta_{dp} +\ell_{\rm I}E} \biggr)^2 \frac{J_{\rm H}}{4(U_d-U_p +\Delta_{dp}+\ell_{\rm I}E)^2-{J_{\rm H}}^2}
    \biggr]
    \notag \\
    &\quad + 2|t'|^4 \biggl[\biggl(\frac{1}{U_d-U_p +\Delta_{dp}+ \ell_{\rm II}E} \biggr)^2 \biggl(\frac{2}{2U_d-U_p+2\Delta_{dp} + 2\ell_{\rm II}E} + \frac{1}{U_d} \biggr)
    \notag \\
    &\qquad 
    - \biggl(\frac{1}{U_d-U_p +\Delta_{dp}
    - \ell_{\rm II}E} \biggr)^2 \biggl(\frac{2}{2U_d-U_p +2\Delta_{dp}-2\ell_{\rm II}E} + \frac{1}{U_d} \biggr)
    \biggr].
    \label{dJ1_chain}
\end{align}
\end{widetext}
We emphasize that Eqs.~\eqref{J1_chain} and \eqref{dJ1_chain} rely on the assumption that the superexchange mediated by the O$_{\rm I}$ ion is the ferromagnetic one [Eq.~\eqref{JF}] and the one by the O$_{\rm II}$ ion is the antiferromagnetic one [Eq.~\eqref{JA}].
$J_1(\bm E)$ and $\delta J_1(\bm E)$ are even and odd functions of $E$, respectively (Figs.~\ref{fig:bondalt_chain_AFM} and \ref{fig:bondalt_chain_FM}).
This $E$ dependence is consistent with the twofold screw symmetry of the CuO$_2$ chain.
For instance, since $R_\pi$ maps $\bm E=E\bm e_\perp$ to $-\bm E$, the $T_1R_\pi$ operator maps the bond alternation $\delta J_1(\bm E)\sum_j (-1)^j \bm S_j \cdot \bm S_{j+1}$ to $\delta J_1(-\bm E) \sum_j (-1)^j \bm S_{j+1}\cdot \bm S_{j+2} = \delta J_1(\bm E) \sum_j (-1)^{j+1}\bm S_{j+1} \cdot \bm S_{j+2}$.
Hence, the $\bm E$-induced bond alternation has the twofold screw symmetry though it breaks the $T_1$ symmetry.

Note that the right-hand side of Eq.~\eqref{dJ1_chain} does not vanish for $\ell_{\rm I} = \ell_{\rm II}$, though it should from the symmetry viewpoint.
This inconsistency comes from the initial assumption that we made. We assumed that the superexchange process via the O$_{\rm I}$ ion and that via the O$_{\rm II}$ ion are supposed to be nonequivalent from the beginning.
Accordingly, Eqs.~\eqref{J1_chain} and \eqref{dJ1_chain} hold when the difference $\ell_{\rm I}- \ell_{\rm II}$ is large enough to justify the nonequivalent superexchange processes.
This condition $\ell_{\rm I, II}$ is convenient for our purpose of the DC electric-field control.

$J_2(\bm E)$ depends on $\bm E$ in a complex manner.
However, its DC electric-field dependence is less important than the induction of the bond alternation because the latter is much more relevant in the renormalization-group sense~\cite{giamarchi_book, gogolin_textbook}.

Before describing the $\bm E$-induced bond alternation effects, we comment on the field direction.
If we apply the DC electric field along the chain direction ($\bm e_\parallel$ of Fig.~\ref{fig:CuO2_chain}), the DC electric field does not yield the bond alternation because it keeps the balance of the electric potentials at the two oxygen sites.
The DC electric field $\bm E \parallel \bm e_\parallel$ purely changes the ratio $J_1(\bm E)/J_2(\bm E)$, keeping $\delta J_1(\bm E)=0$, which will be relevant to studies of the quasi-long-range spin-nematic phase of $J_1-J_2$ spin chains.

\subsection{Dimerization induced by DC electric fields}

\subsubsection{$J_1>0$: singlet dimers}

When both $J_1$ and $J_2$ are positive, the ground-state phase diagram for $E=h=0$ contains two phases: a Tomonaga-Luttinger (TL) liquid phase~\cite{giamarchi_book, gogolin_textbook} and a spontaneously dimerized phase.
A quantum critical point, $J_2/J_1 = \alpha_c\approx 0.2411$~\cite{okamoto_J1J2_critical, white_J1J2_chain}, separates these two phases:
the TL-liquid phase for $J_2/J_1 \le \alpha_c$ and the spontaneously dimerized phase for $\alpha_c < J_2/J_1$.
The former is gapless, and the latter is gapped.
Turning on the DC electric field \eqref{E_perp}, we can introduce the bond alternation to the $J_1$--$J_2$ spin chain.
Trivially, the bond alternation turns the spontaneously dimerized phase into an induced dimerized phase, which we call an $\bm E$-induced dimerized phase, by lifting the ground-state degeneracy of the spontaneously dimerized phase [Fig.~\ref{fig:CuO2_chain}~(c2)].
More interestingly, the bond alternation drives the TL liquid into the $\bm E$-induced dimerized phase.

The dimer order parameter $D(\bm E)$ characterizes the $\bm E$-induced dimer phase.
When the ground state belongs to the TL-liquid phase for $\bm E=\bm 0$, 
the dimer order parameter is obviously zero: $D(\bm 0)=0$.
As soon as we apply the DC electric field to the TL liquid, we can open the excitation gap.
The gap opening due to the bond alternation is well described by the sine-Gordon theory with the following Hamiltonian~\cite{giamarchi_book}.
\begin{align}
    \mathcal H_{\rm chain}
    &\approx \frac{v}{2}\int dx \{(\partial_x\theta)^2 + (\partial_x\phi)^2\}
    \notag \\
    &\qquad + g(\bm E) \int dx \, \cos(\sqrt{2\pi}\phi),
    \label{H_SG}
\end{align}
where $v$ is the spinon velocity and $g(\bm E) \propto \delta J_1(\bm E)\propto |E|$ is the coupling constant that leads to the spin gap.
$\phi$ and $\theta$ are related to the spin operator $\bm S_j$ as~\cite{giamarchi_book, gogolin_textbook, hikihara_coeff_a1b0b1}
\begin{align}
    S_j^z &= \frac 1{\sqrt{2\pi}} \partial_x \phi + (-1)^j a_1 \cos(\sqrt{2\pi}\phi), \\
    S_j^+ &= e^{-i\sqrt{2\pi}\theta} [b_0(-1)^j + b_1 \cos(\sqrt{2\pi}\phi)],
\end{align}
with nonuniversal constants $a_1$, $b_0$, and $b_1$.
It is exactly known that the sine-Gordon theory \eqref{H_SG} gives the following the dimer order parameter $D(\bm E)$~\cite{lukyanov_mass}:
\begin{align}
    D(\bm E) &= \frac{1}{L} \sum_{j=1}^L \braket{(-1)^j \bm S_j \cdot \bm S_{j+1}}
    \notag \\
    &= 3d_z \int dx \, \braket{\cos(\sqrt{2\pi}\phi)}
    \notag \\
    &= 3d_z \biggl[\frac{\Delta(\bm E) \sqrt{\pi} \Gamma(2/3)}{v \Gamma(1/6)} \biggr]^{1/2} \exp\biggl[\int_0^\infty \frac{dt}{t} \biggl\{ - \frac 12 e^{-2t}
    \notag \\
    &\qquad + \frac{\sinh^2(t/2)}{2\sinh(t/4) \sinh t \cosh(3t/4)}
    \biggr\}
    \biggr].
    \label{D_SG}
\end{align}
where the parameter $d_z$ relates the spin operator and the $\phi$ boson via $(-1)^j S_j^p S_{j+1}^p = d_z \sin(\sqrt{2\pi}\phi) + \cdots$ for $p=x,y,z$ in the absence of the DC electric field~\cite{takayoshi_coeff,hikihara_coeff_dimer} and $\Gamma(z)$ is the gamma function.
The parameter $\Delta(\bm E)$ represents the lowest-energy excitation gap induced by $\bm E$.
Note that $v$ and $d_z$ are calculated in the model with $\bm E=\bm 0$ since the $\bm E$ dependence of these quantities merely leads to hardly observable corrections to the right hand side of Eq.~\eqref{D_SG}, which we will discuss later.
The field-theoretical result \eqref{D_SG} becomes  accurate when the ratio $J_1(\bm 0)/J_2(\bm 0)$ is close to the critical value $\alpha_c$.
When $J_1(\bm 0)/J_2(\bm 0) \approx \alpha_c$, the parameters $v$ and $d_z$ are given by $v=1.174J_1(\bm 0)$~\cite{okamoto_J1J2} and $d_z = 0.182$~\cite{takayoshi_coeff}.

\begin{figure}[b!]
    \centering
    \includegraphics[viewport= 0 0 1000 1100, width=\linewidth]{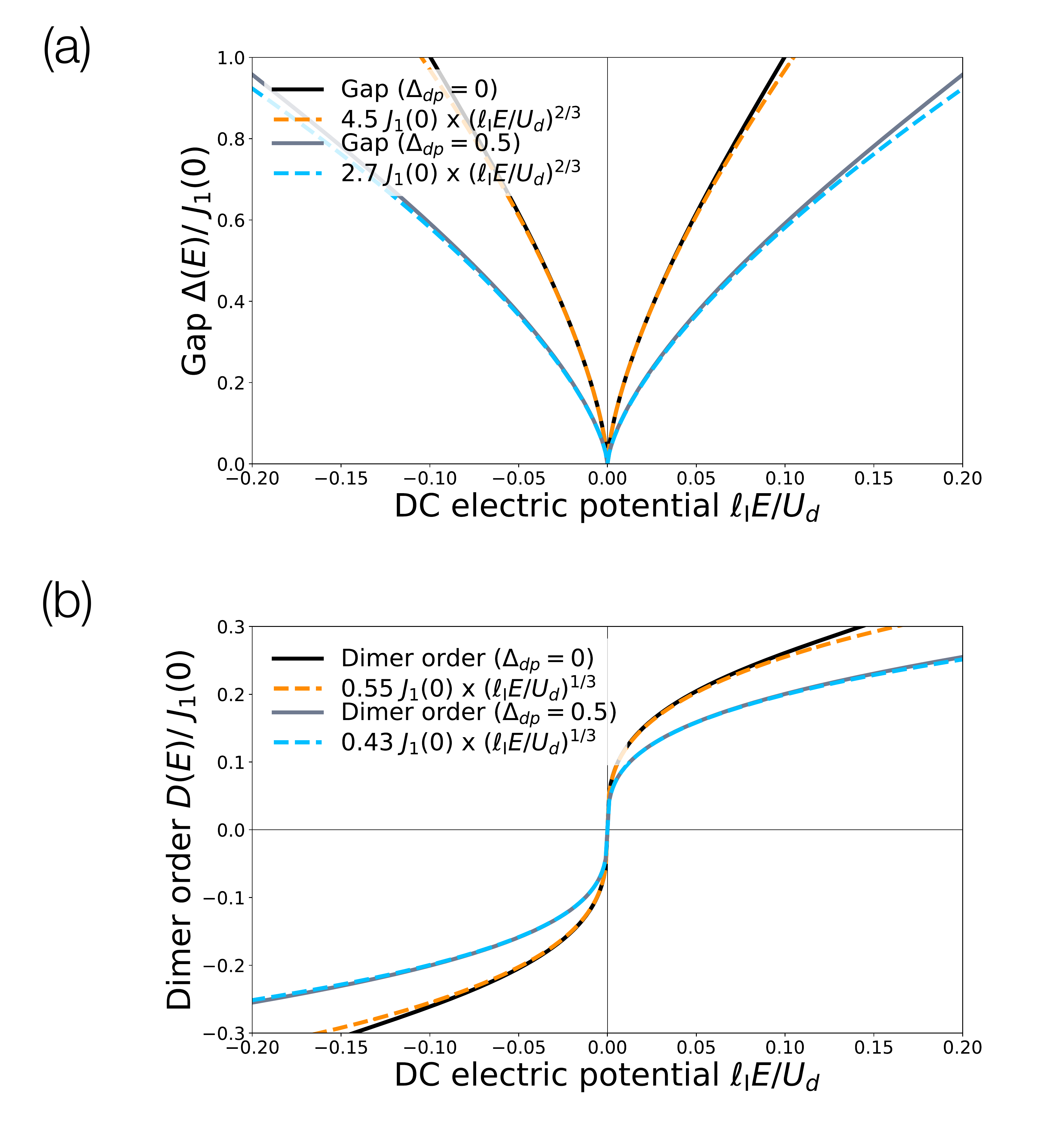}
    \caption{(a) The $\bm E$-induced excitation gap $\Delta(\bm E)$ \eqref{gap_SG} \eqref{E_perp} (solid lines) is plotted and compared with a simple power-law function $\propto (\ell_{\rm I}E/U_d)^{2/3}$ (dashed lines). (b) The $\bm E$-induced dimer order parameter \eqref{D_SG} (the solid lines) is plotted and compared with a simple power-law function $\propto (\ell_{\rm I}E/U_d)^{1/3}$ (the dashed lines).
    We used the same parameters as those in Fig.~\ref{fig:bondalt_chain_AFM}.
    }
    \label{fig:gap_dimer_chain}
\end{figure}

One can obtain the exact excitation gap, $\Delta(\bm E)$, of the sine-Gordon theory \eqref{H_SG}~\cite{lukyanov_mass,lukyanov_sg,zamolodchikov_mass}:
\begin{align}
    \Delta (\bm E)
    &= \frac{2v}{\sqrt{\pi}} \frac{\Gamma(1/6)}{\Gamma(2/3)} \biggl( \frac{3d_z\pi}{2} \frac{\Gamma(3/4)}{\Gamma(1/4)} \frac{|\delta J_1(\bm E)| }{v} \biggr)^{2/3}
    \notag \\
    &\propto \biggl(\frac{|\delta J_1(\bm E)|}{J_1(\bm 0)} \biggr)^{2/3}.
    \label{gap_SG}
\end{align}
Equation~\eqref{dJ1_chain} tells us that the $\bm E$-induced bond alternation $\delta J_1(\bm E)$ is proportional to $E$ for $|E| \ll (U_d-U_p)/\ell_{\rm I}$.
Therefore, we obtain
\begin{align}
    \Delta(\bm E) &\propto |E|^{2/3},
    \label{gap_E2/3} \\
    D(\bm E) &\propto |E|^{1/3}.
    \label{D_E1/3}
\end{align}
The gap \eqref{gap_SG} and the dimer order \eqref{D_SG} are plotted in Fig.~\ref{fig:gap_dimer_chain}.
We can see the power-law behaviors \eqref{gap_E2/3} and \eqref{D_E1/3} near $\bm E=\bm 0$.
The power laws \eqref{gap_E2/3} and \eqref{D_E1/3} hold near the origin but break down at some field strength, say, for $|\ell_{\rm I} E/U_d| \approx 0.1$ when $\Delta_{dp}=0$ (Fig.~\ref{fig:gap_dimer_chain}), because of a nonlinear $E$ dependence of $J_1(\bm E)$ and  $\delta J_1(\bm E)$ [Eqs.~\eqref{J1_chain} and \eqref{dJ1_chain}].
Since $U_d$ is typically $\sim 5~\mathrm{eV}$ for Cu~\cite{antonides_3dTM,yin_3dTM,fujimori_3dTM}, we can expect this nonlinear effect for $|E| \gtrsim 34~\mathrm{MV/cm}$, which is extremely strong.
The nonlinear effect will be hardly observable.
At the same time, this estimation of $|E|$ justifies our perturbative treatment of the DC electric field in the field-theoretical argument such as Eq.~\eqref{D_SG}.

This $\bm E$-induced dimerization differs significantly from a magnetic-field-induced dimerization that one of the authors found recently~\cite{furuya_screw}.
Though a fourfold screw symmetry is essential to cause the magnetic-field-induced dimerization, such a complication is not required in the $\bm E$-induced dimerization.

\subsubsection{$J_1<0$: Haldane dimers}

When $J_1<0<J_2$, the ground-state phase diagram at $E=h=0$ contains the ferromagnetically ordered phase, a vector-chiral phase, and the Haldane-dimer phase~\cite{furukawa_J1J2_XXZ}.
When $0<J_1/J_2 \ll 1$, the ground state is expected to belong to the Haldane-dimer phase, where the triplet dimer is formed on the nearest-neighbor bond.
Reference~\cite{furukawa_J1J2_XXZ} also discusses the effects of the bond alternation on the ground-state phase diagram of the $J_1$--$J_2$ chain.
The bond alternation weakens the geometrical frustration between the nearest-neighbor and the next-nearest-neighbor exchange interactions.
While the $J_1$--$J_2$ spin chain with $J_1<0<J_2$ has the spontaneous Haldane-dimer phase with the twofold ground-state degeneracy, the $J_1$--$J_2$--$\delta J_1$ chain has the induced Haldane-dimer phase with the unique and gapped ground state.
In the latter phase, the triplet dimer is formed on the bond with the stronger nearest-neighbor exchange interaction, namely, on the $J_1+\delta J_1$ bond for $E>0$ and on the $J_1-\delta J_1$ bond for $E<0$ [Fig.~\ref{fig:CuO2_chain}~(c3)].

Differently from the $J_1>0$ case, the DC electric field does not open the spin gap for large $|J_1|$.
When $J_1<0$ and $|J_1|$ is large enough, the ground state has the spontaneous ferromagnetic order accompanied by a gapless nonrelativistic Nambu-Goldstone mode, which is robust against the small bond alternation.
When $|J_1|$ is much smaller than $J_2$, the DC electric field opens the gap exactly in the same way for both the $J_1<0$ and $J_1>0$ cases.

When $\max\{|J_1|, |\delta J_1|\} \ll J_2$, we can regard the $J_1$--$J_2$ spin chain as weakly coupled spin chains.
Bosonizing each spin chain, we obtain an excitation gap~\cite{shelton_ladder},
\begin{align}
    \Delta(\bm E) \propto |\delta J_1(\bm E)| \propto |E|,
\end{align}
instead of Eq.~\eqref{gap_SG}.
Accordingly, the dimer order parameter follows
\begin{align}
    D(\bm E) \approx |\delta J_1(\bm E)|^{1/2}.
\end{align}
Therefore, we find for weak $E \ll (U_d-U_p)/\ell_{\rm I}$ that
\begin{align}
    \Delta(\bm E) &\propto |E|,
    \label{gap_E1}
    \\
    D(\bm E) &\propto |E|^{1/2}.
    \label{D_E1/2}
\end{align}
Though we obtained different power laws from Eqs.~\eqref{gap_E2/3} and \eqref{D_E1/3}, it is not attributed to the sign of $J_1$.
In fact, even when $J_1>0$, we will find the power laws \eqref{gap_E1} and \eqref{D_E1/2} if $J_1$ and $|\delta J_1|$ are much smaller than $J_2$.
In this weak $|J_1|$ region, we can find the difference due to the sign of $J_1$ only in the nature of the dimer order whether it is the Haldane dimer or the singlet one.

\subsection{Proposals for experiments}

The following are our proposals for experiments.
The CuO$_2$ chain of Fig.~\ref{fig:CuO2_chain} is realized, for example, in $A\mathrm{CuMoO_4(OH)}$ ($A=\mathrm{Na, K}$)~\cite{nawa_1d_J1J2_InorgChem, nawa_1d_J1J2_jpsj,nawa_1d_J1J2_prb}.
$(J_1,J_2)$ are given by $(-51~\mathrm{K}, \, 36~\mathrm{K})$ for $A=\mathrm{Na}$ and $(238~\mathrm{K}, \, 0)$ for $A=\mathrm{K}$.
We can expect the $\bm E$-induced dimerization for $A=\mathrm{K}$ and the $\bm E$-induced Haldane-dimer phase for $A=\mathrm{Na}$

For $A=\mathrm{K}$, it will be intriguing to observe the $2/3$-power-law dependence \eqref{gap_E2/3} of the excitation gap $\Delta(\bm E)$.
We expect that the DC electric field with $1~\mathrm{MV/cm}$ opens the gap $\sim 22~\mathrm{K}$ in $\mathrm{KCuMoO_4(OH)}$, which reaches $8.6~\%$ of $J_1(\bm 0)$.
Spectroscopic methods such as the electron spin resonance (ESR)~\cite{zvyagin_cugeo3,glazkov_ntenp}, the inelastic neutron scattering~\cite{nishi_cugeo3_ins}, and the Raman scattering~\cite{sato_raman, prosnikov_raman} will be suitable candidates for observing those characteristics.

There is another interesting phenomenon specific to $\mathrm{KCuMoO_4(OH)}$.
The compound $\mathrm{KCuMoO_4(OH)}$ has a staggered Dzyaloshinskii-Moriya (DM) interaction, $\sum_j (-1)^j \bm D \cdot \bm S_j \times \bm S_{j+1}$, in addition to the Hamiltonian \eqref{H_chain_spin_E}.
The DM interaction originates from the spin-orbit coupling.
Though we did not take the spin-orbit coupling into account thus far in this paper, we can approximately apply our theory to this compound since the DM interaction is weak.
As far as the leading effects caused by the perturbative DM interaction and the perturbative electric field are concerned, we can still adapt our theory to those systems with the weak DM interaction.
If one wishes to discuss DC electric-field controls of DM interactions, one has to take into account the spin-orbit coupling in the electric models such as Eq.~\eqref{HA} and \eqref{HF}, which goes beyond the scope of this paper.

The staggered DM interaction yields a staggered magnetic-field interaction (say, $-h_s \sum_j (-1)^j S_j^x$) when the external DC magnetic field $h$ is applied in a direction perpendicular to  $\bm D$~\cite{oshikawa_staggered_field_prl,oshikawa_staggered_field_prb}.

\begin{figure}[t!]
    \centering
    \includegraphics[viewport = 0 0 1000 500, width=\linewidth]{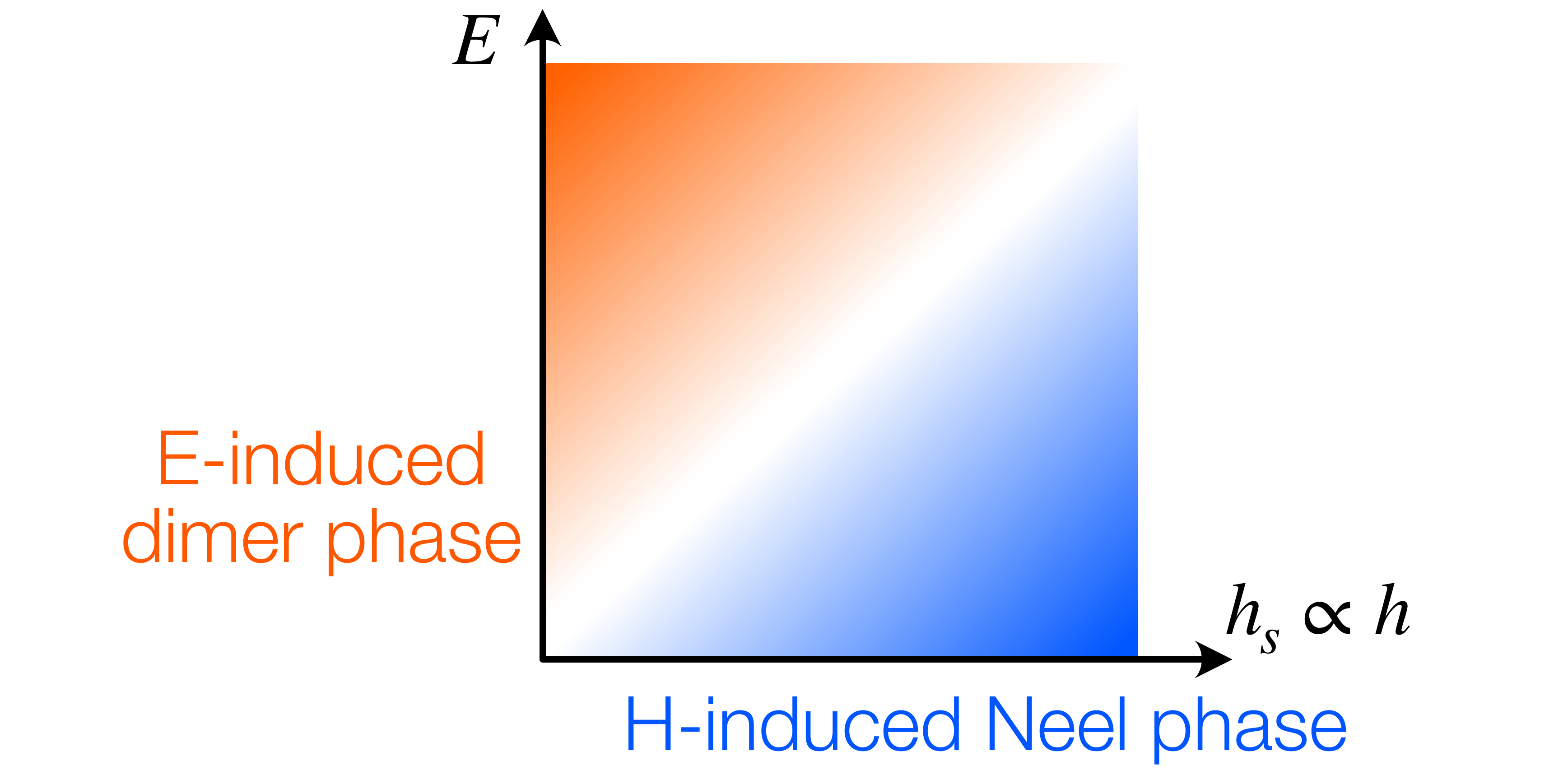}
    \caption{Possible $E$-$H$ ground-state phase diagram of $\mathrm{KCuMoO_4(OH)}$. When $E\not=0$ and $h_s=0$ $(=h)$, the $E$-induced singlet-dimer phase is realized. By contrast, when $E=0$ and $h_s\not=0$, the $h_s$-induced N\'eel phase is realized. These ground states will be smoothly connected to each other because the symmetries protecting these gapped quantum phases are explicitly broken.}
    \label{fig:EvsH}
\end{figure}

The resultant staggered magnetic field is proportional to $h$ and $|\bm D|$: $h_s\propto |\bm D|h/J$.
As well as the bond alternation $\delta J_1(\bm E)$,
the staggered magnetic field $h_s$ also generates the excitation gap.
The staggered magnetic field induces the N\'eel order in a direction perpendicular to the external DC magnetic field~\cite{oshikawa_staggered_field_prb}.
When the DC electric and magnetic fields are applied to the spin chain simultaneously and are tuned, a crossover will occur between the $\bm E$-induced dimerized phase and the $h_s$-induced N\'eel phase.
The entanglement of the singlet dimer is protected by one of the time-reversal, the $D_2$ spin-rotation, and the bond-centered inversion symmetries~\cite{chen_spt_2011a,chen_spt_2011b,nakagawa_pump}.
All these symmetries are explicitly broken in the simultaneous presence of the DC electric and magnetic fields.
Namely, nothing forbids the smooth deformation of the $E$-induced dimer-ordered ground state into the $H$-induced N\'eel ordered ground state.
It will be interesting to look into whether the crossover actually occurs on the two-dimensional parameter space shown in Fig.~\ref{fig:EvsH}.

Due to the staggered DM interaction, the field dependence of the excitation gap will be tractable in ESR measurements~\cite{sakai_nep, oshikawa_esr_prl, furuya_bbs}.
One significant difference of these two phases lies in a localized excited state at the chain end, called the boundary bound state~\cite{furuya_bbs}, exists only in the $h_s$-induced N\'eel phase~\cite{furuya_bbs}.
With an increase of $|E|$,
the boundary bound state will be eventually lost though it will survive for a while in the vicinity of the horizontal axis of Fig.~\ref{fig:EvsH}. 

For $A=\mathrm{Na}$, the nearest-neighbor interaction is ferromagnetic.
The ratio $J_1(\bm 0)/J_2(\bm 0)\approx -1.4$ implies that the ground state at zero magnetic field belongs to the Haldane-dimer phase (Fig. 1 of Ref.~\cite{furukawa_J1J2_XXZ}).
Experimentally, DC electric-field effects on the Haldane-dimer phase can be observed as an increase of the excitation gap similarly to the $J_1>0$ case.
We can also observe an $S=1/2$ edge state.
The spin-1 Haldane phase is a topological phase protected by the $\mathbb Z_2\times \mathbb Z_2$ spin-rotation symmetry~\cite{pollmann_haldane2010,pollmann_haldane2012}.
Since the DC electric fields keep the $\mathbb Z_2\times \mathbb Z_2$ symmetry, the $\bm E$-induced Haldane-dimer phase has doubly degenerate edge states on each chain end, that is, an $S=1/2$ edge spin.
The edge-spin degrees of freedom can be observed by, for example, the ESR spectroscopy~\cite{hagiwara_haldane_esr,yoshida_haldane_esr}.
An increase in the excitation gap immediately means a decrease in the correlation length.
The decrease of the correlation length would affect the ESR spectrum of the spin chain with a finite chain length~\cite{yoshida_haldane_esr}.

\section{Other DC electric-field effects}~\label{sec:other_eff}

This section is devoted to discussions on other major DC electric-field effects that we have not dealt with in this paper.
We can incorporate some effects into our model with slight modifications and some with substantial changes.
The renormalization of the dielectric constant and the structural distortion fall into the former.
The spin-orbit coupling falls into the latter and requires the substantial change of the model.
In what follows, we briefly discuss these three effects and another important effect of the THz laser pulse.

\subsection{Dielectric constant}

We implicitly assumed that the electrons feel the external DC electric field itself.
In real materials, the electron is surrounded by various charges that can screen or enhance the external electric field.
Generally, the dielectric function represents how the external DC electric field is screened or enhanced.
In particular, the dielectric constant in the material gets shifted from its vacuum value.
We can incorporate this effect into our model by regarding $\bm E(\bm r)$ in our model as the actual DC electric field that the electrons in materials actually feel.

\subsection{Structural distortion}

The strong DC electric field can possibly distort the crystal structure.
Still, the $\bm E$-induced change of the on-site potential $V_j$ turns out to be dominant, as shown below.
The structural distortion will make the hopping amplitude $t_j$ depend on $\bm E$ by modifying the lattice spacing.
We did not include this effect in our analysis thus far but can immediately include it without any problem.
Namely, we just replace the constant $t_j$ by an $\bm E$-dependent function $t_j(\bm E)$.
If one wishes to predict the precise $\bm E$ dependence of the hopping amplitude, one needs to model how the DC electric field distorts the lattice.

Besides, the structural distortion can lower the crystalline symmetry.
The symmetry lowering leads to an important effect of switching on hoppings between $d$ and $p$ orbitals that were forbidden by symmetries in the absence of the DC electric field.
Then, we need to include $d$- or $p$-orbital degeneracy explicitly into the model Hamiltonian.
However, since such a newly introduced hopping amplitude is proportional to the DC electric-field strength, the inclusion of the degenerate $d$ or $p$ orbitals will become important only under extremely strong DC electric fields.
Let us denote the additional hopping amplitude as $t'_j(\bm E)$ for $j=0,1$.
Note that $t'_j(\bm 0) = 0$ by definition.
It is straightforward to include $t'$ into our results.
We can obtain the correction by $t'_j(\bm E)$ to Eqs.~\eqref{JA} and \eqref{JF} by replacing $t_j$ in these relations to $t'_j$ totally or partially.
The fourth-order perturbation expansion shows
that this correction to the superexchange coupling is an even order of $\ell |\bm E|/U_d$ because an electron in a $p$ orbital hopped to a $d$ orbital must come back to the same $p$ orbital from the same $d$ orbital in the Mott-insulating phase.
$t'_j(\bm E)$ gives corrections of $O((t'_j(\bm E)/U_d)^4) \le O((\ell |\bm E|/U_d)^4)$ to Eq.~\eqref{JA}, that is, fourth order about the DC electric field.
On the other hand, $t'_j(\bm E)$ gives corrections of $O((t'_j(\bm E)/U_d)^2) \le O((\ell |\bm E|/U_d)^2)$ to Eq.~\eqref{JF}.
In any case, the correction is nonlinear about the DC electric field.

As we saw throughout the paper, the strong DC electric field of $O(1)$~MV/cm is already required to change the superexchange interaction by a few percent.
Accordingly, we need much stronger DC electric field to observe the nonlinear change of the superexchange.
Under such situations, the second- or fourth-order effect will be relevant only for $O(10)$~MV/cm for inorganic materials and for $O(1)$~MV/cm for organic ones.
There will be a chance in organic materials to observe the nonlinear field effects including that by the symmetry-lowering structural distortion.
However, it will be difficult to distinguish the structural distortion effect from other nonlinear terms in Eqs.~\eqref{JA} and \eqref{JF}.
Therefore, the inclusion of the structural distortion keeps our result intact unless the subleading nonlinear field effects are concerned.

Finally, we comment on a possible spontaneous formation of the electric polarization due to the $\bm E$-driven structural phase transition.
Such a spontaneous polarization can become large even if $\ell |\bm E|/U_d \ll 1$.
This large polarization can potentially lead to a large correction to the superexchange coupling.
It will be interesting in the future to investigate such $\bm E$-driven phase-transition effects.

\subsection{Spin-orbit coupling}

The spin-orbit coupling can dramatically change the spin Hamiltonian by adding to the spin Hamiltonian anisotropic interactions such as the DM interaction.
It needs straightforward but lengthy calculations to discuss the effects of the spin-orbit coupling on electric-field controls of magnetism~\cite{furuya_dm}.
We will discuss in a subsequent paper~\cite{furuya_dm} the combination effect of the spin-orbit coupling and the DC electric field.

\subsection{THz laser pulse}\label{sec:thz}

The single-cycle THz laser pulse can be deemed the effective DC electric field when the relevant time scale of the spin system is fast enough.
We need this assumption of the fast spin dynamics to guarantee that the quantum spin system quickly reaches the equilibrium before the pulse laser disappears.
The typical time scale ranges from $\sim 0.1$~ps to $\sim 1$~ns~\cite{kirilyuk_rmp,oshikawa-affleck,furuya_width,beaurepaire_1996,koopmans_2000,mashkovich_2019,Tzschaschel_2019,lenz_2006,vittoria_2010}.
For example, the spin dynamics with the time scale $\sim 0.1-1$~ps is much faster than the single cycle ($\sim 10$~ps) of the THz laser pulse with the $0.1$~THz frequency~\cite{hirori_laser_2011,mukai_laser_2016,nicoletti_laser_2016}.
Then, we can regard the single-cycle THz laser pulse as the effective DC electric field.
On the other hand, if we apply the laser pulse with the $10$~THz frequency to the spin system with the time scale $\gtrsim 10$~ps, the pulse ($\sim 0.1$~ps) disappears much faster than the equilibration of the spin system.
Then, the single-cycle THz laser pulse can be approximated as a delta-function electric field, $\bm E(t) = \bm E_0\delta (t)$.
Even when the spin dynamics is much slower than the THz laser pulse width, the electron dynamics can be much faster than the pulse width.
Indeed, the hopping amplitude $t\sim O(1)$~eV gives the time scale $O(1)~\mathrm{fs}=O(10^{-3})~\mathrm{ps}$.
Therefore, we can incorporate the THz laser pulse as an effective DC electric field into our model even when the pulse width is too short to equilibrate the spin system.
We then notice an interesting possibility.
If we derive the effective spin model, the THz laser pulse turns into the delta-function potential, $\delta J(\bm E_0) \delta(t)$, that instantaneously modifies the exchange interaction at the time $t=0$.
It is an interesting direction of future studies to investigate dynamical effects caused by the instantaneous potential.

To conclude, we emphasize two points.
The THz laser pulse can be deemed the effective DC electric field.
The THz laser pulse can equilibrate the spin system and otherwise induces the instantaneous modification of the exchange interaction, which depends on materials.

\section{Summary}\label{sec:summary}

This paper discussed DC electric-field controls of superexchange interactions in Mott insulators.
We first presented the generic results \eqref{JA} and \eqref{JF} about the superexchange interaction by the fourth-order degenerate perturbation expansion of the two basic electron models.
We considered the antiferromagnetic and ferromagnetic superexchange interactions and obtained their $\bm E$ dependence.

The other part of the main text was devoted to applications of the generic results to basic geometrically frustrated quantum spin systems.
While the results of Sec.~\ref{sec:results_gen} applies to various quantum spin systems regardless of the lattice and the dimensionality, we give our attention to low-dimensional quantum spin systems in this application part to investigate how DC electric fields applied perpendicular to the system control the superexchange interactions without interfering with the spatial anisotropy.
In contrast to this paper, the previous paper~\cite{takasan_dc} discussed how to control the ``direct superexchange'' interaction~\cite{Note1} and introduce the spatial anistoropy to the system through the DC electric field parallel to the system.
These two theories complement each other.

As the first application, we considered $J_1$--$J_2$ frustrated quantum spin systems on the square lattice in Sec.~\ref{sec:2d}.
We first demonstrated in the toy model how the DC electric field adjusts the parameter $J_1(\bm E)/J_2(\bm E)$ that determines the fate of the ground state.
Next, we applied the generic results of Sec.~\ref{sec:results_gen} to the more realistic model that emulates two compounds $\bacd$~\cite{nath_bacdvopo4_2008} and $\amo$~\cite{ishikawa}.
Combined to their crystal structures, the DC electric field breaks the inversion symmetry that introduces the nonequivalence of the next-nearest-neighbor interactions, $J_2(\bm E) \pm \delta J_2(\bm E)$ though the original model exactly has $\delta J_2(\bm 0) = 0$ at $\bm E=\bm 0$.
An increase of the nonequivalence by $\delta J_2(\bm E)$ turns the frustrated square-lattice quantum antiferromagnet eventually into the deformed triangular-lattice antiferromagnet, which will offer a unique experimental method to change the lattice structure effectively. 

We also discussed applications to geometrically frustrated one-dimensional quantum spin systems, the $J_1$--$J_2$ spin chains.
We assumed that there are two oxygen sites between the two nearest-neighbor magnetic ion sites, such as the CuO$_2$ chain.
When two oxygen sites feel different electric potentials, the DC electric field breaks the one-site translation symmetry down to the two-site one.
In other words, the number of spins per unit cell is doubled.
The DC electric field then yields the bond alternation $\delta J_1(\bm E)\sum_j (-1)^j \bm S_j \cdot \bm S_{j+1}$ while $\delta J_1(\bm 0) =0$.
The appearance of the bond alternation drives the quantum critical phase of the spin chain into a unique gapped quantum phase with a dimerization, which is the first proposal of the DC electric-field-induced dimerization.

The $\bm E$-induced dimer phase is the singlet-dimer one for $J_1>0$ or the Haldane-dimer (triplet-dimer) one for $J_1<0$.
The $\bm E$ dependence of the excitation gap will be experimentally visible in, for example, the ESR, the inelastic neutron scattering, and the Raman scattering experiments, which give evidence of the growth of the $E$-induced dimer orders.

We also briefly investigated other major DC electric-field effects that were not incorporated in our analyses.
We hope that this paper stimulates further theoretical and experimental studies on electric-field controls of quantum magnetism.

\section*{Acknowledgments}

The authors are grateful to Tsutomu Momoi and Yasushi Shinohara for stimulating discussions.
S.C.F. and M.S. are supported by a Grant-in-Aid for Scientific Research on Innovative Areas "Quantum Liquid Crystals" (Grant No. JP19H05825).
M.S. is also supported by JSPS KAKENHI (Grant No. 17K05513 and No. 20H01830).
K.T. is supported by the U.S. Department of Energy (DOE), Office of Science, Basic Energy Sciences (BES), under Contract No. DE-AC02-05CH11231 within the Ultrafast Materials Science Program (KC2203). K.T. also acknowledges the support from the JSPS Overseas Research Fellowship.

\appendix 

\section{Fourth-order degenerate perturbations}

This appendix is devoted to derivations of the generic formulas\eqref{JA} and \eqref{JF} of the superexchange coupling.

\subsection{Antiferromagnetic superexchange interaction}\label{app:JA}

\begin{figure}[b!]
    \centering
    \includegraphics[viewport = 0 0 1600 900, width=\linewidth]{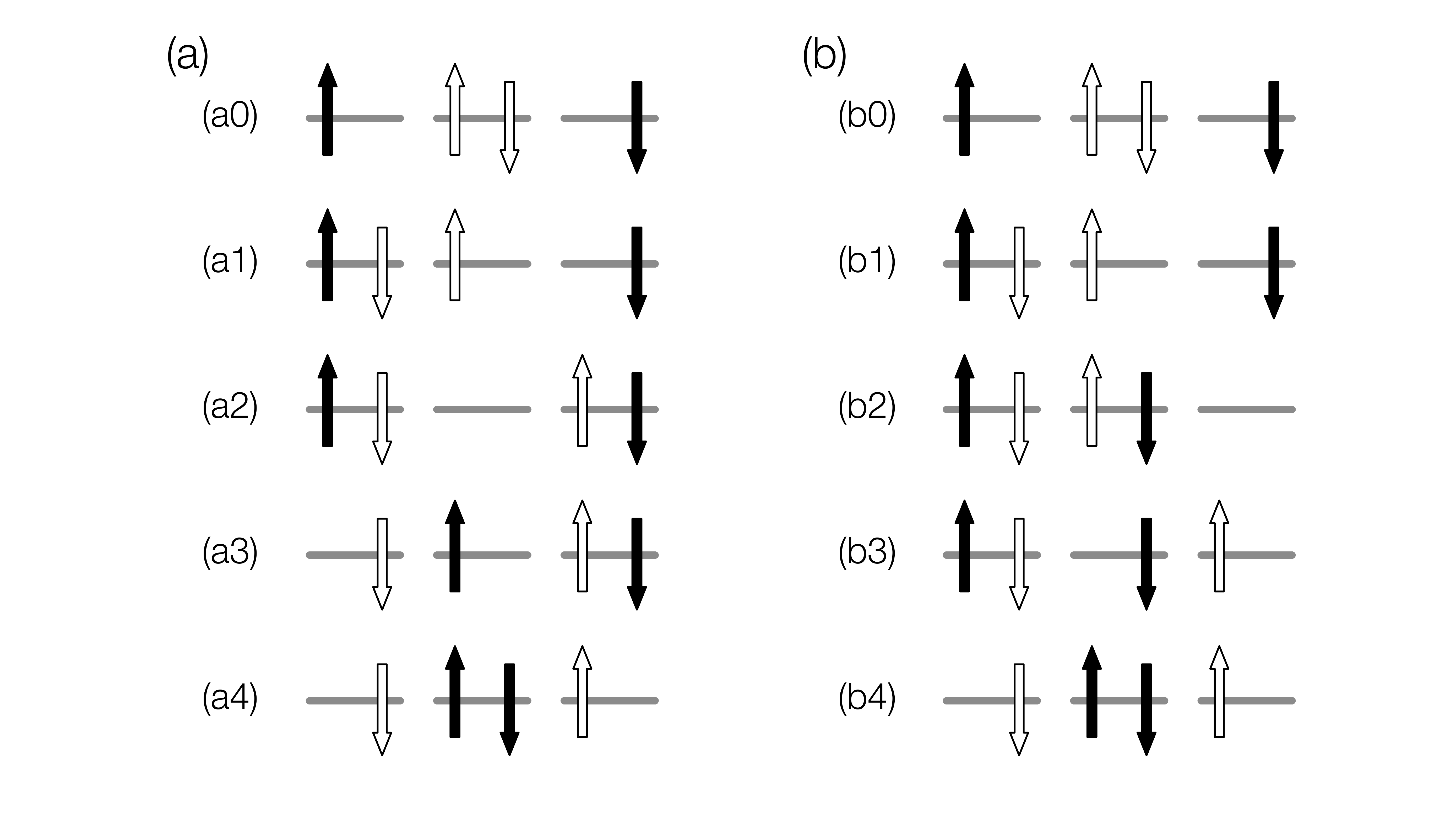}
    \caption{Two typical fourth-order processes, (a) and (b), of the perturbative expansion for the model \eqref{HA}.
    Filled (empty) arrows represent electron spins at $d$ ($p$, respectively) orbitals in the initial state (a0).
    (a0) and (a4) are initial and final states of the process (a) in Eq.~\eqref{H_eff_4}, respectively. (a1), (a2), (a3) are intermediate ones.
    Applying the perturbation $\mathcal H_t$ to the state of (a$n$), we obtain the state of (a($n+1$)).
    The same rules apply to the process (b).
    }
    \label{fig:4th_HA}
\end{figure}

There are two kinds of processes that have nontrivial contributions to the fourth-order term $\mathcal H_4^{\rm eff}$ of Eq.~\eqref{H_eff_4}.
Figure~\ref{fig:4th_HA} depicts two representatives of the fourth-order perturbation processes.
While the $p$ orbitals are temporally empty in the process (a), one of the $p$ orbitals are always filled in the process (b).
Every fourth-order process fits into either the process (a) or the process (b) with minor differences of the spin orientation.
All we have to do is to complete the calculations of the processes (a) and (b) with generic spin orientations.

Let us denote contributions of those processes as
\begin{align}
    \mathcal H_4^{\rm eff}
    &= \mathcal H_4^{\rm eff;a} + \mathcal H_4^{\rm eff;b}.
    \label{H_eff_4_ab}
\end{align}
The contribution from the process (a) exemplified by Fig.~\ref{fig:4th_HA}~(a) is the following.
\begin{widetext}
\begin{align}
    \mathcal H_4^{\rm eff;a}
    &= -P \biggl(\frac{t_0^\ast t_1^\ast}{\Delta_1+U_d-U_p} \sum_{\sigma'} p_{\frac 12, -\sigma'}^\dag d_{1,-\sigma'} p_{\frac 12,\sigma'}^\dag d_{0,\sigma'} + \frac{t_1^\ast t_0^\ast}{\Delta_0 + U_d -U_p} \sum_{\sigma'} p_{\frac 12,-\sigma'}^\dag d_{0,-\sigma'} p_{\frac 12,\sigma'}^\dag d_{1,\sigma'} \biggr) 
    \notag \\
    &\quad \cdot \biggl(
    \frac{t_1t_0}{(\Delta_0+\Delta_1 + 2U_d-U_p)(\Delta_0+ U_d-U_p)} \sum_{\sigma} d_{1,-\sigma}^\dag p_{\frac 12,-\sigma} d_{0,\sigma}^\dag p_{\frac 12, \sigma}
    \notag \\
    &\qquad +  \frac{t_0t_1}{(\Delta_0+\Delta_1+2U_d-U_p)(\Delta_1 + U_d - U_p)} \sum_{\sigma} d_{0,-\sigma}^\dag p_{\frac 12, -\sigma} d_{1,\sigma}^\dag p_{\frac 12, \sigma} 
    \biggr)P.
    \label{HA_4a_el}
\end{align}
Since the $p$ orbitals are fully occupied within the subspace spanned by the unperturbed ground state, the projection $P$ onto this subspace enables us to remove operators of the $p$ orbitals:
\begin{align}
    P p_{\frac 12, -\sigma'}^\dag  p_{\frac 12, \sigma'}^\dag p_{\frac 12, -\sigma} p_{\frac 12, \sigma} P &= \delta_{\sigma', \sigma} - \delta_{\sigma', -\sigma}.
    \label{id_p_HA}
\end{align}
In addition, some products of operators at $d$ orbitals are rewritten in terms of the $S=1/2$ spin operators $\bm S_j = \frac 12 \sum_{\sigma, \sigma'} d_{j,\sigma}^\dag \bm \sigma^{\sigma\sigma'} d_{j,\sigma'}$.
The contribution \eqref{HA_4a_el} of the process (a) is thus reduced to a simple form,
\begin{align}
    \mathcal H_4^{\rm eff;a}
    &= P\biggl[2|t_0t_1|^2 \biggl( \frac{1}{\Delta_0 + U_d-U_p} + \frac{1}{\Delta_1 + U_d-U_p} \biggr)^2 \frac{1}{\Delta_0 + \Delta_1 + 2U_d-U_p} \bm S_0 \cdot \bm S_1 + \mathrm{const.}\biggr]P.
    \label{HA_4a_spin}
\end{align}
We can obtain $\mathcal H_4^{\rm eff;b}$ similarly:
\begin{align}
    \mathcal H_4^{\rm eff;b}
    &= -P \biggl( \frac{t_0^\ast t_1}{\Delta_0 + U_d-U_p} \sum_{\sigma'} p_{\frac 12, \sigma'}^\dag d_{0,\sigma'} d_{1,\sigma'}^\dag p_{\frac 12, \sigma'} \frac{t_1^\ast t_0}{(\Delta_1-\Delta_0 +U_d)(\Delta_0+U_d-U_p)} \sum_\sigma p_{\frac 12, \sigma}^\dag d_{1,\sigma} d_{0,\sigma}^\dag p_{\frac 12, \sigma}
    \notag \\
    &\qquad + \frac{t_t^\ast t_0}{\Delta_1 + U_d-U_p} \sum_{\sigma'} p_{\frac 12, \sigma'}^\dag d_{1,\sigma'} d_{0,\sigma'}^\dag p_{\frac 12, \sigma'} \frac{t_0^\ast t_1}{(\Delta_0 - \Delta_1 + U_d)(\Delta_1+U_d-U_p)} \sum_\sigma p_{\frac 12,\sigma}^\dag d_{0,\sigma} d_{1,\sigma}^\dag p_{\frac 12, \sigma}\biggr)P
    \notag \\
    &= P\biggl[2|t_0t_1|^2 \biggl( \frac{1}{(\Delta_0 + U_d-U_p)^2} \frac{1}{\Delta_0 - \Delta_1 + U_d} + \frac{1}{(\Delta_1 +U_d-U_p)^2} \frac{1}{\Delta_1 - \Delta_0 +U_d} \biggr) \bm S_0 \cdot \bm S_1 + \mathrm{const.} \biggr]P.
    \label{HA_4b_spin}
\end{align}
Combining Eqs.~\eqref{HA_4a_spin} and \eqref{HA_4b_spin}, we reach the final result of the antiferromagnetic superexchange coupling~\eqref{JA}, which is consistent with the special case of $\Delta_0 = \Delta_1$ and $t_0=t_1\in \mathbb R$~\cite{koch_exchange}.
\end{widetext}

\subsection{Ferromagnetic superexchange interaction}\label{app:JF}

The ferromagnetic superexchange coupling \eqref{JF} is similarly derived.
Two processes contribute to the fourth-order term \eqref{H_eff_4} as shown in Fig.~\ref{fig:4th_HF}.
The process (a) of Fig.~\ref{fig:4th_HF} exchanges spins at $j=0$ and $j=1$ sites.
On the other hand, the process (b) of Fig.~\ref{fig:4th_HF} does not.
Nevertheless, the latter must be taken into account because it gives rise to a term $S_0^zS_1^z$, as we will see soon.

A major difference of the present model \eqref{HF} from the previously dealt one \eqref{HA} comes from the Coulomb exchange (the ferromagnetic direct exchange), $J_{\rm H}$, that reconstructs the unperturbed eigenstates of the degenerate $p$ orbitals only when they are half occupied.
We assumed the presence of two degenerate $p$ orbitals $p_x$ and $p_y$.
Without the hopping terms, the eigenstates are given by a product state $\ket{\phi_d}\ket{\phi_p}$, where $\ket{\phi_\mu}$ denotes the eigenstate at the $\mu=d,p$ orbital.
$\ket{\phi_d}$ is further split into a product of local states at two $d$-orbital sites.
The same applies to $\ket{\phi_p}$ except for the case mentioned above.
When the $p$ orbitals are half occupied, their eigenstates are reconstructed as the singlet $\ket{\phi_p}=(\ket{\uparrow_x \downarrow_y}-\ket{\downarrow_x\uparrow_y})/\sqrt{2}$ and triplets, $\ket{\phi_p} = \ket{\uparrow_x\downarrow_y}$, $(\ket{\uparrow_x \downarrow_y}+\ket{\downarrow_x\uparrow_y})/\sqrt{2}$,  and $\ket{\downarrow_x\downarrow_y}$.
Here, $\ket{\sigma_x\sigma_y}$ denotes the eigenstate of the $p$ orbitals.
Note that the triplets have the eigenenergy lower than the singlet by $2J_{\rm H}$ ($>0$).

Paying attention to the reconstruction of $p$ orbitals, we can calculate the fourth-order term \eqref{H_eff_4}.
Let us inherit the notation of Eq.~\eqref{H_eff_4_ab}.
Here, the processes (a) and (b) are replaced by those of Fig.~\ref{fig:4th_HF}.
The process (a) of Fig.~\ref{fig:4th_HF} leads to
\begin{widetext}

\begin{figure}[t!]
    \centering
    \includegraphics[viewport = 0 0 1600 900, width=0.5\linewidth]{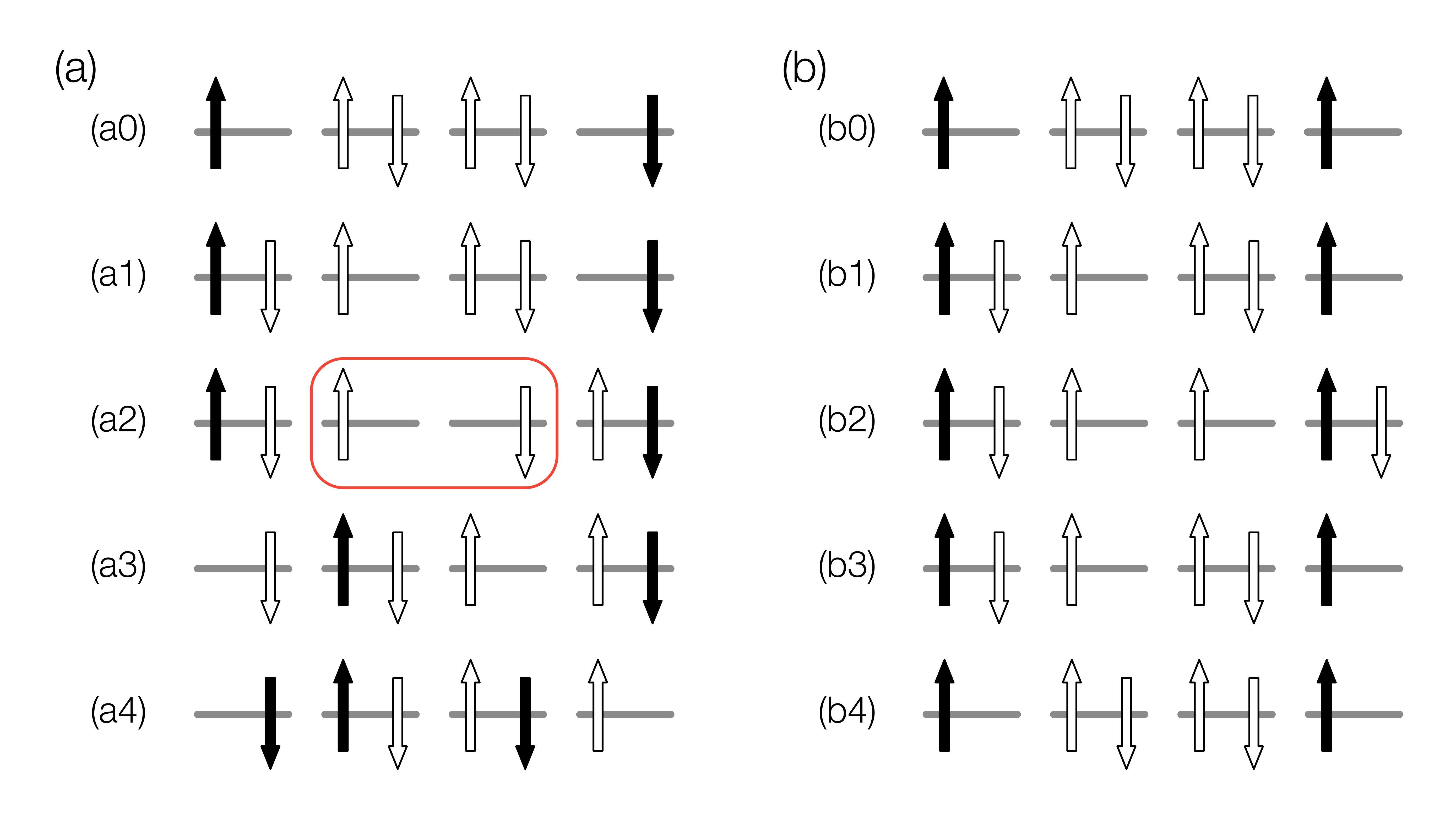}
    \caption{Two typical fourth-order processes, (a) and (b), of the perturbative expansion for the model \eqref{HF}, similarly drawn to Fig.~\ref{fig:4th_HA}.
    The red rounded square represents a reconstruction of the $p$-orbital eigenstates by the Coulomb exchange $J_{\rm H}$ [cf. Fig.~\ref{fig:HF_levels}~(b)].
    The $p$-orbital states surrounded by this curve is either symmetrized or antisymmetrized.
    The former corresponds to one of the triplet states and the latter to the singlet state.
    This reconstruction permits hoppings, for instance, from the $j=0$ site to the $y$ orbital at the $j=1/2$ site, that were forbidden for $J_{\rm H}=0$.
    }
    \label{fig:4th_HF}
\end{figure}

\begin{align}
    \mathcal H_4^{\rm eff;a}
    &= -P \biggl(\frac{t_1^\ast t_0^\ast}{\Delta_1 + U_d-U_p} \sum_{\sigma'} p_{\frac 12,-\sigma',y}^\dag d_{1,-\sigma'} p_{\frac 12, \sigma',x}^\dag d_{0,\sigma'} + \frac{t_0^\ast t_1^\ast}{\Delta_0 +U_d-U_p} \sum_{\sigma'} p_{\frac 12,-\sigma',x}^\dag d_{0,-\sigma'}p_{\frac 12,\sigma',y}^\dag d_{1,\sigma'}
    \biggr)
    \notag \\
    &\quad \cdot
    \biggr[\frac 12 \biggl(\frac{t_1}{\Delta_0 + \Delta_1 +2(U_d-U_p) +J_{\rm H}}  + \frac{t_1}{\Delta_0 + \Delta_1 +2(U_d-U_p) -J_{\rm H}} 
    \biggr) \frac{t_0}{\Delta_0 + U_d-U_p} \sum_{\sigma} d_{1,-\sigma}^\dag p_{\frac 12,-\sigma,y} d_{0,\sigma}^\dag p_{\frac 12, \sigma}
    \notag \\
    &\quad + \frac 12 \biggl(\frac{t_0}{\Delta_0 + \Delta_1 +2(U_d-U_p) +J_{\rm H}}  + \frac{t_0}{\Delta_0 + \Delta_1 +2(U_d-U_p) -J_{\rm H}} 
    \biggr) 
    \frac{t_1}{\Delta_1+U_d-U_p} \sum_{\sigma} d_{0,\sigma}^\dag p_{\frac 12,-\sigma, x} d_{1,\sigma}^\dag p_{\frac 12, \sigma, y}
    \notag \\
    & \quad + \frac 12 \biggl( \frac{t_1}{\Delta_0 + \Delta_1 +2(U_d-U_p) +J_{\rm H}} - \frac{t_1}{\Delta_0 + \Delta_1 + 2(U_d-U_p) - J_{\rm H}} \biggr) \frac{t_0}{\Delta_0 +U_d-U_p} \sum_{\sigma} d_{1,-\sigma}^\dag p_{\frac 12, \sigma, y} d_{0,\sigma}^\dag p_{\frac 12,-\sigma,x}
    \notag \\
    &\quad +\frac 12 \biggl( \frac{t_0}{\Delta_0 + \Delta_1 +2(U_d-U_p) +J_{\rm H}} - \frac{t_0}{\Delta_0 + \Delta_1 + 2(U_d-U_p) - J_{\rm H}} \biggr) \frac{t_1}{\Delta_1+U_d-U_p} \sum_{\sigma} d_{0,-\sigma}^\dag p_{\frac 12, \sigma, x} d_{1,\sigma}^\dag p_{\frac 12, -\sigma, y}
    \biggr]P.
\end{align}
The last two terms represent spin-dependent hoppings that appear as a direct consequence of nonzero $J_{\rm H}$.
They vanish in the $J_{\rm H} \to 0$ limit.
Discarding the $p$ operators and translating the $d$ operators into spin operators with the aid of the projection $P$, we obtain
\begin{align}
    \mathcal H_4^{\rm eff;a}
    &= P \biggl[ |t_0t_1|^2 \biggl(\frac{1}{\Delta_0 +U_d-U_p} + \frac{1}{\Delta_1 + U_d-U_p} \biggr)^2 \frac{\Delta_0 + \Delta_1 +2(U_d-U_p)}{[\Delta_0 + \Delta_1 + 2(U_d-U_p)]^2-{J_{\rm H}}^2} (1+2S_0^zS_1^z)
    \notag \\
    &\qquad -|t_0t_1|^2 \biggl( \frac{1}{\Delta_0 + U_d-U_p} + \frac{1}{\Delta_1 + U_d-U_p} \biggr)^2 \frac{J_{\rm H}}{[\Delta_0 + \Delta_1 + 2(U_d-U_p)]^2-{J_{\rm H}}^2} (S_0^+S_1^-+S_0^-S_1^+)
    \biggr]P
    \notag \\
    &= P \biggl[
    -2|t_0t_1|^2  \biggl( \frac{1}{\Delta_0 + U_d-U_p} + \frac{1}{\Delta_1 + U_d-U_p} \biggr)^2 \frac{J_{\rm H}}{[\Delta_0 + \Delta_1 + 2(U_d-U_p)]^2-{J_{\rm H}}^2} \bm S_0 \cdot \bm S_1
    \notag \\
    &\qquad +2|t_0t_1|^2 \biggl(\frac{1}{\Delta_0 +U_d-U_p} + \frac{1}{\Delta_1 + U_d-U_p} \biggr)^2 \frac{1}{\Delta_0 + \Delta_1 + 2(U_d-U_p)-J_{\rm H}} S_0^zS_1^z
    +\mathrm{const.}\biggr]P.
    \label{HF_4a_spin}
\end{align}
Eq.~\eqref{HF_4a_spin} turns out to break the $\mathrm{SU(2)}$ spin-rotation symmetry that the model possesses.
The process (b) yields a compensating $S_0^zS_1^z$ term and restores the $\mathrm{SU (2)}$ spin-rotation symmetry.
The process (b) leads to
\begin{align}
    \mathcal H_4^{\rm eff;b}
    &= -P \biggl(\frac{t_1^\ast t_0^\ast}{\Delta_1 + U_d-U_p} \sum_{\sigma'} p_{\frac 12, \sigma',y}^\dag d_{1,\sigma'} p_{\frac 12,\sigma',x}^\dag d_{0,\sigma'}
    + \frac{t_0^\ast t_1^\ast}{\Delta_0 +U_d-U_p} \sum_{\sigma'} p_{\frac 12,\sigma',x}^\dag d_{0,\sigma'} p_{\frac 12, \sigma',y}^\dag d_{1,\sigma'} \biggr)
    \notag \\
    &\qquad \cdot \biggl( \frac{t_1}{\Delta_0 + \Delta_1 +2(U_d-U_p) -J_{\rm H}} \frac{t_0}{\Delta_0 +U_d-U_p} \sum_{\sigma} d_{1,\sigma}^\dag p_{\frac 12, \sigma, y} d_{0,\sigma}^\dag p_{\frac 12, \sigma, x}
    \notag \\
    &\qquad + \frac{t_0}{\Delta_0 + \Delta_1 +2(U_d-U_p) -J_{\rm H}} \frac{t_1}{\Delta_1 + U_d-U_p} \sum_{\sigma} d_{0,\sigma}^\dag p_{\frac 12, \sigma, x} d_{1,\sigma}^\dag p_{\frac 12, \sigma, y}
    \biggr)P
    \notag \\
    &=P \biggl[ -2|t_0t_1|^2 \biggl(\frac{1}{\Delta_0 + U_d-U_p} + \frac{1}{\Delta_1 + U_d-U_p} \biggr)^2 \frac{1}{\Delta_0 + \Delta_1 + 2(U_d-U_p) -J_{\rm H}} S_0^z S_1^z + \mathrm{const.} \biggr]P.
    \label{HF_4b_spin}
\end{align}
The last line of Eq.~\eqref{HF_4b_spin} indeed cancels the anisotropic term of Eq.~\eqref{HF_4a_spin}.
We thus end up with the isotropic ferromagnetic exchange coupling \eqref{JF}.

\end{widetext}

\bibliography{ref.bib}

\begin{thebibliography}{157}%
\makeatletter
\providecommand \@ifxundefined [1]{%
 \@ifx{#1\undefined}
}%
\providecommand \@ifnum [1]{%
 \ifnum #1\expandafter \@firstoftwo
 \else \expandafter \@secondoftwo
 \fi
}%
\providecommand \@ifx [1]{%
 \ifx #1\expandafter \@firstoftwo
 \else \expandafter \@secondoftwo
 \fi
}%
\providecommand \natexlab [1]{#1}%
\providecommand \enquote  [1]{``#1''}%
\providecommand \bibnamefont  [1]{#1}%
\providecommand \bibfnamefont [1]{#1}%
\providecommand \citenamefont [1]{#1}%
\providecommand \href@noop [0]{\@secondoftwo}%
\providecommand \href [0]{\begingroup \@sanitize@url \@href}%
\providecommand \@href[1]{\@@startlink{#1}\@@href}%
\providecommand \@@href[1]{\endgroup#1\@@endlink}%
\providecommand \@sanitize@url [0]{\catcode `\\12\catcode `\$12\catcode
  `\&12\catcode `\#12\catcode `\^12\catcode `\_12\catcode `\%12\relax}%
\providecommand \@@startlink[1]{}%
\providecommand \@@endlink[0]{}%
\providecommand \url  [0]{\begingroup\@sanitize@url \@url }%
\providecommand \@url [1]{\endgroup\@href {#1}{\urlprefix }}%
\providecommand \urlprefix  [0]{URL }%
\providecommand \Eprint [0]{\href }%
\providecommand \doibase [0]{http://dx.doi.org/}%
\providecommand \selectlanguage [0]{\@gobble}%
\providecommand \bibinfo  [0]{\@secondoftwo}%
\providecommand \bibfield  [0]{\@secondoftwo}%
\providecommand \translation [1]{[#1]}%
\providecommand \BibitemOpen [0]{}%
\providecommand \bibitemStop [0]{}%
\providecommand \bibitemNoStop [0]{.\EOS\space}%
\providecommand \EOS [0]{\spacefactor3000\relax}%
\providecommand \BibitemShut  [1]{\csname bibitem#1\endcsname}%
\let\auto@bib@innerbib\@empty
\bibitem [{\citenamefont {Anderson}(1973)}]{anderson_spinliquid}%
  \BibitemOpen
  \bibfield  {author} {\bibinfo {author} {\bibfnamefont {P.W.}\ \bibnamefont
  {Anderson}},\ }\bibfield  {title} {\enquote {\bibinfo {title} {Resonating
  valence bonds: A new kind of insulator?}}\ }\href {\doibase
  https://doi.org/10.1016/0025-5408(73)90167-0} {\bibfield  {journal} {\bibinfo
   {journal} {Materials Research Bulletin}\ }\textbf {\bibinfo {volume} {8}},\
  \bibinfo {pages} {153 -- 160} (\bibinfo {year} {1973})}\BibitemShut {NoStop}%
\bibitem [{\citenamefont {Savary}\ and\ \citenamefont
  {Balents}(2016)}]{savary_review_2016}%
  \BibitemOpen
  \bibfield  {author} {\bibinfo {author} {\bibfnamefont {Lucile}\ \bibnamefont
  {Savary}}\ and\ \bibinfo {author} {\bibfnamefont {Leon}\ \bibnamefont
  {Balents}},\ }\bibfield  {title} {\enquote {\bibinfo {title} {Quantum spin
  liquids: a review},}\ }\href {\doibase 10.1088/0034-4885/80/1/016502}
  {\bibfield  {journal} {\bibinfo  {journal} {Reports on Progress in Physics}\
  }\textbf {\bibinfo {volume} {80}},\ \bibinfo {pages} {016502} (\bibinfo
  {year} {2016})}\BibitemShut {NoStop}%
\bibitem [{\citenamefont {Zhou}\ \emph {et~al.}(2017)\citenamefont {Zhou},
  \citenamefont {Kanoda},\ and\ \citenamefont {Ng}}]{zhou_review_spinliquid}%
  \BibitemOpen
  \bibfield  {author} {\bibinfo {author} {\bibfnamefont {Yi}~\bibnamefont
  {Zhou}}, \bibinfo {author} {\bibfnamefont {Kazushi}\ \bibnamefont {Kanoda}},
  \ and\ \bibinfo {author} {\bibfnamefont {Tai-Kai}\ \bibnamefont {Ng}},\
  }\bibfield  {title} {\enquote {\bibinfo {title} {Quantum spin liquid
  states},}\ }\href {\doibase 10.1103/RevModPhys.89.025003} {\bibfield
  {journal} {\bibinfo  {journal} {Rev. Mod. Phys.}\ }\textbf {\bibinfo {volume}
  {89}},\ \bibinfo {pages} {025003} (\bibinfo {year} {2017})}\BibitemShut
  {NoStop}%
\bibitem [{\citenamefont {Knolle}\ and\ \citenamefont
  {Moessner}(2019)}]{knolle_spinliquid_review}%
  \BibitemOpen
  \bibfield  {author} {\bibinfo {author} {\bibfnamefont {J.}~\bibnamefont
  {Knolle}}\ and\ \bibinfo {author} {\bibfnamefont {R.}~\bibnamefont
  {Moessner}},\ }\bibfield  {title} {\enquote {\bibinfo {title} {{A Field Guide
  to Spin Liquids}},}\ }\href {\doibase
  10.1146/annurev-conmatphys-031218-013401} {\bibfield  {journal} {\bibinfo
  {journal} {Annual Review of Condensed Matter Physics}\ }\textbf {\bibinfo
  {volume} {10}},\ \bibinfo {pages} {451--472} (\bibinfo {year}
  {2019})}\BibitemShut {NoStop}%
\bibitem [{\citenamefont {Jiang}\ \emph {et~al.}(2012)\citenamefont {Jiang},
  \citenamefont {Yao},\ and\ \citenamefont {Balents}}]{jiang_qsl_sq}%
  \BibitemOpen
  \bibfield  {author} {\bibinfo {author} {\bibfnamefont {Hong-Chen}\
  \bibnamefont {Jiang}}, \bibinfo {author} {\bibfnamefont {Hong}\ \bibnamefont
  {Yao}}, \ and\ \bibinfo {author} {\bibfnamefont {Leon}\ \bibnamefont
  {Balents}},\ }\bibfield  {title} {\enquote {\bibinfo {title} {{Spin liquid
  ground state of the spin-$\frac{1}{2}$ square ${J}_{1}$-${J}_{2}$ Heisenberg
  model}},}\ }\href {\doibase 10.1103/PhysRevB.86.024424} {\bibfield  {journal}
  {\bibinfo  {journal} {Phys. Rev. B}\ }\textbf {\bibinfo {volume} {86}},\
  \bibinfo {pages} {024424} (\bibinfo {year} {2012})}\BibitemShut {NoStop}%
\bibitem [{\citenamefont {Metavitsiadis}\ \emph {et~al.}(2014)\citenamefont
  {Metavitsiadis}, \citenamefont {Sellmann},\ and\ \citenamefont
  {Eggert}}]{eggert_j1j2}%
  \BibitemOpen
  \bibfield  {author} {\bibinfo {author} {\bibfnamefont {Alexandros}\
  \bibnamefont {Metavitsiadis}}, \bibinfo {author} {\bibfnamefont {Daniel}\
  \bibnamefont {Sellmann}}, \ and\ \bibinfo {author} {\bibfnamefont
  {Sebastian}\ \bibnamefont {Eggert}},\ }\bibfield  {title} {\enquote {\bibinfo
  {title} {{Spin-liquid versus dimer phases in an anisotropic
  ${J}_{1}$-${J}_{2}$ frustrated square antiferromagnet}},}\ }\href {\doibase
  10.1103/PhysRevB.89.241104} {\bibfield  {journal} {\bibinfo  {journal} {Phys.
  Rev. B}\ }\textbf {\bibinfo {volume} {89}},\ \bibinfo {pages} {241104}
  (\bibinfo {year} {2014})}\BibitemShut {NoStop}%
\bibitem [{\citenamefont {Chubukov}(1991)}]{chubukov_nematic_91}%
  \BibitemOpen
  \bibfield  {author} {\bibinfo {author} {\bibfnamefont {Andrey~V.}\
  \bibnamefont {Chubukov}},\ }\bibfield  {title} {\enquote {\bibinfo {title}
  {Chiral, nematic, and dimer states in quantum spin chains},}\ }\href
  {\doibase 10.1103/PhysRevB.44.4693} {\bibfield  {journal} {\bibinfo
  {journal} {Phys. Rev. B}\ }\textbf {\bibinfo {volume} {44}},\ \bibinfo
  {pages} {4693--4696} (\bibinfo {year} {1991})}\BibitemShut {NoStop}%
\bibitem [{\citenamefont {Shannon}\ \emph {et~al.}(2006)\citenamefont
  {Shannon}, \citenamefont {Momoi},\ and\ \citenamefont
  {Sindzingre}}]{shannon_sq_nematic}%
  \BibitemOpen
  \bibfield  {author} {\bibinfo {author} {\bibfnamefont {Nic}\ \bibnamefont
  {Shannon}}, \bibinfo {author} {\bibfnamefont {Tsutomu}\ \bibnamefont
  {Momoi}}, \ and\ \bibinfo {author} {\bibfnamefont {Philippe}\ \bibnamefont
  {Sindzingre}},\ }\bibfield  {title} {\enquote {\bibinfo {title} {{Nematic
  Order in Square Lattice Frustrated Ferromagnets}},}\ }\href {\doibase
  10.1103/PhysRevLett.96.027213} {\bibfield  {journal} {\bibinfo  {journal}
  {Phys. Rev. Lett.}\ }\textbf {\bibinfo {volume} {96}},\ \bibinfo {pages}
  {027213} (\bibinfo {year} {2006})}\BibitemShut {NoStop}%
\bibitem [{\citenamefont {L\"auchli}\ \emph {et~al.}(2006)\citenamefont
  {L\"auchli}, \citenamefont {Mila},\ and\ \citenamefont
  {Penc}}]{lauchli_blbq}%
  \BibitemOpen
  \bibfield  {author} {\bibinfo {author} {\bibfnamefont {Andreas}\ \bibnamefont
  {L\"auchli}}, \bibinfo {author} {\bibfnamefont {Fr\'ed\'eric}\ \bibnamefont
  {Mila}}, \ and\ \bibinfo {author} {\bibfnamefont {Karlo}\ \bibnamefont
  {Penc}},\ }\bibfield  {title} {\enquote {\bibinfo {title} {{Quadrupolar
  Phases of the $S=1$ Bilinear-Biquadratic Heisenberg Model on the Triangular
  Lattice}},}\ }\href {\doibase 10.1103/PhysRevLett.97.087205} {\bibfield
  {journal} {\bibinfo  {journal} {Phys. Rev. Lett.}\ }\textbf {\bibinfo
  {volume} {97}},\ \bibinfo {pages} {087205} (\bibinfo {year}
  {2006})}\BibitemShut {NoStop}%
\bibitem [{\citenamefont {Hikihara}\ \emph {et~al.}(2008)\citenamefont
  {Hikihara}, \citenamefont {Kecke}, \citenamefont {Momoi},\ and\ \citenamefont
  {Furusaki}}]{hikihara_nematic_2008}%
  \BibitemOpen
  \bibfield  {author} {\bibinfo {author} {\bibfnamefont {Toshiya}\ \bibnamefont
  {Hikihara}}, \bibinfo {author} {\bibfnamefont {Lars}\ \bibnamefont {Kecke}},
  \bibinfo {author} {\bibfnamefont {Tsutomu}\ \bibnamefont {Momoi}}, \ and\
  \bibinfo {author} {\bibfnamefont {Akira}\ \bibnamefont {Furusaki}},\
  }\bibfield  {title} {\enquote {\bibinfo {title} {Vector chiral and multipolar
  orders in the spin-$\frac{1}{2}$ frustrated ferromagnetic chain in magnetic
  field},}\ }\href {\doibase 10.1103/PhysRevB.78.144404} {\bibfield  {journal}
  {\bibinfo  {journal} {Phys. Rev. B}\ }\textbf {\bibinfo {volume} {78}},\
  \bibinfo {pages} {144404} (\bibinfo {year} {2008})}\BibitemShut {NoStop}%
\bibitem [{\citenamefont {Sudan}\ \emph {et~al.}(2009)\citenamefont {Sudan},
  \citenamefont {L\"uscher},\ and\ \citenamefont
  {L\"auchli}}]{sudan_J1J2_chain}%
  \BibitemOpen
  \bibfield  {author} {\bibinfo {author} {\bibfnamefont {Julien}\ \bibnamefont
  {Sudan}}, \bibinfo {author} {\bibfnamefont {Andreas}\ \bibnamefont
  {L\"uscher}}, \ and\ \bibinfo {author} {\bibfnamefont {Andreas~M.}\
  \bibnamefont {L\"auchli}},\ }\bibfield  {title} {\enquote {\bibinfo {title}
  {{Emergent multipolar spin correlations in a fluctuating spiral: The
  frustrated ferromagnetic spin-$\frac{1}{2}$ Heisenberg chain in a magnetic
  field}},}\ }\href {\doibase 10.1103/PhysRevB.80.140402} {\bibfield  {journal}
  {\bibinfo  {journal} {Phys. Rev. B}\ }\textbf {\bibinfo {volume} {80}},\
  \bibinfo {pages} {140402} (\bibinfo {year} {2009})}\BibitemShut {NoStop}%
\bibitem [{\citenamefont {Penc}\ and\ \citenamefont
  {L\''auchli}(2011)}]{nematic_review}%
  \BibitemOpen
  \bibfield  {author} {\bibinfo {author} {\bibfnamefont {K.}~\bibnamefont
  {Penc}}\ and\ \bibinfo {author} {\bibfnamefont {A.~M.}\ \bibnamefont
  {L\''auchli}},\ }\href@noop {} {\emph {\bibinfo {title} {Introduction to
  frustrated magnetism: materials, experiments, theory}}},\ Vol.\ \bibinfo
  {volume} {164}\ (\bibinfo  {publisher} {Springer, Berlin},\ \bibinfo {year}
  {2011})\ p.\ \bibinfo {pages} {331}\BibitemShut {NoStop}%
\bibitem [{\citenamefont {Giamarchi}\ \emph {et~al.}(2008)\citenamefont
  {Giamarchi}, \citenamefont {R{\"u}egg},\ and\ \citenamefont
  {Tchernyshyov}}]{giamarchi_bec}%
  \BibitemOpen
  \bibfield  {author} {\bibinfo {author} {\bibfnamefont {Thierry}\ \bibnamefont
  {Giamarchi}}, \bibinfo {author} {\bibfnamefont {Christian}\ \bibnamefont
  {R{\"u}egg}}, \ and\ \bibinfo {author} {\bibfnamefont {Oleg}\ \bibnamefont
  {Tchernyshyov}},\ }\bibfield  {title} {\enquote {\bibinfo {title}
  {{Bose--Einstein condensation in magnetic insulators}},}\ }\href
  {https://doi.org/10.1038/nphys893} {\bibfield  {journal} {\bibinfo  {journal}
  {Nature Physics}\ }\textbf {\bibinfo {volume} {4}},\ \bibinfo {pages}
  {198--204} (\bibinfo {year} {2008})}\BibitemShut {NoStop}%
\bibitem [{\citenamefont {Zayed}\ \emph {et~al.}(2017)\citenamefont {Zayed},
  \citenamefont {R{\"u}egg}, \citenamefont {L{\"a}uchli}, \citenamefont
  {Panagopoulos}, \citenamefont {Saxena}, \citenamefont {Ellerby},
  \citenamefont {McMorrow}, \citenamefont {Str{\"a}ssle}, \citenamefont
  {Klotz}, \citenamefont {Hamel} \emph {et~al.}}]{zayed_scbo}%
  \BibitemOpen
  \bibfield  {author} {\bibinfo {author} {\bibfnamefont {ME}~\bibnamefont
  {Zayed}}, \bibinfo {author} {\bibfnamefont {Ch}~\bibnamefont {R{\"u}egg}},
  \bibinfo {author} {\bibfnamefont {AM}~\bibnamefont {L{\"a}uchli}}, \bibinfo
  {author} {\bibfnamefont {C}~\bibnamefont {Panagopoulos}}, \bibinfo {author}
  {\bibfnamefont {SS}~\bibnamefont {Saxena}}, \bibinfo {author} {\bibfnamefont
  {M}~\bibnamefont {Ellerby}}, \bibinfo {author} {\bibfnamefont
  {DF}~\bibnamefont {McMorrow}}, \bibinfo {author} {\bibfnamefont
  {Th}~\bibnamefont {Str{\"a}ssle}}, \bibinfo {author} {\bibfnamefont
  {S}~\bibnamefont {Klotz}}, \bibinfo {author} {\bibfnamefont {G}~\bibnamefont
  {Hamel}},  \emph {et~al.},\ }\bibfield  {title} {\enquote {\bibinfo {title}
  {{4-spin plaquette singlet state in the Shastry--Sutherland compound
  SrCu$_2$(BO$_3$)$_2$}},}\ }\href {https://doi.org/10.1038/nphys4190}
  {\bibfield  {journal} {\bibinfo  {journal} {Nature physics}\ }\textbf
  {\bibinfo {volume} {13}},\ \bibinfo {pages} {962--966} (\bibinfo {year}
  {2017})}\BibitemShut {NoStop}%
\bibitem [{\citenamefont {Sakurai}\ \emph {et~al.}(2018)\citenamefont
  {Sakurai}, \citenamefont {Hirao}, \citenamefont {Hijii}, \citenamefont
  {Okubo}, \citenamefont {Ohta}, \citenamefont {Uwatoko}, \citenamefont
  {Kudo},\ and\ \citenamefont {Koike}}]{sakurai_scbo}%
  \BibitemOpen
  \bibfield  {author} {\bibinfo {author} {\bibfnamefont {Takahiro}\
  \bibnamefont {Sakurai}}, \bibinfo {author} {\bibfnamefont {Yuki}\
  \bibnamefont {Hirao}}, \bibinfo {author} {\bibfnamefont {Keigo}\ \bibnamefont
  {Hijii}}, \bibinfo {author} {\bibfnamefont {Susumu}\ \bibnamefont {Okubo}},
  \bibinfo {author} {\bibfnamefont {Hitoshi}\ \bibnamefont {Ohta}}, \bibinfo
  {author} {\bibfnamefont {Yoshiya}\ \bibnamefont {Uwatoko}}, \bibinfo {author}
  {\bibfnamefont {Kazutaka}\ \bibnamefont {Kudo}}, \ and\ \bibinfo {author}
  {\bibfnamefont {Yoji}\ \bibnamefont {Koike}},\ }\bibfield  {title} {\enquote
  {\bibinfo {title} {Direct observation of the quantum phase transition of
  {SrCu}2({BO}3)2 by high-pressure and terahertz electron spin resonance},}\
  }\href {\doibase 10.7566/jpsj.87.033701} {\bibfield  {journal} {\bibinfo
  {journal} {Journal of the Physical Society of Japan}\ }\textbf {\bibinfo
  {volume} {87}},\ \bibinfo {pages} {033701} (\bibinfo {year}
  {2018})}\BibitemShut {NoStop}%
\bibitem [{\citenamefont {Zvyagin}\ \emph {et~al.}(2019)\citenamefont
  {Zvyagin}, \citenamefont {Graf}, \citenamefont {Sakurai}, \citenamefont
  {Kimura}, \citenamefont {Nojiri}, \citenamefont {Wosnitza}, \citenamefont
  {Ohta}, \citenamefont {Ono},\ and\ \citenamefont
  {Tanaka}}]{zvyagin_cs2cucl4}%
  \BibitemOpen
  \bibfield  {author} {\bibinfo {author} {\bibfnamefont {SA}~\bibnamefont
  {Zvyagin}}, \bibinfo {author} {\bibfnamefont {D}~\bibnamefont {Graf}},
  \bibinfo {author} {\bibfnamefont {T}~\bibnamefont {Sakurai}}, \bibinfo
  {author} {\bibfnamefont {S}~\bibnamefont {Kimura}}, \bibinfo {author}
  {\bibfnamefont {H}~\bibnamefont {Nojiri}}, \bibinfo {author} {\bibfnamefont
  {J}~\bibnamefont {Wosnitza}}, \bibinfo {author} {\bibfnamefont
  {H}~\bibnamefont {Ohta}}, \bibinfo {author} {\bibfnamefont {T}~\bibnamefont
  {Ono}}, \ and\ \bibinfo {author} {\bibfnamefont {H}~\bibnamefont {Tanaka}},\
  }\bibfield  {title} {\enquote {\bibinfo {title} {{Pressure-tuning the quantum
  spin Hamiltonian of the triangular lattice antiferromagnet
  Cs$_2$CuCl$_4$}},}\ }\href {https://doi.org/10.1038/s41467-019-09071-7}
  {\bibfield  {journal} {\bibinfo  {journal} {Nature communications}\ }\textbf
  {\bibinfo {volume} {10}},\ \bibinfo {pages} {1064} (\bibinfo {year}
  {2019})}\BibitemShut {NoStop}%
\bibitem [{\citenamefont {Oka}\ and\ \citenamefont
  {Aoki}(2009)}]{oka_photovolatic}%
  \BibitemOpen
  \bibfield  {author} {\bibinfo {author} {\bibfnamefont {Takashi}\ \bibnamefont
  {Oka}}\ and\ \bibinfo {author} {\bibfnamefont {Hideo}\ \bibnamefont {Aoki}},\
  }\bibfield  {title} {\enquote {\bibinfo {title} {{Photovoltaic Hall effect in
  graphene}},}\ }\href {\doibase 10.1103/PhysRevB.79.081406} {\bibfield
  {journal} {\bibinfo  {journal} {Phys. Rev. B}\ }\textbf {\bibinfo {volume}
  {79}},\ \bibinfo {pages} {081406} (\bibinfo {year} {2009})}\BibitemShut
  {NoStop}%
\bibitem [{\citenamefont {Kitagawa}\ \emph {et~al.}(2011)\citenamefont
  {Kitagawa}, \citenamefont {Oka}, \citenamefont {Brataas}, \citenamefont
  {Fu},\ and\ \citenamefont {Demler}}]{kitagawa_floquet}%
  \BibitemOpen
  \bibfield  {author} {\bibinfo {author} {\bibfnamefont {Takuya}\ \bibnamefont
  {Kitagawa}}, \bibinfo {author} {\bibfnamefont {Takashi}\ \bibnamefont {Oka}},
  \bibinfo {author} {\bibfnamefont {Arne}\ \bibnamefont {Brataas}}, \bibinfo
  {author} {\bibfnamefont {Liang}\ \bibnamefont {Fu}}, \ and\ \bibinfo {author}
  {\bibfnamefont {Eugene}\ \bibnamefont {Demler}},\ }\bibfield  {title}
  {\enquote {\bibinfo {title} {Transport properties of nonequilibrium systems
  under the application of light: Photoinduced quantum hall insulators without
  landau levels},}\ }\href {\doibase 10.1103/PhysRevB.84.235108} {\bibfield
  {journal} {\bibinfo  {journal} {Phys. Rev. B}\ }\textbf {\bibinfo {volume}
  {84}},\ \bibinfo {pages} {235108} (\bibinfo {year} {2011})}\BibitemShut
  {NoStop}%
\bibitem [{\citenamefont {Wang}\ \emph {et~al.}(2013)\citenamefont {Wang},
  \citenamefont {Steinberg}, \citenamefont {Jarillo-Herrero},\ and\
  \citenamefont {Gedik}}]{wang_floquet-bloch}%
  \BibitemOpen
  \bibfield  {author} {\bibinfo {author} {\bibfnamefont {Y.~H.}\ \bibnamefont
  {Wang}}, \bibinfo {author} {\bibfnamefont {H.}~\bibnamefont {Steinberg}},
  \bibinfo {author} {\bibfnamefont {P.}~\bibnamefont {Jarillo-Herrero}}, \ and\
  \bibinfo {author} {\bibfnamefont {N.}~\bibnamefont {Gedik}},\ }\bibfield
  {title} {\enquote {\bibinfo {title} {{Observation of Floquet-Bloch States on
  the Surface of a Topological Insulator}},}\ }\href {\doibase
  10.1126/science.1239834} {\bibfield  {journal} {\bibinfo  {journal}
  {Science}\ }\textbf {\bibinfo {volume} {342}},\ \bibinfo {pages} {453--457}
  (\bibinfo {year} {2013})}\BibitemShut {NoStop}%
\bibitem [{\citenamefont {Jotzu}\ \emph {et~al.}(2014)\citenamefont {Jotzu},
  \citenamefont {Messer}, \citenamefont {Desbuquois}, \citenamefont {Lebrat},
  \citenamefont {Uehlinger}, \citenamefont {Greif},\ and\ \citenamefont
  {Esslinger}}]{jotzu_floquet_haldane}%
  \BibitemOpen
  \bibfield  {author} {\bibinfo {author} {\bibfnamefont {Gregor}\ \bibnamefont
  {Jotzu}}, \bibinfo {author} {\bibfnamefont {Michael}\ \bibnamefont {Messer}},
  \bibinfo {author} {\bibfnamefont {R{\'e}mi}\ \bibnamefont {Desbuquois}},
  \bibinfo {author} {\bibfnamefont {Martin}\ \bibnamefont {Lebrat}}, \bibinfo
  {author} {\bibfnamefont {Thomas}\ \bibnamefont {Uehlinger}}, \bibinfo
  {author} {\bibfnamefont {Daniel}\ \bibnamefont {Greif}}, \ and\ \bibinfo
  {author} {\bibfnamefont {Tilman}\ \bibnamefont {Esslinger}},\ }\bibfield
  {title} {\enquote {\bibinfo {title} {{Experimental realization of the
  topological Haldane model with ultracold fermions}},}\ }\href@noop {}
  {\bibfield  {journal} {\bibinfo  {journal} {Nature}\ }\textbf {\bibinfo
  {volume} {515}},\ \bibinfo {pages} {237--240} (\bibinfo {year}
  {2014})}\BibitemShut {NoStop}%
\bibitem [{\citenamefont {Shirley}(1965)}]{Shirley_floquet}%
  \BibitemOpen
  \bibfield  {author} {\bibinfo {author} {\bibfnamefont {Jon~H.}\ \bibnamefont
  {Shirley}},\ }\bibfield  {title} {\enquote {\bibinfo {title} {{Solution of
  the Schr\"odinger Equation with a Hamiltonian Periodic in Time}},}\ }\href
  {\doibase 10.1103/PhysRev.138.B979} {\bibfield  {journal} {\bibinfo
  {journal} {Phys. Rev.}\ }\textbf {\bibinfo {volume} {138}},\ \bibinfo {pages}
  {B979--B987} (\bibinfo {year} {1965})}\BibitemShut {NoStop}%
\bibitem [{\citenamefont {Sambe}(1973)}]{sambe_floquet}%
  \BibitemOpen
  \bibfield  {author} {\bibinfo {author} {\bibfnamefont {Hideo}\ \bibnamefont
  {Sambe}},\ }\bibfield  {title} {\enquote {\bibinfo {title} {{Steady States
  and Quasienergies of a Quantum-Mechanical System in an Oscillating Field}},}\
  }\href {\doibase 10.1103/PhysRevA.7.2203} {\bibfield  {journal} {\bibinfo
  {journal} {Phys. Rev. A}\ }\textbf {\bibinfo {volume} {7}},\ \bibinfo {pages}
  {2203--2213} (\bibinfo {year} {1973})}\BibitemShut {NoStop}%
\bibitem [{\citenamefont {Bukov}\ \emph {et~al.}(2015)\citenamefont {Bukov},
  \citenamefont {D'Alessio},\ and\ \citenamefont
  {Polkovnikov}}]{bukov_floquet}%
  \BibitemOpen
  \bibfield  {author} {\bibinfo {author} {\bibfnamefont {Marin}\ \bibnamefont
  {Bukov}}, \bibinfo {author} {\bibfnamefont {Luca}\ \bibnamefont {D'Alessio}},
  \ and\ \bibinfo {author} {\bibfnamefont {Anatoli}\ \bibnamefont
  {Polkovnikov}},\ }\bibfield  {title} {\enquote {\bibinfo {title} {Universal
  high-frequency behavior of periodically driven systems: from dynamical
  stabilization to floquet engineering},}\ }\href {\doibase
  10.1080/00018732.2015.1055918} {\bibfield  {journal} {\bibinfo  {journal}
  {Advances in Physics}\ }\textbf {\bibinfo {volume} {64}},\ \bibinfo {pages}
  {139--226} (\bibinfo {year} {2015})}\BibitemShut {NoStop}%
\bibitem [{\citenamefont {Eckardt}(2017)}]{eckardt_rmp}%
  \BibitemOpen
  \bibfield  {author} {\bibinfo {author} {\bibfnamefont {Andr\'e}\ \bibnamefont
  {Eckardt}},\ }\bibfield  {title} {\enquote {\bibinfo {title} {Colloquium:
  Atomic quantum gases in periodically driven optical lattices},}\ }\href
  {\doibase 10.1103/RevModPhys.89.011004} {\bibfield  {journal} {\bibinfo
  {journal} {Rev. Mod. Phys.}\ }\textbf {\bibinfo {volume} {89}},\ \bibinfo
  {pages} {011004} (\bibinfo {year} {2017})}\BibitemShut {NoStop}%
\bibitem [{\citenamefont {Oka}\ and\ \citenamefont
  {Kitamura}(2019)}]{oka_kitamura_floquet}%
  \BibitemOpen
  \bibfield  {author} {\bibinfo {author} {\bibfnamefont {Takashi}\ \bibnamefont
  {Oka}}\ and\ \bibinfo {author} {\bibfnamefont {Sota}\ \bibnamefont
  {Kitamura}},\ }\bibfield  {title} {\enquote {\bibinfo {title} {Floquet
  engineering of quantum materials},}\ }\href {\doibase
  10.1146/annurev-conmatphys-031218-013423} {\bibfield  {journal} {\bibinfo
  {journal} {Annual Review of Condensed Matter Physics}\ }\textbf {\bibinfo
  {volume} {10}},\ \bibinfo {pages} {387--408} (\bibinfo {year}
  {2019})}\BibitemShut {NoStop}%
\bibitem [{\citenamefont {Sato}(2021)}]{sato_floquet_book}%
  \BibitemOpen
  \bibfield  {author} {\bibinfo {author} {\bibfnamefont {Masahiro}\
  \bibnamefont {Sato}},\ }\bibfield  {title} {\enquote {\bibinfo {title}
  {Floquet theory and ultrafast control of magnetism},}\ }\href@noop {}
  {\bibfield  {journal} {\bibinfo  {journal} {Chirality, Magnetism and
  Magnetoelectricity: Separate Phenomena and Joint Effects in Metamaterial
  Structures}\ }\textbf {\bibinfo {volume} {138}},\ \bibinfo {pages} {265--286}
  (\bibinfo {year} {2021})}\BibitemShut {NoStop}%
\bibitem [{\citenamefont {Takayoshi}\ \emph
  {et~al.}(2014{\natexlab{a}})\citenamefont {Takayoshi}, \citenamefont {Aoki},\
  and\ \citenamefont {Oka}}]{takayoshi_laser_2014a}%
  \BibitemOpen
  \bibfield  {author} {\bibinfo {author} {\bibfnamefont {Shintaro}\
  \bibnamefont {Takayoshi}}, \bibinfo {author} {\bibfnamefont {Hideo}\
  \bibnamefont {Aoki}}, \ and\ \bibinfo {author} {\bibfnamefont {Takashi}\
  \bibnamefont {Oka}},\ }\bibfield  {title} {\enquote {\bibinfo {title}
  {Magnetization and phase transition induced by circularly polarized laser in
  quantum magnets},}\ }\href {\doibase 10.1103/PhysRevB.90.085150} {\bibfield
  {journal} {\bibinfo  {journal} {Phys. Rev. B}\ }\textbf {\bibinfo {volume}
  {90}},\ \bibinfo {pages} {085150} (\bibinfo {year}
  {2014}{\natexlab{a}})}\BibitemShut {NoStop}%
\bibitem [{\citenamefont {Takayoshi}\ \emph
  {et~al.}(2014{\natexlab{b}})\citenamefont {Takayoshi}, \citenamefont {Sato},\
  and\ \citenamefont {Oka}}]{takayoshi_laser_2014b}%
  \BibitemOpen
  \bibfield  {author} {\bibinfo {author} {\bibfnamefont {Shintaro}\
  \bibnamefont {Takayoshi}}, \bibinfo {author} {\bibfnamefont {Masahiro}\
  \bibnamefont {Sato}}, \ and\ \bibinfo {author} {\bibfnamefont {Takashi}\
  \bibnamefont {Oka}},\ }\bibfield  {title} {\enquote {\bibinfo {title}
  {Laser-induced magnetization curve},}\ }\href {\doibase
  10.1103/PhysRevB.90.214413} {\bibfield  {journal} {\bibinfo  {journal} {Phys.
  Rev. B}\ }\textbf {\bibinfo {volume} {90}},\ \bibinfo {pages} {214413}
  (\bibinfo {year} {2014}{\natexlab{b}})}\BibitemShut {NoStop}%
\bibitem [{\citenamefont {Sato}\ \emph {et~al.}(2016)\citenamefont {Sato},
  \citenamefont {Takayoshi},\ and\ \citenamefont {Oka}}]{sato_laser_dm}%
  \BibitemOpen
  \bibfield  {author} {\bibinfo {author} {\bibfnamefont {Masahiro}\
  \bibnamefont {Sato}}, \bibinfo {author} {\bibfnamefont {Shintaro}\
  \bibnamefont {Takayoshi}}, \ and\ \bibinfo {author} {\bibfnamefont {Takashi}\
  \bibnamefont {Oka}},\ }\bibfield  {title} {\enquote {\bibinfo {title}
  {{Laser-Driven Multiferroics and Ultrafast Spin Current Generation}},}\
  }\href {\doibase 10.1103/PhysRevLett.117.147202} {\bibfield  {journal}
  {\bibinfo  {journal} {Phys. Rev. Lett.}\ }\textbf {\bibinfo {volume} {117}},\
  \bibinfo {pages} {147202} (\bibinfo {year} {2016})}\BibitemShut {NoStop}%
\bibitem [{\citenamefont {Mentink}\ \emph {et~al.}(2015)\citenamefont
  {Mentink}, \citenamefont {Balzer},\ and\ \citenamefont
  {Eckstein}}]{mentink_ultrafast}%
  \BibitemOpen
  \bibfield  {author} {\bibinfo {author} {\bibfnamefont {JH}~\bibnamefont
  {Mentink}}, \bibinfo {author} {\bibfnamefont {Karsten}\ \bibnamefont
  {Balzer}}, \ and\ \bibinfo {author} {\bibfnamefont {Martin}\ \bibnamefont
  {Eckstein}},\ }\bibfield  {title} {\enquote {\bibinfo {title} {Ultrafast and
  reversible control of the exchange interaction in mott insulators},}\ }\href
  {https://www.nature.com/articles/ncomms7708} {\bibfield  {journal} {\bibinfo
  {journal} {Nature communications}\ }\textbf {\bibinfo {volume} {6}},\
  \bibinfo {pages} {6708} (\bibinfo {year} {2015})}\BibitemShut {NoStop}%
\bibitem [{\citenamefont {Kitamura}\ \emph {et~al.}(2017)\citenamefont
  {Kitamura}, \citenamefont {Oka},\ and\ \citenamefont {Aoki}}]{kitamura_mott}%
  \BibitemOpen
  \bibfield  {author} {\bibinfo {author} {\bibfnamefont {Sota}\ \bibnamefont
  {Kitamura}}, \bibinfo {author} {\bibfnamefont {Takashi}\ \bibnamefont {Oka}},
  \ and\ \bibinfo {author} {\bibfnamefont {Hideo}\ \bibnamefont {Aoki}},\
  }\bibfield  {title} {\enquote {\bibinfo {title} {{Probing and controlling
  spin chirality in Mott insulators by circularly polarized laser}},}\ }\href
  {\doibase 10.1103/PhysRevB.96.014406} {\bibfield  {journal} {\bibinfo
  {journal} {Phys. Rev. B}\ }\textbf {\bibinfo {volume} {96}},\ \bibinfo
  {pages} {014406} (\bibinfo {year} {2017})}\BibitemShut {NoStop}%
\bibitem [{\citenamefont {Takasan}\ \emph {et~al.}(2017)\citenamefont
  {Takasan}, \citenamefont {Nakagawa},\ and\ \citenamefont
  {Kawakami}}]{takasan_kondo}%
  \BibitemOpen
  \bibfield  {author} {\bibinfo {author} {\bibfnamefont {Kazuaki}\ \bibnamefont
  {Takasan}}, \bibinfo {author} {\bibfnamefont {Masaya}\ \bibnamefont
  {Nakagawa}}, \ and\ \bibinfo {author} {\bibfnamefont {Norio}\ \bibnamefont
  {Kawakami}},\ }\bibfield  {title} {\enquote {\bibinfo {title}
  {{Laser-irradiated Kondo insulators: Controlling the Kondo effect and
  topological phases}},}\ }\href {\doibase 10.1103/PhysRevB.96.115120}
  {\bibfield  {journal} {\bibinfo  {journal} {Phys. Rev. B}\ }\textbf {\bibinfo
  {volume} {96}},\ \bibinfo {pages} {115120} (\bibinfo {year}
  {2017})}\BibitemShut {NoStop}%
\bibitem [{\citenamefont {Claassen}\ \emph {et~al.}(2017)\citenamefont
  {Claassen}, \citenamefont {Jiang}, \citenamefont {Moritz},\ and\
  \citenamefont {Devereaux}}]{claassen_mott}%
  \BibitemOpen
  \bibfield  {author} {\bibinfo {author} {\bibfnamefont {Martin}\ \bibnamefont
  {Claassen}}, \bibinfo {author} {\bibfnamefont {Hong-Chen}\ \bibnamefont
  {Jiang}}, \bibinfo {author} {\bibfnamefont {Brian}\ \bibnamefont {Moritz}}, \
  and\ \bibinfo {author} {\bibfnamefont {Thomas~P}\ \bibnamefont {Devereaux}},\
  }\bibfield  {title} {\enquote {\bibinfo {title} {Dynamical time-reversal
  symmetry breaking and photo-induced chiral spin liquids in frustrated mott
  insulators},}\ }\href {https://www.nature.com/articles/s41467-017-00876-y}
  {\bibfield  {journal} {\bibinfo  {journal} {Nature communications}\ }\textbf
  {\bibinfo {volume} {8}},\ \bibinfo {pages} {1192} (\bibinfo {year}
  {2017})}\BibitemShut {NoStop}%
\bibitem [{\citenamefont {Chaudhary}\ \emph {et~al.}(2019)\citenamefont
  {Chaudhary}, \citenamefont {Hsieh},\ and\ \citenamefont
  {Refael}}]{chaudhary_orbital_floquet}%
  \BibitemOpen
  \bibfield  {author} {\bibinfo {author} {\bibfnamefont {Swati}\ \bibnamefont
  {Chaudhary}}, \bibinfo {author} {\bibfnamefont {David}\ \bibnamefont
  {Hsieh}}, \ and\ \bibinfo {author} {\bibfnamefont {Gil}\ \bibnamefont
  {Refael}},\ }\bibfield  {title} {\enquote {\bibinfo {title} {Orbital floquet
  engineering of exchange interactions in magnetic materials},}\ }\href
  {\doibase 10.1103/PhysRevB.100.220403} {\bibfield  {journal} {\bibinfo
  {journal} {Phys. Rev. B}\ }\textbf {\bibinfo {volume} {100}},\ \bibinfo
  {pages} {220403} (\bibinfo {year} {2019})}\BibitemShut {NoStop}%
\bibitem [{\citenamefont {Sato}\ \emph {et~al.}(2014)\citenamefont {Sato},
  \citenamefont {Sasaki},\ and\ \citenamefont {Oka}}]{sato_floquet_majorana}%
  \BibitemOpen
  \bibfield  {author} {\bibinfo {author} {\bibfnamefont {Masahiro}\
  \bibnamefont {Sato}}, \bibinfo {author} {\bibfnamefont {Yuki}\ \bibnamefont
  {Sasaki}}, \ and\ \bibinfo {author} {\bibfnamefont {Takashi}\ \bibnamefont
  {Oka}},\ }\bibfield  {title} {\enquote {\bibinfo {title} {{Floquet Majorana
  Edge Mode and Non-Abelian Anyons in a Driven Kitaev Model}},}\ }\href
  {https://arxiv.org/abs/1404.2010} {\bibfield  {journal} {\bibinfo  {journal}
  {arXiv preprint arXiv:1404.2010}\ } (\bibinfo {year} {2014})}\BibitemShut
  {NoStop}%
\bibitem [{\citenamefont {Higashikawa}\ \emph {et~al.}(2018)\citenamefont
  {Higashikawa}, \citenamefont {Fujita},\ and\ \citenamefont
  {Sato}}]{higashikawa_floquet}%
  \BibitemOpen
  \bibfield  {author} {\bibinfo {author} {\bibfnamefont {Sho}\ \bibnamefont
  {Higashikawa}}, \bibinfo {author} {\bibfnamefont {Hiroyuki}\ \bibnamefont
  {Fujita}}, \ and\ \bibinfo {author} {\bibfnamefont {Masahiro}\ \bibnamefont
  {Sato}},\ }\bibfield  {title} {\enquote {\bibinfo {title} {Floquet
  engineering of classical systems},}\ }\href
  {https://arxiv.org/abs/1810.01103} {\bibfield  {journal} {\bibinfo  {journal}
  {arXiv preprint arXiv:1810.01103}\ } (\bibinfo {year} {2018})}\BibitemShut
  {NoStop}%
\bibitem [{\citenamefont {Ueno}\ \emph {et~al.}(2014)\citenamefont {Ueno},
  \citenamefont {Shimotani}, \citenamefont {Yuan}, \citenamefont {Ye},
  \citenamefont {Kawasaki},\ and\ \citenamefont {Iwasa}}]{ueno_fet}%
  \BibitemOpen
  \bibfield  {author} {\bibinfo {author} {\bibfnamefont {Kazunori}\
  \bibnamefont {Ueno}}, \bibinfo {author} {\bibfnamefont {Hidekazu}\
  \bibnamefont {Shimotani}}, \bibinfo {author} {\bibfnamefont {Hongtao}\
  \bibnamefont {Yuan}}, \bibinfo {author} {\bibfnamefont {Jianting}\
  \bibnamefont {Ye}}, \bibinfo {author} {\bibfnamefont {Masashi}\ \bibnamefont
  {Kawasaki}}, \ and\ \bibinfo {author} {\bibfnamefont {Yoshihiro}\
  \bibnamefont {Iwasa}},\ }\bibfield  {title} {\enquote {\bibinfo {title}
  {Field-induced superconductivity in electric double layer transistors},}\
  }\href {\doibase 10.7566/JPSJ.83.032001} {\bibfield  {journal} {\bibinfo
  {journal} {Journal of the Physical Society of Japan}\ }\textbf {\bibinfo
  {volume} {83}},\ \bibinfo {pages} {032001} (\bibinfo {year}
  {2014})}\BibitemShut {NoStop}%
\bibitem [{\citenamefont {Bisri}\ \emph {et~al.}(2017)\citenamefont {Bisri},
  \citenamefont {Shimizu}, \citenamefont {Nakano},\ and\ \citenamefont
  {Iwasa}}]{bisri_fet}%
  \BibitemOpen
  \bibfield  {author} {\bibinfo {author} {\bibfnamefont {Satria~Zulkarnaen}\
  \bibnamefont {Bisri}}, \bibinfo {author} {\bibfnamefont {Sunao}\ \bibnamefont
  {Shimizu}}, \bibinfo {author} {\bibfnamefont {Masaki}\ \bibnamefont
  {Nakano}}, \ and\ \bibinfo {author} {\bibfnamefont {Yoshihiro}\ \bibnamefont
  {Iwasa}},\ }\bibfield  {title} {\enquote {\bibinfo {title} {Endeavor of
  iontronics: From fundamentals to applications of ion-controlled
  electronics},}\ }\href {\doibase https://doi.org/10.1002/adma.201607054}
  {\bibfield  {journal} {\bibinfo  {journal} {Advanced Materials}\ }\textbf
  {\bibinfo {volume} {29}},\ \bibinfo {pages} {1607054} (\bibinfo {year}
  {2017})}\BibitemShut {NoStop}%
\bibitem [{\citenamefont {Romming}\ \emph {et~al.}(2013)\citenamefont
  {Romming}, \citenamefont {Hanneken}, \citenamefont {Menzel}, \citenamefont
  {Bickel}, \citenamefont {Wolter}, \citenamefont {von Bergmann}, \citenamefont
  {Kubetzka},\ and\ \citenamefont {Wiesendanger}}]{Romming2013}%
  \BibitemOpen
  \bibfield  {author} {\bibinfo {author} {\bibfnamefont {Niklas}\ \bibnamefont
  {Romming}}, \bibinfo {author} {\bibfnamefont {Christian}\ \bibnamefont
  {Hanneken}}, \bibinfo {author} {\bibfnamefont {Matthias}\ \bibnamefont
  {Menzel}}, \bibinfo {author} {\bibfnamefont {Jessica~E.}\ \bibnamefont
  {Bickel}}, \bibinfo {author} {\bibfnamefont {Boris}\ \bibnamefont {Wolter}},
  \bibinfo {author} {\bibfnamefont {Kirsten}\ \bibnamefont {von Bergmann}},
  \bibinfo {author} {\bibfnamefont {Andr{\'{e}}}\ \bibnamefont {Kubetzka}}, \
  and\ \bibinfo {author} {\bibfnamefont {Roland}\ \bibnamefont
  {Wiesendanger}},\ }\bibfield  {title} {\enquote {\bibinfo {title} {Writing
  and deleting single magnetic skyrmions},}\ }\href {\doibase
  10.1126/science.1240573} {\bibfield  {journal} {\bibinfo  {journal}
  {Science}\ }\textbf {\bibinfo {volume} {341}},\ \bibinfo {pages} {636--639}
  (\bibinfo {year} {2013})}\BibitemShut {NoStop}%
\bibitem [{\citenamefont {Hsu}\ \emph {et~al.}(2016)\citenamefont {Hsu},
  \citenamefont {Kubetzka}, \citenamefont {Finco}, \citenamefont {Romming},
  \citenamefont {von Bergmann},\ and\ \citenamefont {Wiesendanger}}]{Hsu2016}%
  \BibitemOpen
  \bibfield  {author} {\bibinfo {author} {\bibfnamefont {Pin-Jui}\ \bibnamefont
  {Hsu}}, \bibinfo {author} {\bibfnamefont {Andr{\'{e}}}\ \bibnamefont
  {Kubetzka}}, \bibinfo {author} {\bibfnamefont {Aurore}\ \bibnamefont
  {Finco}}, \bibinfo {author} {\bibfnamefont {Niklas}\ \bibnamefont {Romming}},
  \bibinfo {author} {\bibfnamefont {Kirsten}\ \bibnamefont {von Bergmann}}, \
  and\ \bibinfo {author} {\bibfnamefont {Roland}\ \bibnamefont
  {Wiesendanger}},\ }\bibfield  {title} {\enquote {\bibinfo {title}
  {Electric-field-driven switching of individual magnetic skyrmions},}\ }\href
  {\doibase 10.1038/nnano.2016.234} {\bibfield  {journal} {\bibinfo  {journal}
  {Nature Nanotechnology}\ }\textbf {\bibinfo {volume} {12}},\ \bibinfo {pages}
  {123--126} (\bibinfo {year} {2016})}\BibitemShut {NoStop}%
\bibitem [{Note1()}]{Note1}%
  \BibitemOpen
  \bibinfo {note} {In this paper, we define superexchange interactions as
  exchange interactions originating from hoppings between magnetic ions and
  intermediate nonmagnetic ions. We refer to the exchange interactions that
  originate from direct hoppings between magnetic ions as ``direct
  superexchange'' interactions to distinguish it from the above-mentioned
  superexchange and the direct exchange interaction~\cite
  {koch_exchange}.}\BibitemShut {Stop}%
\bibitem [{\citenamefont {Takasan}\ and\ \citenamefont
  {Sato}(2019)}]{takasan_dc}%
  \BibitemOpen
  \bibfield  {author} {\bibinfo {author} {\bibfnamefont {Kazuaki}\ \bibnamefont
  {Takasan}}\ and\ \bibinfo {author} {\bibfnamefont {Masahiro}\ \bibnamefont
  {Sato}},\ }\bibfield  {title} {\enquote {\bibinfo {title} {Control of
  magnetic and topological orders with a dc electric field},}\ }\href {\doibase
  10.1103/PhysRevB.100.060408} {\bibfield  {journal} {\bibinfo  {journal}
  {Phys. Rev. B}\ }\textbf {\bibinfo {volume} {100}},\ \bibinfo {pages}
  {060408} (\bibinfo {year} {2019})}\BibitemShut {NoStop}%
\bibitem [{\citenamefont {Matsukura}\ \emph {et~al.}(2015)\citenamefont
  {Matsukura}, \citenamefont {Tokura},\ and\ \citenamefont
  {Ohno}}]{matsukura_nat_nanotech_2015}%
  \BibitemOpen
  \bibfield  {author} {\bibinfo {author} {\bibfnamefont {Fumihiro}\
  \bibnamefont {Matsukura}}, \bibinfo {author} {\bibfnamefont {Yoshinori}\
  \bibnamefont {Tokura}}, \ and\ \bibinfo {author} {\bibfnamefont {Hideo}\
  \bibnamefont {Ohno}},\ }\bibfield  {title} {\enquote {\bibinfo {title}
  {Control of magnetism by electric fields},}\ }\href
  {https://www.nature.com/articles/nnano.2015.22} {\bibfield  {journal}
  {\bibinfo  {journal} {Nature nanotechnology}\ }\textbf {\bibinfo {volume}
  {10}},\ \bibinfo {pages} {209--220} (\bibinfo {year} {2015})}\BibitemShut
  {NoStop}%
\bibitem [{\citenamefont {Chen}\ \emph {et~al.}(2015)\citenamefont {Chen},
  \citenamefont {Matsukura},\ and\ \citenamefont {Ohno}}]{matsukura_GaMnAs}%
  \BibitemOpen
  \bibfield  {author} {\bibinfo {author} {\bibfnamefont {Lin}\ \bibnamefont
  {Chen}}, \bibinfo {author} {\bibfnamefont {Fumihiro}\ \bibnamefont
  {Matsukura}}, \ and\ \bibinfo {author} {\bibfnamefont {Hideo}\ \bibnamefont
  {Ohno}},\ }\bibfield  {title} {\enquote {\bibinfo {title} {{Electric-Field
  Modulation of Damping Constant in a Ferromagnetic Semiconductor
  (Ga,Mn)As}},}\ }\href {\doibase 10.1103/PhysRevLett.115.057204} {\bibfield
  {journal} {\bibinfo  {journal} {Phys. Rev. Lett.}\ }\textbf {\bibinfo
  {volume} {115}},\ \bibinfo {pages} {057204} (\bibinfo {year}
  {2015})}\BibitemShut {NoStop}%
\bibitem [{\citenamefont {Hirori}\ \emph {et~al.}(2011)\citenamefont {Hirori},
  \citenamefont {Doi}, \citenamefont {Blanchard},\ and\ \citenamefont
  {Tanaka}}]{hirori_laser_2011}%
  \BibitemOpen
  \bibfield  {author} {\bibinfo {author} {\bibfnamefont {H.}~\bibnamefont
  {Hirori}}, \bibinfo {author} {\bibfnamefont {A.}~\bibnamefont {Doi}},
  \bibinfo {author} {\bibfnamefont {F.}~\bibnamefont {Blanchard}}, \ and\
  \bibinfo {author} {\bibfnamefont {K.}~\bibnamefont {Tanaka}},\ }\bibfield
  {title} {\enquote {\bibinfo {title} {Single-cycle terahertz pulses with
  amplitudes exceeding 1 {MV}/cm generated by optical rectification in
  {LiNbO}3},}\ }\href {\doibase 10.1063/1.3560062} {\bibfield  {journal}
  {\bibinfo  {journal} {Applied Physics Letters}\ }\textbf {\bibinfo {volume}
  {98}},\ \bibinfo {pages} {091106} (\bibinfo {year} {2011})}\BibitemShut
  {NoStop}%
\bibitem [{\citenamefont {Mukai}\ \emph {et~al.}(2016)\citenamefont {Mukai},
  \citenamefont {Hirori}, \citenamefont {Yamamoto}, \citenamefont {Kageyama},\
  and\ \citenamefont {Tanaka}}]{mukai_laser_2016}%
  \BibitemOpen
  \bibfield  {author} {\bibinfo {author} {\bibfnamefont {Y}~\bibnamefont
  {Mukai}}, \bibinfo {author} {\bibfnamefont {H}~\bibnamefont {Hirori}},
  \bibinfo {author} {\bibfnamefont {T}~\bibnamefont {Yamamoto}}, \bibinfo
  {author} {\bibfnamefont {H}~\bibnamefont {Kageyama}}, \ and\ \bibinfo
  {author} {\bibfnamefont {K}~\bibnamefont {Tanaka}},\ }\bibfield  {title}
  {\enquote {\bibinfo {title} {Nonlinear magnetization dynamics of
  antiferromagnetic spin resonance induced by intense terahertz magnetic
  field},}\ }\href {\doibase 10.1088/1367-2630/18/1/013045} {\bibfield
  {journal} {\bibinfo  {journal} {New Journal of Physics}\ }\textbf {\bibinfo
  {volume} {18}},\ \bibinfo {pages} {013045} (\bibinfo {year}
  {2016})}\BibitemShut {NoStop}%
\bibitem [{\citenamefont {Nicoletti}\ and\ \citenamefont
  {Cavalleri}(2016)}]{nicoletti_laser_2016}%
  \BibitemOpen
  \bibfield  {author} {\bibinfo {author} {\bibfnamefont {Daniele}\ \bibnamefont
  {Nicoletti}}\ and\ \bibinfo {author} {\bibfnamefont {Andrea}\ \bibnamefont
  {Cavalleri}},\ }\bibfield  {title} {\enquote {\bibinfo {title} {Nonlinear
  light{\textendash}matter interaction at terahertz frequencies},}\ }\href
  {\doibase 10.1364/aop.8.000401} {\bibfield  {journal} {\bibinfo  {journal}
  {Advances in Optics and Photonics}\ }\textbf {\bibinfo {volume} {8}},\
  \bibinfo {pages} {401} (\bibinfo {year} {2016})}\BibitemShut {NoStop}%
\bibitem [{Note2()}]{Note2}%
  \BibitemOpen
  \bibinfo {note} {Generic results in Sec.~\ref {sec:results_gen} will hold
  also for the spatially nonuniform DC electric fields whereas we will mainly
  focus on the spatially uniform one throughout the paper.}\BibitemShut {Stop}%
\bibitem [{\citenamefont {Nath}\ \emph {et~al.}(2008)\citenamefont {Nath},
  \citenamefont {Tsirlin}, \citenamefont {Rosner},\ and\ \citenamefont
  {Geibel}}]{nath_bacdvopo4_2008}%
  \BibitemOpen
  \bibfield  {author} {\bibinfo {author} {\bibfnamefont {R.}~\bibnamefont
  {Nath}}, \bibinfo {author} {\bibfnamefont {A.~A.}\ \bibnamefont {Tsirlin}},
  \bibinfo {author} {\bibfnamefont {H.}~\bibnamefont {Rosner}}, \ and\ \bibinfo
  {author} {\bibfnamefont {C.}~\bibnamefont {Geibel}},\ }\bibfield  {title}
  {\enquote {\bibinfo {title} {Magnetic properties of
  $\text{BaCdVO}{({\text{PO}}_{4})}_{2}$: A strongly frustrated
  spin-$\frac{1}{2}$ square lattice close to the quantum critical regime},}\
  }\href {\doibase 10.1103/PhysRevB.78.064422} {\bibfield  {journal} {\bibinfo
  {journal} {Phys. Rev. B}\ }\textbf {\bibinfo {volume} {78}},\ \bibinfo
  {pages} {064422} (\bibinfo {year} {2008})}\BibitemShut {NoStop}%
\bibitem [{\citenamefont {Schrieffer}\ and\ \citenamefont
  {Wolff}(1966)}]{schriefferwolf_transf}%
  \BibitemOpen
  \bibfield  {author} {\bibinfo {author} {\bibfnamefont {J.~R.}\ \bibnamefont
  {Schrieffer}}\ and\ \bibinfo {author} {\bibfnamefont {P.~A.}\ \bibnamefont
  {Wolff}},\ }\bibfield  {title} {\enquote {\bibinfo {title} {{Relation between
  the Anderson and Kondo Hamiltonians}},}\ }\href {\doibase
  10.1103/PhysRev.149.491} {\bibfield  {journal} {\bibinfo  {journal} {Phys.
  Rev.}\ }\textbf {\bibinfo {volume} {149}},\ \bibinfo {pages} {491--492}
  (\bibinfo {year} {1966})}\BibitemShut {NoStop}%
\bibitem [{\citenamefont {Slagle}\ and\ \citenamefont
  {Kim}(2017)}]{slagle_deg_pert}%
  \BibitemOpen
  \bibfield  {author} {\bibinfo {author} {\bibfnamefont {Kevin}\ \bibnamefont
  {Slagle}}\ and\ \bibinfo {author} {\bibfnamefont {Yong~Baek}\ \bibnamefont
  {Kim}},\ }\bibfield  {title} {\enquote {\bibinfo {title} {Fracton topological
  order from nearest-neighbor two-spin interactions and dualities},}\ }\href
  {\doibase 10.1103/PhysRevB.96.165106} {\bibfield  {journal} {\bibinfo
  {journal} {Phys. Rev. B}\ }\textbf {\bibinfo {volume} {96}},\ \bibinfo
  {pages} {165106} (\bibinfo {year} {2017})}\BibitemShut {NoStop}%
\bibitem [{\citenamefont {Koch}(2012)}]{koch_exchange}%
  \BibitemOpen
  \bibfield  {author} {\bibinfo {author} {\bibfnamefont {Erik}\ \bibnamefont
  {Koch}},\ }\bibfield  {title} {\enquote {\bibinfo {title} {Exchange
  mechanisms},}\ }\href@noop {} {\bibfield  {journal} {\bibinfo  {journal}
  {Correlated electrons: from models to materials}\ }\textbf {\bibinfo {volume}
  {2}},\ \bibinfo {pages} {1--31} (\bibinfo {year} {2012})}\BibitemShut
  {NoStop}%
\bibitem [{\citenamefont {Kim}\ \emph {et~al.}(2017)\citenamefont {Kim},
  \citenamefont {Khmelevskyi}, \citenamefont {Mohn},\ and\ \citenamefont
  {Franchini}}]{kim_sr2vo4}%
  \BibitemOpen
  \bibfield  {author} {\bibinfo {author} {\bibfnamefont {Bongjae}\ \bibnamefont
  {Kim}}, \bibinfo {author} {\bibfnamefont {Sergii}\ \bibnamefont
  {Khmelevskyi}}, \bibinfo {author} {\bibfnamefont {Peter}\ \bibnamefont
  {Mohn}}, \ and\ \bibinfo {author} {\bibfnamefont {Cesare}\ \bibnamefont
  {Franchini}},\ }\bibfield  {title} {\enquote {\bibinfo {title} {{Competing
  magnetic interactions in a spin-$\frac{1}{2}$ square lattice: Hidden order in
  ${\mathrm{Sr}}_{2}{\mathrm{VO}}_{4}$}},}\ }\href {\doibase
  10.1103/PhysRevB.96.180405} {\bibfield  {journal} {\bibinfo  {journal} {Phys.
  Rev. B}\ }\textbf {\bibinfo {volume} {96}},\ \bibinfo {pages} {180405}
  (\bibinfo {year} {2017})}\BibitemShut {NoStop}%
\bibitem [{Note3()}]{Note3}%
  \BibitemOpen
  \bibinfo {note} {Even when the spins $\protect \bm {S}_0$ and $\protect \bm
  {S}_1$ come from $p$ orbitals, our conclusion will not change since we hardly
  rely on the fact that the magnetic moment is attributed to the $d$ orbital.
  We will come back to this point later in Sec.~\ref
  {sec:toy2_spin}.}\BibitemShut {Stop}%
\bibitem [{\citenamefont {Nakamura}\ \emph {et~al.}(2009)\citenamefont
  {Nakamura}, \citenamefont {Yoshimoto}, \citenamefont {Kosugi}, \citenamefont
  {Arita},\ and\ \citenamefont {Imada}}]{nakamura_k-et}%
  \BibitemOpen
  \bibfield  {author} {\bibinfo {author} {\bibfnamefont {Kazuma}\ \bibnamefont
  {Nakamura}}, \bibinfo {author} {\bibfnamefont {Yoshihide}\ \bibnamefont
  {Yoshimoto}}, \bibinfo {author} {\bibfnamefont {Taichi}\ \bibnamefont
  {Kosugi}}, \bibinfo {author} {\bibfnamefont {Ryotaro}\ \bibnamefont {Arita}},
  \ and\ \bibinfo {author} {\bibfnamefont {Masatoshi}\ \bibnamefont {Imada}},\
  }\bibfield  {title} {\enquote {\bibinfo {title} {{Ab initio Derivation of
  Low-Energy Model for $\kappa$-ET Type Organic Conductors}},}\ }\href
  {\doibase 10.1143/JPSJ.78.083710} {\bibfield  {journal} {\bibinfo  {journal}
  {Journal of the Physical Society of Japan}\ }\textbf {\bibinfo {volume}
  {78}},\ \bibinfo {pages} {083710} (\bibinfo {year} {2009})}\BibitemShut
  {NoStop}%
\bibitem [{\citenamefont {Shimizu}\ \emph {et~al.}(2003)\citenamefont
  {Shimizu}, \citenamefont {Miyagawa}, \citenamefont {Kanoda}, \citenamefont
  {Maesato},\ and\ \citenamefont {Saito}}]{shimizu_organic_triangular}%
  \BibitemOpen
  \bibfield  {author} {\bibinfo {author} {\bibfnamefont {Y.}~\bibnamefont
  {Shimizu}}, \bibinfo {author} {\bibfnamefont {K.}~\bibnamefont {Miyagawa}},
  \bibinfo {author} {\bibfnamefont {K.}~\bibnamefont {Kanoda}}, \bibinfo
  {author} {\bibfnamefont {M.}~\bibnamefont {Maesato}}, \ and\ \bibinfo
  {author} {\bibfnamefont {G.}~\bibnamefont {Saito}},\ }\bibfield  {title}
  {\enquote {\bibinfo {title} {{Spin Liquid State in an Organic Mott Insulator
  with a Triangular Lattice}},}\ }\href {\doibase
  10.1103/PhysRevLett.91.107001} {\bibfield  {journal} {\bibinfo  {journal}
  {Phys. Rev. Lett.}\ }\textbf {\bibinfo {volume} {91}},\ \bibinfo {pages}
  {107001} (\bibinfo {year} {2003})}\BibitemShut {NoStop}%
\bibitem [{Note4()}]{Note4}%
  \BibitemOpen
  \bibinfo {note} {The THz laser pulse typically has the $O(1)$~ps pulse width,
  which is much longer than the typical time scale of hoppings. The hopping
  amplitude is $O(10^{-1}U_d)=O(10^{-1})$~eV. Accordingly, the time scale of
  hoppings is $O(10^1)$~fs $=O(10^{-2})$~ps. In other words, electrons can hop
  $O(10^2)$ times during the single-cycle THz laser pulse is applied to the
  system. Under such circumstances, we may regard the THz laser pulse as an
  effective DC electric field.}\BibitemShut {Stop}%
\bibitem [{\citenamefont {F\"ul\"op}\ \emph {et~al.}(2020)\citenamefont
  {F\"ul\"op}, \citenamefont {Tzortzakis},\ and\ \citenamefont
  {Kampfrath}}]{fulop_terahertz}%
  \BibitemOpen
  \bibfield  {author} {\bibinfo {author} {\bibfnamefont {J\'azsef~Andr\'as}\
  \bibnamefont {F\"ul\"op}}, \bibinfo {author} {\bibfnamefont {Stelios}\
  \bibnamefont {Tzortzakis}}, \ and\ \bibinfo {author} {\bibfnamefont {Tobias}\
  \bibnamefont {Kampfrath}},\ }\bibfield  {title} {\enquote {\bibinfo {title}
  {{Laser-Driven Strong-Field Terahertz Sources}},}\ }\href {\doibase
  https://doi.org/10.1002/adom.201900681} {\bibfield  {journal} {\bibinfo
  {journal} {Advanced Optical Materials}\ }\textbf {\bibinfo {volume} {8}},\
  \bibinfo {pages} {1900681} (\bibinfo {year} {2020})}\BibitemShut {NoStop}%
\bibitem [{\citenamefont {Schmidt}\ \emph {et~al.}(2007)\citenamefont
  {Schmidt}, \citenamefont {Thalmeier},\ and\ \citenamefont
  {Shannon}}]{schmidt_square}%
  \BibitemOpen
  \bibfield  {author} {\bibinfo {author} {\bibfnamefont {B.}~\bibnamefont
  {Schmidt}}, \bibinfo {author} {\bibfnamefont {P.}~\bibnamefont {Thalmeier}},
  \ and\ \bibinfo {author} {\bibfnamefont {Nic}\ \bibnamefont {Shannon}},\
  }\bibfield  {title} {\enquote {\bibinfo {title} {Magnetocaloric effect in the
  frustrated square lattice ${J}_{1}\text{\ensuremath{-}}{J}_{2}$ model},}\
  }\href {\doibase 10.1103/PhysRevB.76.125113} {\bibfield  {journal} {\bibinfo
  {journal} {Phys. Rev. B}\ }\textbf {\bibinfo {volume} {76}},\ \bibinfo
  {pages} {125113} (\bibinfo {year} {2007})}\BibitemShut {NoStop}%
\bibitem [{\citenamefont {Chandra}\ and\ \citenamefont
  {Doucot}(1988)}]{chandra_sq_afm_1988}%
  \BibitemOpen
  \bibfield  {author} {\bibinfo {author} {\bibfnamefont {P.}~\bibnamefont
  {Chandra}}\ and\ \bibinfo {author} {\bibfnamefont {B.}~\bibnamefont
  {Doucot}},\ }\bibfield  {title} {\enquote {\bibinfo {title} {{Possible
  spin-liquid state at large $S$ for the frustrated square Heisenberg
  lattice}},}\ }\href {\doibase 10.1103/PhysRevB.38.9335} {\bibfield  {journal}
  {\bibinfo  {journal} {Phys. Rev. B}\ }\textbf {\bibinfo {volume} {38}},\
  \bibinfo {pages} {9335--9338} (\bibinfo {year} {1988})}\BibitemShut {NoStop}%
\bibitem [{\citenamefont {Dagotto}\ and\ \citenamefont
  {Moreo}(1989)}]{dagotto_sq_afm_1989}%
  \BibitemOpen
  \bibfield  {author} {\bibinfo {author} {\bibfnamefont {Elbio}\ \bibnamefont
  {Dagotto}}\ and\ \bibinfo {author} {\bibfnamefont {Adriana}\ \bibnamefont
  {Moreo}},\ }\bibfield  {title} {\enquote {\bibinfo {title} {{Phase diagram of
  the frustrated spin-1/2 Heisenberg antiferromagnet in 2 dimensions}},}\
  }\href {\doibase 10.1103/PhysRevLett.63.2148} {\bibfield  {journal} {\bibinfo
   {journal} {Phys. Rev. Lett.}\ }\textbf {\bibinfo {volume} {63}},\ \bibinfo
  {pages} {2148--2151} (\bibinfo {year} {1989})}\BibitemShut {NoStop}%
\bibitem [{\citenamefont {Read}\ and\ \citenamefont
  {Sachdev}(1991)}]{read_sq_afm_1991}%
  \BibitemOpen
  \bibfield  {author} {\bibinfo {author} {\bibfnamefont {N.}~\bibnamefont
  {Read}}\ and\ \bibinfo {author} {\bibfnamefont {Subir}\ \bibnamefont
  {Sachdev}},\ }\bibfield  {title} {\enquote {\bibinfo {title} {{Large-N
  expansion for frustrated quantum antiferromagnets}},}\ }\href {\doibase
  10.1103/PhysRevLett.66.1773} {\bibfield  {journal} {\bibinfo  {journal}
  {Phys. Rev. Lett.}\ }\textbf {\bibinfo {volume} {66}},\ \bibinfo {pages}
  {1773--1776} (\bibinfo {year} {1991})}\BibitemShut {NoStop}%
\bibitem [{\citenamefont {Nomura}\ and\ \citenamefont
  {Okamoto}(1994)}]{nomura_zigzag_1994}%
  \BibitemOpen
  \bibfield  {author} {\bibinfo {author} {\bibfnamefont {K}~\bibnamefont
  {Nomura}}\ and\ \bibinfo {author} {\bibfnamefont {K}~\bibnamefont
  {Okamoto}},\ }\bibfield  {title} {\enquote {\bibinfo {title} {{Critical
  properties of S= 1/2 antiferromagnetic {XXZ} chain with
  next-nearest-neighbour interactions}},}\ }\href {\doibase
  10.1088/0305-4470/27/17/012} {\bibfield  {journal} {\bibinfo  {journal}
  {Journal of Physics A: Mathematical and General}\ }\textbf {\bibinfo {volume}
  {27}},\ \bibinfo {pages} {5773--5788} (\bibinfo {year} {1994})}\BibitemShut
  {NoStop}%
\bibitem [{\citenamefont {Singh}\ \emph {et~al.}(1999)\citenamefont {Singh},
  \citenamefont {Weihong}, \citenamefont {Hamer},\ and\ \citenamefont
  {Oitmaa}}]{singh_sq_afm_1999}%
  \BibitemOpen
  \bibfield  {author} {\bibinfo {author} {\bibfnamefont {Rajiv R.~P.}\
  \bibnamefont {Singh}}, \bibinfo {author} {\bibfnamefont {Zheng}\ \bibnamefont
  {Weihong}}, \bibinfo {author} {\bibfnamefont {C.~J.}\ \bibnamefont {Hamer}},
  \ and\ \bibinfo {author} {\bibfnamefont {J.}~\bibnamefont {Oitmaa}},\
  }\bibfield  {title} {\enquote {\bibinfo {title} {Dimer order with striped
  correlations in the ${J}_{1}{\ensuremath{-}j}_{2}$ heisenberg model},}\
  }\href {\doibase 10.1103/PhysRevB.60.7278} {\bibfield  {journal} {\bibinfo
  {journal} {Phys. Rev. B}\ }\textbf {\bibinfo {volume} {60}},\ \bibinfo
  {pages} {7278--7283} (\bibinfo {year} {1999})}\BibitemShut {NoStop}%
\bibitem [{\citenamefont {Capriotti}\ and\ \citenamefont
  {Sorella}(2000)}]{capriotti_sq_afm_2000}%
  \BibitemOpen
  \bibfield  {author} {\bibinfo {author} {\bibfnamefont {Luca}\ \bibnamefont
  {Capriotti}}\ and\ \bibinfo {author} {\bibfnamefont {Sandro}\ \bibnamefont
  {Sorella}},\ }\bibfield  {title} {\enquote {\bibinfo {title} {{Spontaneous
  Plaquette Dimerization in the ${\mathit{J}}_{1}--{\mathit{J}}_{2}$ Heisenberg
  Model}},}\ }\href {\doibase 10.1103/PhysRevLett.84.3173} {\bibfield
  {journal} {\bibinfo  {journal} {Phys. Rev. Lett.}\ }\textbf {\bibinfo
  {volume} {84}},\ \bibinfo {pages} {3173--3176} (\bibinfo {year}
  {2000})}\BibitemShut {NoStop}%
\bibitem [{\citenamefont {Sirker}\ \emph {et~al.}(2006)\citenamefont {Sirker},
  \citenamefont {Weihong}, \citenamefont {Sushkov},\ and\ \citenamefont
  {Oitmaa}}]{sirker_sq_afm_2006}%
  \BibitemOpen
  \bibfield  {author} {\bibinfo {author} {\bibfnamefont {J.}~\bibnamefont
  {Sirker}}, \bibinfo {author} {\bibfnamefont {Zheng}\ \bibnamefont {Weihong}},
  \bibinfo {author} {\bibfnamefont {O.~P.}\ \bibnamefont {Sushkov}}, \ and\
  \bibinfo {author} {\bibfnamefont {J.}~\bibnamefont {Oitmaa}},\ }\bibfield
  {title} {\enquote {\bibinfo {title} {{$J1\text{\ensuremath{-}}J2$ model:
  First-order phase transition versus deconfinement of spinons}},}\ }\href
  {\doibase 10.1103/PhysRevB.73.184420} {\bibfield  {journal} {\bibinfo
  {journal} {Phys. Rev. B}\ }\textbf {\bibinfo {volume} {73}},\ \bibinfo
  {pages} {184420} (\bibinfo {year} {2006})}\BibitemShut {NoStop}%
\bibitem [{\citenamefont {Ueda}\ and\ \citenamefont
  {Totsuka}(2007)}]{ueda_j1j2}%
  \BibitemOpen
  \bibfield  {author} {\bibinfo {author} {\bibfnamefont {Hiroaki~T.}\
  \bibnamefont {Ueda}}\ and\ \bibinfo {author} {\bibfnamefont {Keisuke}\
  \bibnamefont {Totsuka}},\ }\bibfield  {title} {\enquote {\bibinfo {title}
  {Ground-state phase diagram and magnetic properties of a tetramerized
  spin-$1/2$ ${J}_{1}\text{\ensuremath{-}}{J}_{2}$ model: Bec of bound magnons
  and absence of the transverse magnetization},}\ }\href {\doibase
  10.1103/PhysRevB.76.214428} {\bibfield  {journal} {\bibinfo  {journal} {Phys.
  Rev. B}\ }\textbf {\bibinfo {volume} {76}},\ \bibinfo {pages} {214428}
  (\bibinfo {year} {2007})}\BibitemShut {NoStop}%
\bibitem [{\citenamefont {Hu}\ \emph {et~al.}(2013)\citenamefont {Hu},
  \citenamefont {Becca}, \citenamefont {Parola},\ and\ \citenamefont
  {Sorella}}]{hu_sq_afm_2013}%
  \BibitemOpen
  \bibfield  {author} {\bibinfo {author} {\bibfnamefont {Wen-Jun}\ \bibnamefont
  {Hu}}, \bibinfo {author} {\bibfnamefont {Federico}\ \bibnamefont {Becca}},
  \bibinfo {author} {\bibfnamefont {Alberto}\ \bibnamefont {Parola}}, \ and\
  \bibinfo {author} {\bibfnamefont {Sandro}\ \bibnamefont {Sorella}},\
  }\bibfield  {title} {\enquote {\bibinfo {title} {Direct evidence for a
  gapless ${Z}_{2}$ spin liquid by frustrating n\'eel antiferromagnetism},}\
  }\href {\doibase 10.1103/PhysRevB.88.060402} {\bibfield  {journal} {\bibinfo
  {journal} {Phys. Rev. B}\ }\textbf {\bibinfo {volume} {88}},\ \bibinfo
  {pages} {060402} (\bibinfo {year} {2013})}\BibitemShut {NoStop}%
\bibitem [{\citenamefont {Wang}\ \emph {et~al.}(2016)\citenamefont {Wang},
  \citenamefont {Gu}, \citenamefont {Verstraete},\ and\ \citenamefont
  {Wen}}]{wang_sq_afm_2016}%
  \BibitemOpen
  \bibfield  {author} {\bibinfo {author} {\bibfnamefont {Ling}\ \bibnamefont
  {Wang}}, \bibinfo {author} {\bibfnamefont {Zheng-Cheng}\ \bibnamefont {Gu}},
  \bibinfo {author} {\bibfnamefont {Frank}\ \bibnamefont {Verstraete}}, \ and\
  \bibinfo {author} {\bibfnamefont {Xiao-Gang}\ \bibnamefont {Wen}},\
  }\bibfield  {title} {\enquote {\bibinfo {title} {Tensor-product state
  approach to spin-$\frac{1}{2}$ square ${J}_{1}\text{\ensuremath{-}}{J}_{2}$
  antiferromagnetic heisenberg model: Evidence for deconfined quantum
  criticality},}\ }\href {\doibase 10.1103/PhysRevB.94.075143} {\bibfield
  {journal} {\bibinfo  {journal} {Phys. Rev. B}\ }\textbf {\bibinfo {volume}
  {94}},\ \bibinfo {pages} {075143} (\bibinfo {year} {2016})}\BibitemShut
  {NoStop}%
\bibitem [{\citenamefont {Wang}\ and\ \citenamefont
  {Sandvik}(2018)}]{wang_sq_afm_2018}%
  \BibitemOpen
  \bibfield  {author} {\bibinfo {author} {\bibfnamefont {Ling}\ \bibnamefont
  {Wang}}\ and\ \bibinfo {author} {\bibfnamefont {Anders~W.}\ \bibnamefont
  {Sandvik}},\ }\bibfield  {title} {\enquote {\bibinfo {title} {Critical level
  crossings and gapless spin liquid in the square-lattice spin-$1/2$
  ${J}_{1}\ensuremath{-}{J}_{2}$ heisenberg antiferromagnet},}\ }\href
  {\doibase 10.1103/PhysRevLett.121.107202} {\bibfield  {journal} {\bibinfo
  {journal} {Phys. Rev. Lett.}\ }\textbf {\bibinfo {volume} {121}},\ \bibinfo
  {pages} {107202} (\bibinfo {year} {2018})}\BibitemShut {NoStop}%
\bibitem [{\citenamefont {Shindou}\ \emph {et~al.}(2011)\citenamefont
  {Shindou}, \citenamefont {Yunoki},\ and\ \citenamefont
  {Momoi}}]{shindou_sq_fm_2011}%
  \BibitemOpen
  \bibfield  {author} {\bibinfo {author} {\bibfnamefont {Ryuichi}\ \bibnamefont
  {Shindou}}, \bibinfo {author} {\bibfnamefont {Seiji}\ \bibnamefont {Yunoki}},
  \ and\ \bibinfo {author} {\bibfnamefont {Tsutomu}\ \bibnamefont {Momoi}},\
  }\bibfield  {title} {\enquote {\bibinfo {title} {Projective studies of spin
  nematics in a quantum frustrated ferromagnet},}\ }\href {\doibase
  10.1103/PhysRevB.84.134414} {\bibfield  {journal} {\bibinfo  {journal} {Phys.
  Rev. B}\ }\textbf {\bibinfo {volume} {84}},\ \bibinfo {pages} {134414}
  (\bibinfo {year} {2011})}\BibitemShut {NoStop}%
\bibitem [{\citenamefont {Iqbal}\ \emph {et~al.}(2016)\citenamefont {Iqbal},
  \citenamefont {Ghosh}, \citenamefont {Narayanan}, \citenamefont {Kumar},
  \citenamefont {Reuther},\ and\ \citenamefont {Thomale}}]{iqbal_sq_fm_2016}%
  \BibitemOpen
  \bibfield  {author} {\bibinfo {author} {\bibfnamefont {Yasir}\ \bibnamefont
  {Iqbal}}, \bibinfo {author} {\bibfnamefont {Pratyay}\ \bibnamefont {Ghosh}},
  \bibinfo {author} {\bibfnamefont {Rajesh}\ \bibnamefont {Narayanan}},
  \bibinfo {author} {\bibfnamefont {Brijesh}\ \bibnamefont {Kumar}}, \bibinfo
  {author} {\bibfnamefont {Johannes}\ \bibnamefont {Reuther}}, \ and\ \bibinfo
  {author} {\bibfnamefont {Ronny}\ \bibnamefont {Thomale}},\ }\bibfield
  {title} {\enquote {\bibinfo {title} {Intertwined nematic orders in a
  frustrated ferromagnet},}\ }\href {\doibase 10.1103/PhysRevB.94.224403}
  {\bibfield  {journal} {\bibinfo  {journal} {Phys. Rev. B}\ }\textbf {\bibinfo
  {volume} {94}},\ \bibinfo {pages} {224403} (\bibinfo {year}
  {2016})}\BibitemShut {NoStop}%
\bibitem [{\citenamefont {Sato}\ and\ \citenamefont
  {Morisaku}(2020)}]{morisaku_nematic_laser}%
  \BibitemOpen
  \bibfield  {author} {\bibinfo {author} {\bibfnamefont {Masahiro}\
  \bibnamefont {Sato}}\ and\ \bibinfo {author} {\bibfnamefont {Yoshitaka}\
  \bibnamefont {Morisaku}},\ }\bibfield  {title} {\enquote {\bibinfo {title}
  {Two-photon driven magnon-pair resonance as a signature of spin-nematic
  order},}\ }\href {\doibase 10.1103/PhysRevB.102.060401} {\bibfield  {journal}
  {\bibinfo  {journal} {Phys. Rev. B}\ }\textbf {\bibinfo {volume} {102}},\
  \bibinfo {pages} {060401} (\bibinfo {year} {2020})}\BibitemShut {NoStop}%
\bibitem [{\citenamefont {Melzi}\ \emph {et~al.}(2001)\citenamefont {Melzi},
  \citenamefont {Aldrovandi}, \citenamefont {Tedoldi}, \citenamefont
  {Carretta}, \citenamefont {Millet},\ and\ \citenamefont
  {Mila}}]{melzi_J1J2_sq_2001}%
  \BibitemOpen
  \bibfield  {author} {\bibinfo {author} {\bibfnamefont {R.}~\bibnamefont
  {Melzi}}, \bibinfo {author} {\bibfnamefont {S.}~\bibnamefont {Aldrovandi}},
  \bibinfo {author} {\bibfnamefont {F.}~\bibnamefont {Tedoldi}}, \bibinfo
  {author} {\bibfnamefont {P.}~\bibnamefont {Carretta}}, \bibinfo {author}
  {\bibfnamefont {P.}~\bibnamefont {Millet}}, \ and\ \bibinfo {author}
  {\bibfnamefont {F.}~\bibnamefont {Mila}},\ }\bibfield  {title} {\enquote
  {\bibinfo {title} {{Magnetic and thermodynamic properties of
  ${\mathrm{Li}}_{2}{\mathrm{VOSiO}}_{4}:$ A two-dimensional $S=1/2$ frustrated
  antiferromagnet on a square lattice}},}\ }\href {\doibase
  10.1103/PhysRevB.64.024409} {\bibfield  {journal} {\bibinfo  {journal} {Phys.
  Rev. B}\ }\textbf {\bibinfo {volume} {64}},\ \bibinfo {pages} {024409}
  (\bibinfo {year} {2001})}\BibitemShut {NoStop}%
\bibitem [{\citenamefont {Rosner}\ \emph {et~al.}(2003)\citenamefont {Rosner},
  \citenamefont {Singh}, \citenamefont {Zheng}, \citenamefont {Oitmaa},\ and\
  \citenamefont {Pickett}}]{Rosner_J1J2_sq_2003}%
  \BibitemOpen
  \bibfield  {author} {\bibinfo {author} {\bibfnamefont {H.}~\bibnamefont
  {Rosner}}, \bibinfo {author} {\bibfnamefont {R.~R.~P.}\ \bibnamefont
  {Singh}}, \bibinfo {author} {\bibfnamefont {W.~H.}\ \bibnamefont {Zheng}},
  \bibinfo {author} {\bibfnamefont {J.}~\bibnamefont {Oitmaa}}, \ and\ \bibinfo
  {author} {\bibfnamefont {W.~E.}\ \bibnamefont {Pickett}},\ }\bibfield
  {title} {\enquote {\bibinfo {title} {{High-temperature expansions for the
  ${J}_{1}-{J}_{2}$ Heisenberg models: Applications to ab initio calculated
  models for ${\mathrm{Li}}_{2}{\mathrm{VOSiO}}_{4}$ and
  ${\mathrm{Li}}_{2}{\mathrm{VOGeO}}_{4}$}},}\ }\href {\doibase
  10.1103/PhysRevB.67.014416} {\bibfield  {journal} {\bibinfo  {journal} {Phys.
  Rev. B}\ }\textbf {\bibinfo {volume} {67}},\ \bibinfo {pages} {014416}
  (\bibinfo {year} {2003})}\BibitemShut {NoStop}%
\bibitem [{\citenamefont {Kaul}\ \emph {et~al.}(2004)\citenamefont {Kaul},
  \citenamefont {Rosner}, \citenamefont {Shannon}, \citenamefont
  {Shpanchenko},\ and\ \citenamefont {Geibel}}]{Kaul_J1J2_sq_2004}%
  \BibitemOpen
  \bibfield  {author} {\bibinfo {author} {\bibfnamefont {E.E.}\ \bibnamefont
  {Kaul}}, \bibinfo {author} {\bibfnamefont {H.}~\bibnamefont {Rosner}},
  \bibinfo {author} {\bibfnamefont {N.}~\bibnamefont {Shannon}}, \bibinfo
  {author} {\bibfnamefont {R.V.}\ \bibnamefont {Shpanchenko}}, \ and\ \bibinfo
  {author} {\bibfnamefont {C.}~\bibnamefont {Geibel}},\ }\bibfield  {title}
  {\enquote {\bibinfo {title} {{Evidence for a frustrated square lattice with
  ferromagnetic nearest-neighbor interaction in the new compound
  Pb2VO(PO4)2}},}\ }\href {\doibase https://doi.org/10.1016/j.jmmm.2003.12.002}
  {\bibfield  {journal} {\bibinfo  {journal} {Journal of Magnetism and Magnetic
  Materials}\ }\textbf {\bibinfo {volume} {272-276}},\ \bibinfo {pages}
  {922--923} (\bibinfo {year} {2004})},\ \bibinfo {note} {proceedings of the
  International Conference on Magnetism (ICM 2003)}\BibitemShut {NoStop}%
\bibitem [{\citenamefont {Bombardi}\ \emph {et~al.}(2004)\citenamefont
  {Bombardi}, \citenamefont {Rodriguez-Carvajal}, \citenamefont {Di~Matteo},
  \citenamefont {de~Bergevin}, \citenamefont {Paolasini}, \citenamefont
  {Carretta}, \citenamefont {Millet},\ and\ \citenamefont
  {Caciuffo}}]{bombardi_j1j2_2004}%
  \BibitemOpen
  \bibfield  {author} {\bibinfo {author} {\bibfnamefont {A.}~\bibnamefont
  {Bombardi}}, \bibinfo {author} {\bibfnamefont {J.}~\bibnamefont
  {Rodriguez-Carvajal}}, \bibinfo {author} {\bibfnamefont {S.}~\bibnamefont
  {Di~Matteo}}, \bibinfo {author} {\bibfnamefont {F.}~\bibnamefont
  {de~Bergevin}}, \bibinfo {author} {\bibfnamefont {L.}~\bibnamefont
  {Paolasini}}, \bibinfo {author} {\bibfnamefont {P.}~\bibnamefont {Carretta}},
  \bibinfo {author} {\bibfnamefont {P.}~\bibnamefont {Millet}}, \ and\ \bibinfo
  {author} {\bibfnamefont {R.}~\bibnamefont {Caciuffo}},\ }\bibfield  {title}
  {\enquote {\bibinfo {title} {{Direct Determination of the Magnetic Ground
  State in the Square Lattice $S=1/2$ Antiferromagnet
  ${\mathrm{L}\mathrm{i}}_{2}{\mathrm{V}\mathrm{O}\mathrm{S}\mathrm{i}\mathrm{O}}_{4}$}},}\
  }\href {\doibase 10.1103/PhysRevLett.93.027202} {\bibfield  {journal}
  {\bibinfo  {journal} {Phys. Rev. Lett.}\ }\textbf {\bibinfo {volume} {93}},\
  \bibinfo {pages} {027202} (\bibinfo {year} {2004})}\BibitemShut {NoStop}%
\bibitem [{\citenamefont {Bombardi}\ \emph {et~al.}(2005)\citenamefont
  {Bombardi}, \citenamefont {Chapon}, \citenamefont {Margiolaki}, \citenamefont
  {Mazzoli}, \citenamefont {Gonthier}, \citenamefont {Duc},\ and\ \citenamefont
  {Radaelli}}]{bombardi_j1j2_2005}%
  \BibitemOpen
  \bibfield  {author} {\bibinfo {author} {\bibfnamefont {A.}~\bibnamefont
  {Bombardi}}, \bibinfo {author} {\bibfnamefont {L.~C.}\ \bibnamefont
  {Chapon}}, \bibinfo {author} {\bibfnamefont {I.}~\bibnamefont {Margiolaki}},
  \bibinfo {author} {\bibfnamefont {C.}~\bibnamefont {Mazzoli}}, \bibinfo
  {author} {\bibfnamefont {S.}~\bibnamefont {Gonthier}}, \bibinfo {author}
  {\bibfnamefont {F.}~\bibnamefont {Duc}}, \ and\ \bibinfo {author}
  {\bibfnamefont {P.~G.}\ \bibnamefont {Radaelli}},\ }\bibfield  {title}
  {\enquote {\bibinfo {title} {{Magnetic order and lattice anomalies in the
  ${J}_{1}\text{\ensuremath{-}}{J}_{2}$ model system
  $\mathrm{V}\mathrm{O}\mathrm{Mo}{\mathrm{O}}_{4}$}},}\ }\href {\doibase
  10.1103/PhysRevB.71.220406} {\bibfield  {journal} {\bibinfo  {journal} {Phys.
  Rev. B}\ }\textbf {\bibinfo {volume} {71}},\ \bibinfo {pages} {220406}
  (\bibinfo {year} {2005})}\BibitemShut {NoStop}%
\bibitem [{\citenamefont {Kageyama}\ \emph {et~al.}(2005)\citenamefont
  {Kageyama}, \citenamefont {Kitano}, \citenamefont {Oba}, \citenamefont
  {Nishi}, \citenamefont {Nagai}, \citenamefont {Hirota}, \citenamefont
  {Viciu}, \citenamefont {Wiley}, \citenamefont {Yasuda}, \citenamefont {Baba},
  \citenamefont {Ajiro},\ and\ \citenamefont {Yoshimura}}]{kageyama_j1j2_2005}%
  \BibitemOpen
  \bibfield  {author} {\bibinfo {author} {\bibfnamefont {H.}~\bibnamefont
  {Kageyama}}, \bibinfo {author} {\bibfnamefont {T.}~\bibnamefont {Kitano}},
  \bibinfo {author} {\bibfnamefont {N.}~\bibnamefont {Oba}}, \bibinfo {author}
  {\bibfnamefont {M.}~\bibnamefont {Nishi}}, \bibinfo {author} {\bibfnamefont
  {S.}~\bibnamefont {Nagai}}, \bibinfo {author} {\bibfnamefont
  {K.}~\bibnamefont {Hirota}}, \bibinfo {author} {\bibfnamefont
  {L.}~\bibnamefont {Viciu}}, \bibinfo {author} {\bibfnamefont {J.~B.}\
  \bibnamefont {Wiley}}, \bibinfo {author} {\bibfnamefont {J.}~\bibnamefont
  {Yasuda}}, \bibinfo {author} {\bibfnamefont {Y.}~\bibnamefont {Baba}},
  \bibinfo {author} {\bibfnamefont {Y.}~\bibnamefont {Ajiro}}, \ and\ \bibinfo
  {author} {\bibfnamefont {K.}~\bibnamefont {Yoshimura}},\ }\bibfield  {title}
  {\enquote {\bibinfo {title} {{Spin-Singlet Ground State in Two-Dimensional
  S=1/2 Frustrated Square Lattice: ({CuCl}){LaNb}2O7}},}\ }\href {\doibase
  10.1143/jpsj.74.1702} {\bibfield  {journal} {\bibinfo  {journal} {Journal of
  the Physical Society of Japan}\ }\textbf {\bibinfo {volume} {74}},\ \bibinfo
  {pages} {1702--1705} (\bibinfo {year} {2005})}\BibitemShut {NoStop}%
\bibitem [{\citenamefont {Oba}\ \emph {et~al.}(2006)\citenamefont {Oba},
  \citenamefont {Kageyama}, \citenamefont {Kitano}, \citenamefont {Yasuda},
  \citenamefont {Baba}, \citenamefont {Nishi}, \citenamefont {Hirota},
  \citenamefont {Narumi}, \citenamefont {Hagiwara}, \citenamefont {Kindo},
  \citenamefont {Saito}, \citenamefont {Ajiro},\ and\ \citenamefont
  {Yoshimura}}]{oba_j1j2_2006}%
  \BibitemOpen
  \bibfield  {author} {\bibinfo {author} {\bibfnamefont {Noriaki}\ \bibnamefont
  {Oba}}, \bibinfo {author} {\bibfnamefont {Hiroshi}\ \bibnamefont {Kageyama}},
  \bibinfo {author} {\bibfnamefont {Taro}\ \bibnamefont {Kitano}}, \bibinfo
  {author} {\bibfnamefont {Jun}\ \bibnamefont {Yasuda}}, \bibinfo {author}
  {\bibfnamefont {Yoichi}\ \bibnamefont {Baba}}, \bibinfo {author}
  {\bibfnamefont {Masakazu}\ \bibnamefont {Nishi}}, \bibinfo {author}
  {\bibfnamefont {Kazuma}\ \bibnamefont {Hirota}}, \bibinfo {author}
  {\bibfnamefont {Yasuo}\ \bibnamefont {Narumi}}, \bibinfo {author}
  {\bibfnamefont {Masayuki}\ \bibnamefont {Hagiwara}}, \bibinfo {author}
  {\bibfnamefont {Koichi}\ \bibnamefont {Kindo}}, \bibinfo {author}
  {\bibfnamefont {Takashi}\ \bibnamefont {Saito}}, \bibinfo {author}
  {\bibfnamefont {Yoshitami}\ \bibnamefont {Ajiro}}, \ and\ \bibinfo {author}
  {\bibfnamefont {Kazuyoshi}\ \bibnamefont {Yoshimura}},\ }\bibfield  {title}
  {\enquote {\bibinfo {title} {{Collinear Order in Frustrated Quantum
  Antiferromagnet on Square Lattice ({CuBr}){LaNb}2O7}},}\ }\href {\doibase
  10.1143/jpsj.75.113601} {\bibfield  {journal} {\bibinfo  {journal} {Journal
  of the Physical Society of Japan}\ }\textbf {\bibinfo {volume} {75}},\
  \bibinfo {pages} {113601} (\bibinfo {year} {2006})}\BibitemShut {NoStop}%
\bibitem [{\citenamefont {Tsirlin}\ \emph {et~al.}(2008)\citenamefont
  {Tsirlin}, \citenamefont {Belik}, \citenamefont {Shpanchenko}, \citenamefont
  {Antipov}, \citenamefont {Takayama-Muromachi},\ and\ \citenamefont
  {Rosner}}]{tsirlin_j1j2_2008}%
  \BibitemOpen
  \bibfield  {author} {\bibinfo {author} {\bibfnamefont {Alexander~A.}\
  \bibnamefont {Tsirlin}}, \bibinfo {author} {\bibfnamefont {Alexei~A.}\
  \bibnamefont {Belik}}, \bibinfo {author} {\bibfnamefont {Roman~V.}\
  \bibnamefont {Shpanchenko}}, \bibinfo {author} {\bibfnamefont {Evgeny~V.}\
  \bibnamefont {Antipov}}, \bibinfo {author} {\bibfnamefont {Eiji}\
  \bibnamefont {Takayama-Muromachi}}, \ and\ \bibinfo {author} {\bibfnamefont
  {Helge}\ \bibnamefont {Rosner}},\ }\bibfield  {title} {\enquote {\bibinfo
  {title} {{Frustrated spin-$1/2$ square lattice in the layered perovskite
  $\mathrm{Pb}\mathrm{V}{\mathrm{O}}_{3}$}},}\ }\href {\doibase
  10.1103/PhysRevB.77.092402} {\bibfield  {journal} {\bibinfo  {journal} {Phys.
  Rev. B}\ }\textbf {\bibinfo {volume} {77}},\ \bibinfo {pages} {092402}
  (\bibinfo {year} {2008})}\BibitemShut {NoStop}%
\bibitem [{\citenamefont {Tsirlin}\ and\ \citenamefont
  {Rosner}(2009)}]{tsirlin_J1J2_sq_2009}%
  \BibitemOpen
  \bibfield  {author} {\bibinfo {author} {\bibfnamefont {Alexander~A.}\
  \bibnamefont {Tsirlin}}\ and\ \bibinfo {author} {\bibfnamefont {Helge}\
  \bibnamefont {Rosner}},\ }\bibfield  {title} {\enquote {\bibinfo {title}
  {{Extension of the spin-$\frac{1}{2}$ frustrated square lattice model: The
  case of layered vanadium phosphates}},}\ }\href {\doibase
  10.1103/PhysRevB.79.214417} {\bibfield  {journal} {\bibinfo  {journal} {Phys.
  Rev. B}\ }\textbf {\bibinfo {volume} {79}},\ \bibinfo {pages} {214417}
  (\bibinfo {year} {2009})}\BibitemShut {NoStop}%
\bibitem [{\citenamefont {Ishikawa}\ \emph {et~al.}(2017)\citenamefont
  {Ishikawa}, \citenamefont {Nakamura}, \citenamefont {Yoshida}, \citenamefont
  {Takigawa}, \citenamefont {Babkevich}, \citenamefont {Qureshi}, \citenamefont
  {R\o{}nnow}, \citenamefont {Yajima},\ and\ \citenamefont {Hiroi}}]{ishikawa}%
  \BibitemOpen
  \bibfield  {author} {\bibinfo {author} {\bibfnamefont {Hajime}\ \bibnamefont
  {Ishikawa}}, \bibinfo {author} {\bibfnamefont {Nanako}\ \bibnamefont
  {Nakamura}}, \bibinfo {author} {\bibfnamefont {Makoto}\ \bibnamefont
  {Yoshida}}, \bibinfo {author} {\bibfnamefont {Masashi}\ \bibnamefont
  {Takigawa}}, \bibinfo {author} {\bibfnamefont {Peter}\ \bibnamefont
  {Babkevich}}, \bibinfo {author} {\bibfnamefont {Navid}\ \bibnamefont
  {Qureshi}}, \bibinfo {author} {\bibfnamefont {Henrik~M.}\ \bibnamefont
  {R\o{}nnow}}, \bibinfo {author} {\bibfnamefont {Takeshi}\ \bibnamefont
  {Yajima}}, \ and\ \bibinfo {author} {\bibfnamefont {Zenji}\ \bibnamefont
  {Hiroi}},\ }\bibfield  {title} {\enquote {\bibinfo {title}
  {{${J}_{1}\text{\ensuremath{-}}{J}_{2}$ square-lattice Heisenberg
  antiferromagnets with $4{d}^{1}$ spins:
  $A\mathrm{MoOP}{\mathrm{O}}_{4}\mathrm{Cl}$
  $(A=\mathrm{K},\,\mathrm{Rb})$}},}\ }\href {\doibase
  10.1103/PhysRevB.95.064408} {\bibfield  {journal} {\bibinfo  {journal} {Phys.
  Rev. B}\ }\textbf {\bibinfo {volume} {95}},\ \bibinfo {pages} {064408}
  (\bibinfo {year} {2017})}\BibitemShut {NoStop}%
\bibitem [{\citenamefont {Kohama}\ \emph {et~al.}(2019)\citenamefont {Kohama},
  \citenamefont {Ishikawa}, \citenamefont {Matsuo}, \citenamefont {Kindo},
  \citenamefont {Shannon},\ and\ \citenamefont
  {Hiroi}}]{kohama_bacdvopo4_2019}%
  \BibitemOpen
  \bibfield  {author} {\bibinfo {author} {\bibfnamefont {Yoshimitsu}\
  \bibnamefont {Kohama}}, \bibinfo {author} {\bibfnamefont {Hajime}\
  \bibnamefont {Ishikawa}}, \bibinfo {author} {\bibfnamefont {Akira}\
  \bibnamefont {Matsuo}}, \bibinfo {author} {\bibfnamefont {Koichi}\
  \bibnamefont {Kindo}}, \bibinfo {author} {\bibfnamefont {Nic}\ \bibnamefont
  {Shannon}}, \ and\ \bibinfo {author} {\bibfnamefont {Zenji}\ \bibnamefont
  {Hiroi}},\ }\bibfield  {title} {\enquote {\bibinfo {title} {Possible
  observation of quantum spin-nematic phase in a frustrated magnet},}\ }\href
  {\doibase 10.1073/pnas.1821969116} {\bibfield  {journal} {\bibinfo  {journal}
  {Proceedings of the National Academy of Sciences}\ }\textbf {\bibinfo
  {volume} {116}},\ \bibinfo {pages} {10686--10690} (\bibinfo {year}
  {2019})}\BibitemShut {NoStop}%
\bibitem [{\citenamefont {Skoulatos}\ \emph {et~al.}(2019)\citenamefont
  {Skoulatos}, \citenamefont {Rucker}, \citenamefont {Nilsen}, \citenamefont
  {Bertin}, \citenamefont {Pomjakushina}, \citenamefont {Ollivier},
  \citenamefont {Schneidewind}, \citenamefont {Georgii}, \citenamefont
  {Zaharko}, \citenamefont {Keller}, \citenamefont {R\"uegg}, \citenamefont
  {Pfleiderer}, \citenamefont {Schmidt}, \citenamefont {Shannon}, \citenamefont
  {Kriele}, \citenamefont {Senyshyn},\ and\ \citenamefont
  {Smerald}}]{skoulatos_bacdvopo4_2019}%
  \BibitemOpen
  \bibfield  {author} {\bibinfo {author} {\bibfnamefont {M.}~\bibnamefont
  {Skoulatos}}, \bibinfo {author} {\bibfnamefont {F.}~\bibnamefont {Rucker}},
  \bibinfo {author} {\bibfnamefont {G.~J.}\ \bibnamefont {Nilsen}}, \bibinfo
  {author} {\bibfnamefont {A.}~\bibnamefont {Bertin}}, \bibinfo {author}
  {\bibfnamefont {E.}~\bibnamefont {Pomjakushina}}, \bibinfo {author}
  {\bibfnamefont {J.}~\bibnamefont {Ollivier}}, \bibinfo {author}
  {\bibfnamefont {A.}~\bibnamefont {Schneidewind}}, \bibinfo {author}
  {\bibfnamefont {R.}~\bibnamefont {Georgii}}, \bibinfo {author} {\bibfnamefont
  {O.}~\bibnamefont {Zaharko}}, \bibinfo {author} {\bibfnamefont
  {L.}~\bibnamefont {Keller}}, \bibinfo {author} {\bibfnamefont {Ch.}\
  \bibnamefont {R\"uegg}}, \bibinfo {author} {\bibfnamefont {C.}~\bibnamefont
  {Pfleiderer}}, \bibinfo {author} {\bibfnamefont {B.}~\bibnamefont {Schmidt}},
  \bibinfo {author} {\bibfnamefont {N.}~\bibnamefont {Shannon}}, \bibinfo
  {author} {\bibfnamefont {A.}~\bibnamefont {Kriele}}, \bibinfo {author}
  {\bibfnamefont {A.}~\bibnamefont {Senyshyn}}, \ and\ \bibinfo {author}
  {\bibfnamefont {A.}~\bibnamefont {Smerald}},\ }\bibfield  {title} {\enquote
  {\bibinfo {title} {{Putative spin-nematic phase in
  $\mathrm{BaCdVO}({\mathrm{PO}}_{4}{)}_{2}$}},}\ }\href {\doibase
  10.1103/PhysRevB.100.014405} {\bibfield  {journal} {\bibinfo  {journal}
  {Phys. Rev. B}\ }\textbf {\bibinfo {volume} {100}},\ \bibinfo {pages}
  {014405} (\bibinfo {year} {2019})}\BibitemShut {NoStop}%
\bibitem [{\citenamefont {Coldea}\ \emph {et~al.}(2001)\citenamefont {Coldea},
  \citenamefont {Tennant}, \citenamefont {Tsvelik},\ and\ \citenamefont
  {Tylczynski}}]{coldea_aniso_tri}%
  \BibitemOpen
  \bibfield  {author} {\bibinfo {author} {\bibfnamefont {R.}~\bibnamefont
  {Coldea}}, \bibinfo {author} {\bibfnamefont {D.~A.}\ \bibnamefont {Tennant}},
  \bibinfo {author} {\bibfnamefont {A.~M.}\ \bibnamefont {Tsvelik}}, \ and\
  \bibinfo {author} {\bibfnamefont {Z.}~\bibnamefont {Tylczynski}},\ }\bibfield
   {title} {\enquote {\bibinfo {title} {Experimental realization of a 2d
  fractional quantum spin liquid},}\ }\href {\doibase
  10.1103/PhysRevLett.86.1335} {\bibfield  {journal} {\bibinfo  {journal}
  {Phys. Rev. Lett.}\ }\textbf {\bibinfo {volume} {86}},\ \bibinfo {pages}
  {1335--1338} (\bibinfo {year} {2001})}\BibitemShut {NoStop}%
\bibitem [{\citenamefont {Starykh}\ and\ \citenamefont
  {Balents}(2007)}]{starykh_aniso_tri}%
  \BibitemOpen
  \bibfield  {author} {\bibinfo {author} {\bibfnamefont {Oleg~A.}\ \bibnamefont
  {Starykh}}\ and\ \bibinfo {author} {\bibfnamefont {Leon}\ \bibnamefont
  {Balents}},\ }\bibfield  {title} {\enquote {\bibinfo {title} {Ordering in
  spatially anisotropic triangular antiferromagnets},}\ }\href {\doibase
  10.1103/PhysRevLett.98.077205} {\bibfield  {journal} {\bibinfo  {journal}
  {Phys. Rev. Lett.}\ }\textbf {\bibinfo {volume} {98}},\ \bibinfo {pages}
  {077205} (\bibinfo {year} {2007})}\BibitemShut {NoStop}%
\bibitem [{\citenamefont {Bishop}\ \emph {et~al.}(2009)\citenamefont {Bishop},
  \citenamefont {Li}, \citenamefont {Farnell},\ and\ \citenamefont
  {Campbell}}]{bishop_aniso_tri}%
  \BibitemOpen
  \bibfield  {author} {\bibinfo {author} {\bibfnamefont {R.~F.}\ \bibnamefont
  {Bishop}}, \bibinfo {author} {\bibfnamefont {P.~H.~Y.}\ \bibnamefont {Li}},
  \bibinfo {author} {\bibfnamefont {D.~J.~J.}\ \bibnamefont {Farnell}}, \ and\
  \bibinfo {author} {\bibfnamefont {C.~E.}\ \bibnamefont {Campbell}},\
  }\bibfield  {title} {\enquote {\bibinfo {title} {Magnetic order in a
  spin-$\frac{1}{2}$ interpolating square-triangle heisenberg
  antiferromagnet},}\ }\href {\doibase 10.1103/PhysRevB.79.174405} {\bibfield
  {journal} {\bibinfo  {journal} {Phys. Rev. B}\ }\textbf {\bibinfo {volume}
  {79}},\ \bibinfo {pages} {174405} (\bibinfo {year} {2009})}\BibitemShut
  {NoStop}%
\bibitem [{\citenamefont {Furukawa}\ \emph
  {et~al.}(2010{\natexlab{a}})\citenamefont {Furukawa}, \citenamefont {Sato},\
  and\ \citenamefont {Onoda}}]{furukawa_J1J2_2010b}%
  \BibitemOpen
  \bibfield  {author} {\bibinfo {author} {\bibfnamefont {Shunsuke}\
  \bibnamefont {Furukawa}}, \bibinfo {author} {\bibfnamefont {Masahiro}\
  \bibnamefont {Sato}}, \ and\ \bibinfo {author} {\bibfnamefont {Shigeki}\
  \bibnamefont {Onoda}},\ }\bibfield  {title} {\enquote {\bibinfo {title}
  {{Chiral Order and Electromagnetic Dynamics in One-Dimensional Multiferroic
  Cuprates}},}\ }\href {\doibase 10.1103/PhysRevLett.105.257205} {\bibfield
  {journal} {\bibinfo  {journal} {Phys. Rev. Lett.}\ }\textbf {\bibinfo
  {volume} {105}},\ \bibinfo {pages} {257205} (\bibinfo {year}
  {2010}{\natexlab{a}})}\BibitemShut {NoStop}%
\bibitem [{\citenamefont {Kecke}\ \emph {et~al.}(2007)\citenamefont {Kecke},
  \citenamefont {Momoi},\ and\ \citenamefont {Furusaki}}]{kecke_multimagnon}%
  \BibitemOpen
  \bibfield  {author} {\bibinfo {author} {\bibfnamefont {Lars}\ \bibnamefont
  {Kecke}}, \bibinfo {author} {\bibfnamefont {Tsutomu}\ \bibnamefont {Momoi}},
  \ and\ \bibinfo {author} {\bibfnamefont {Akira}\ \bibnamefont {Furusaki}},\
  }\bibfield  {title} {\enquote {\bibinfo {title} {Multimagnon bound states in
  the frustrated ferromagnetic one-dimensional chain},}\ }\href {\doibase
  10.1103/PhysRevB.76.060407} {\bibfield  {journal} {\bibinfo  {journal} {Phys.
  Rev. B}\ }\textbf {\bibinfo {volume} {76}},\ \bibinfo {pages} {060407}
  (\bibinfo {year} {2007})}\BibitemShut {NoStop}%
\bibitem [{\citenamefont {Furukawa}\ \emph {et~al.}(2012)\citenamefont
  {Furukawa}, \citenamefont {Sato}, \citenamefont {Onoda},\ and\ \citenamefont
  {Furusaki}}]{furukawa_J1J2_XXZ}%
  \BibitemOpen
  \bibfield  {author} {\bibinfo {author} {\bibfnamefont {Shunsuke}\
  \bibnamefont {Furukawa}}, \bibinfo {author} {\bibfnamefont {Masahiro}\
  \bibnamefont {Sato}}, \bibinfo {author} {\bibfnamefont {Shigeki}\
  \bibnamefont {Onoda}}, \ and\ \bibinfo {author} {\bibfnamefont {Akira}\
  \bibnamefont {Furusaki}},\ }\bibfield  {title} {\enquote {\bibinfo {title}
  {{Ground-state phase diagram of a spin-$\frac{1}{2}$ frustrated ferromagnetic
  XXZ chain: Haldane dimer phase and gapped/gapless chiral phases}},}\ }\href
  {\doibase 10.1103/PhysRevB.86.094417} {\bibfield  {journal} {\bibinfo
  {journal} {Phys. Rev. B}\ }\textbf {\bibinfo {volume} {86}},\ \bibinfo
  {pages} {094417} (\bibinfo {year} {2012})}\BibitemShut {NoStop}%
\bibitem [{\citenamefont {Zhitomirsky}\ and\ \citenamefont
  {Tsunetsugu}(2010)}]{zhitomirsky_licuvo4}%
  \BibitemOpen
  \bibfield  {author} {\bibinfo {author} {\bibfnamefont {M.~E.}\ \bibnamefont
  {Zhitomirsky}}\ and\ \bibinfo {author} {\bibfnamefont {H.}~\bibnamefont
  {Tsunetsugu}},\ }\bibfield  {title} {\enquote {\bibinfo {title} {Magnon
  pairing in quantum spin nematic},}\ }\href {\doibase
  10.1209/0295-5075/92/37001} {\bibfield  {journal} {\bibinfo  {journal} {{EPL}
  (Europhysics Letters)}\ }\textbf {\bibinfo {volume} {92}},\ \bibinfo {pages}
  {37001} (\bibinfo {year} {2010})}\BibitemShut {NoStop}%
\bibitem [{\citenamefont {Sato}\ \emph {et~al.}(2009)\citenamefont {Sato},
  \citenamefont {Momoi},\ and\ \citenamefont
  {Furusaki}}]{sato_nematic_nmr2009}%
  \BibitemOpen
  \bibfield  {author} {\bibinfo {author} {\bibfnamefont {Masahiro}\
  \bibnamefont {Sato}}, \bibinfo {author} {\bibfnamefont {Tsutomu}\
  \bibnamefont {Momoi}}, \ and\ \bibinfo {author} {\bibfnamefont {Akira}\
  \bibnamefont {Furusaki}},\ }\bibfield  {title} {\enquote {\bibinfo {title}
  {{NMR relaxation rate and dynamical structure factors in nematic and
  multipolar liquids of frustrated spin chains under magnetic fields}},}\
  }\href {\doibase 10.1103/PhysRevB.79.060406} {\bibfield  {journal} {\bibinfo
  {journal} {Phys. Rev. B}\ }\textbf {\bibinfo {volume} {79}},\ \bibinfo
  {pages} {060406} (\bibinfo {year} {2009})}\BibitemShut {NoStop}%
\bibitem [{\citenamefont {Sato}\ \emph
  {et~al.}(2011{\natexlab{a}})\citenamefont {Sato}, \citenamefont {Hikihara},\
  and\ \citenamefont {Momoi}}]{sato_nematic_nmr2011}%
  \BibitemOpen
  \bibfield  {author} {\bibinfo {author} {\bibfnamefont {Masahiro}\
  \bibnamefont {Sato}}, \bibinfo {author} {\bibfnamefont {Toshiya}\
  \bibnamefont {Hikihara}}, \ and\ \bibinfo {author} {\bibfnamefont {Tsutomu}\
  \bibnamefont {Momoi}},\ }\bibfield  {title} {\enquote {\bibinfo {title}
  {{Field and temperature dependence of NMR relaxation rate in the magnetic
  quadrupolar liquid phase of spin-$\frac{1}{2}$ frustrated ferromagnetic
  chains}},}\ }\href {\doibase 10.1103/PhysRevB.83.064405} {\bibfield
  {journal} {\bibinfo  {journal} {Phys. Rev. B}\ }\textbf {\bibinfo {volume}
  {83}},\ \bibinfo {pages} {064405} (\bibinfo {year}
  {2011}{\natexlab{a}})}\BibitemShut {NoStop}%
\bibitem [{\citenamefont {Sato}\ \emph
  {et~al.}(2011{\natexlab{b}})\citenamefont {Sato}, \citenamefont {Furukawa},
  \citenamefont {Onoda},\ and\ \citenamefont
  {Furusaki}}]{sato_J1J2_ModPhysLett}%
  \BibitemOpen
  \bibfield  {author} {\bibinfo {author} {\bibfnamefont {Masahiro}\
  \bibnamefont {Sato}}, \bibinfo {author} {\bibfnamefont {Shunsuke}\
  \bibnamefont {Furukawa}}, \bibinfo {author} {\bibfnamefont {Shigeki}\
  \bibnamefont {Onoda}}, \ and\ \bibinfo {author} {\bibfnamefont {Akira}\
  \bibnamefont {Furusaki}},\ }\bibfield  {title} {\enquote {\bibinfo {title}
  {{Competing phases in spin-$1/2$ $J_1-J_2$ chain with easy-plane
  anisotropy}},}\ }\href {\doibase 10.1142/S0217984911026607} {\bibfield
  {journal} {\bibinfo  {journal} {Modern Physics Letters B}\ }\textbf {\bibinfo
  {volume} {25}},\ \bibinfo {pages} {901--908} (\bibinfo {year}
  {2011}{\natexlab{b}})}\BibitemShut {NoStop}%
\bibitem [{\citenamefont {Sato}\ \emph {et~al.}(2013)\citenamefont {Sato},
  \citenamefont {Hikihara},\ and\ \citenamefont
  {Momoi}}]{sato_nematic_quasi1d}%
  \BibitemOpen
  \bibfield  {author} {\bibinfo {author} {\bibfnamefont {Masahiro}\
  \bibnamefont {Sato}}, \bibinfo {author} {\bibfnamefont {Toshiya}\
  \bibnamefont {Hikihara}}, \ and\ \bibinfo {author} {\bibfnamefont {Tsutomu}\
  \bibnamefont {Momoi}},\ }\bibfield  {title} {\enquote {\bibinfo {title}
  {Spin-nematic and spin-density-wave orders in spatially anisotropic
  frustrated magnets in a magnetic field},}\ }\href {\doibase
  10.1103/PhysRevLett.110.077206} {\bibfield  {journal} {\bibinfo  {journal}
  {Phys. Rev. Lett.}\ }\textbf {\bibinfo {volume} {110}},\ \bibinfo {pages}
  {077206} (\bibinfo {year} {2013})}\BibitemShut {NoStop}%
\bibitem [{\citenamefont {Hirobe}\ \emph {et~al.}(2019)\citenamefont {Hirobe},
  \citenamefont {Sato}, \citenamefont {Hagihala}, \citenamefont {Shiomi},
  \citenamefont {Masuda},\ and\ \citenamefont
  {Saitoh}}]{hirobe_nematic_seebeck}%
  \BibitemOpen
  \bibfield  {author} {\bibinfo {author} {\bibfnamefont {Daichi}\ \bibnamefont
  {Hirobe}}, \bibinfo {author} {\bibfnamefont {Masahiro}\ \bibnamefont {Sato}},
  \bibinfo {author} {\bibfnamefont {Masato}\ \bibnamefont {Hagihala}}, \bibinfo
  {author} {\bibfnamefont {Yuki}\ \bibnamefont {Shiomi}}, \bibinfo {author}
  {\bibfnamefont {Takatsugu}\ \bibnamefont {Masuda}}, \ and\ \bibinfo {author}
  {\bibfnamefont {Eiji}\ \bibnamefont {Saitoh}},\ }\bibfield  {title} {\enquote
  {\bibinfo {title} {Magnon pairs and spin-nematic correlation in the spin
  seebeck effect},}\ }\href {\doibase 10.1103/PhysRevLett.123.117202}
  {\bibfield  {journal} {\bibinfo  {journal} {Phys. Rev. Lett.}\ }\textbf
  {\bibinfo {volume} {123}},\ \bibinfo {pages} {117202} (\bibinfo {year}
  {2019})}\BibitemShut {NoStop}%
\bibitem [{\citenamefont {Furuya}(2017)}]{furuya_esr_angle}%
  \BibitemOpen
  \bibfield  {author} {\bibinfo {author} {\bibfnamefont {Shunsuke~C.}\
  \bibnamefont {Furuya}},\ }\bibfield  {title} {\enquote {\bibinfo {title}
  {Angular dependence of electron spin resonance for detecting the quadrupolar
  liquid state of frustrated spin chains},}\ }\href {\doibase
  10.1103/PhysRevB.95.014416} {\bibfield  {journal} {\bibinfo  {journal} {Phys.
  Rev. B}\ }\textbf {\bibinfo {volume} {95}},\ \bibinfo {pages} {014416}
  (\bibinfo {year} {2017})}\BibitemShut {NoStop}%
\bibitem [{\citenamefont {Okunishi}(2008)}]{okunishi_zigzag}%
  \BibitemOpen
  \bibfield  {author} {\bibinfo {author} {\bibfnamefont {Kouichi}\ \bibnamefont
  {Okunishi}},\ }\bibfield  {title} {\enquote {\bibinfo {title} {{On
  Calculation of Vector Spin Chirality for Zigzag Spin Chains}},}\ }\href
  {\doibase 10.1143/JPSJ.77.114004} {\bibfield  {journal} {\bibinfo  {journal}
  {Journal of the Physical Society of Japan}\ }\textbf {\bibinfo {volume}
  {77}},\ \bibinfo {pages} {114004} (\bibinfo {year} {2008})}\BibitemShut
  {NoStop}%
\bibitem [{\citenamefont {Hikihara}\ \emph {et~al.}(2010)\citenamefont
  {Hikihara}, \citenamefont {Momoi}, \citenamefont {Furusaki},\ and\
  \citenamefont {Kawamura}}]{hikihara_zigzag}%
  \BibitemOpen
  \bibfield  {author} {\bibinfo {author} {\bibfnamefont {Toshiya}\ \bibnamefont
  {Hikihara}}, \bibinfo {author} {\bibfnamefont {Tsutomu}\ \bibnamefont
  {Momoi}}, \bibinfo {author} {\bibfnamefont {Akira}\ \bibnamefont {Furusaki}},
  \ and\ \bibinfo {author} {\bibfnamefont {Hikaru}\ \bibnamefont {Kawamura}},\
  }\bibfield  {title} {\enquote {\bibinfo {title} {Magnetic phase diagram of
  the spin-$\frac{1}{2}$ antiferromagnetic zigzag ladder},}\ }\href {\doibase
  10.1103/PhysRevB.81.224433} {\bibfield  {journal} {\bibinfo  {journal} {Phys.
  Rev. B}\ }\textbf {\bibinfo {volume} {81}},\ \bibinfo {pages} {224433}
  (\bibinfo {year} {2010})}\BibitemShut {NoStop}%
\bibitem [{\citenamefont {Furukawa}\ \emph
  {et~al.}(2010{\natexlab{b}})\citenamefont {Furukawa}, \citenamefont {Sato},\
  and\ \citenamefont {Furusaki}}]{furukawa_J1J2_2010a}%
  \BibitemOpen
  \bibfield  {author} {\bibinfo {author} {\bibfnamefont {Shunsuke}\
  \bibnamefont {Furukawa}}, \bibinfo {author} {\bibfnamefont {Masahiro}\
  \bibnamefont {Sato}}, \ and\ \bibinfo {author} {\bibfnamefont {Akira}\
  \bibnamefont {Furusaki}},\ }\bibfield  {title} {\enquote {\bibinfo {title}
  {Unconventional n\'eel and dimer orders in a spin-$\frac{1}{2}$ frustrated
  ferromagnetic chain with easy-plane anisotropy},}\ }\href {\doibase
  10.1103/PhysRevB.81.094430} {\bibfield  {journal} {\bibinfo  {journal} {Phys.
  Rev. B}\ }\textbf {\bibinfo {volume} {81}},\ \bibinfo {pages} {094430}
  (\bibinfo {year} {2010}{\natexlab{b}})}\BibitemShut {NoStop}%
\bibitem [{\citenamefont {Parvej}\ and\ \citenamefont
  {Kumar}(2017)}]{kumar_chain}%
  \BibitemOpen
  \bibfield  {author} {\bibinfo {author} {\bibfnamefont {Aslam}\ \bibnamefont
  {Parvej}}\ and\ \bibinfo {author} {\bibfnamefont {Manoranjan}\ \bibnamefont
  {Kumar}},\ }\bibfield  {title} {\enquote {\bibinfo {title} {Multipolar phase
  in frustrated spin-1/2 and spin-1 chains},}\ }\href {\doibase
  10.1103/PhysRevB.96.054413} {\bibfield  {journal} {\bibinfo  {journal} {Phys.
  Rev. B}\ }\textbf {\bibinfo {volume} {96}},\ \bibinfo {pages} {054413}
  (\bibinfo {year} {2017})}\BibitemShut {NoStop}%
\bibitem [{\citenamefont {Enderle}\ \emph {et~al.}(2005)\citenamefont
  {Enderle}, \citenamefont {Mukherjee}, \citenamefont {F{\aa}k}, \citenamefont
  {Kremer}, \citenamefont {Broto}, \citenamefont {Rosner}, \citenamefont
  {Drechsler}, \citenamefont {Richter}, \citenamefont {Malek}, \citenamefont
  {Prokofiev}, \citenamefont {Assmus}, \citenamefont {Pujol}, \citenamefont
  {Raggazzoni}, \citenamefont {Rakoto}, \citenamefont {Rheinstädter},\ and\
  \citenamefont {R{\o}nnow}}]{Enderle_2005}%
  \BibitemOpen
  \bibfield  {author} {\bibinfo {author} {\bibfnamefont {M}~\bibnamefont
  {Enderle}}, \bibinfo {author} {\bibfnamefont {C}~\bibnamefont {Mukherjee}},
  \bibinfo {author} {\bibfnamefont {B}~\bibnamefont {F{\aa}k}}, \bibinfo
  {author} {\bibfnamefont {R.~K}\ \bibnamefont {Kremer}}, \bibinfo {author}
  {\bibfnamefont {J.-M}\ \bibnamefont {Broto}}, \bibinfo {author}
  {\bibfnamefont {H}~\bibnamefont {Rosner}}, \bibinfo {author} {\bibfnamefont
  {S.-L}\ \bibnamefont {Drechsler}}, \bibinfo {author} {\bibfnamefont
  {J}~\bibnamefont {Richter}}, \bibinfo {author} {\bibfnamefont
  {J}~\bibnamefont {Malek}}, \bibinfo {author} {\bibfnamefont {A}~\bibnamefont
  {Prokofiev}}, \bibinfo {author} {\bibfnamefont {W}~\bibnamefont {Assmus}},
  \bibinfo {author} {\bibfnamefont {S}~\bibnamefont {Pujol}}, \bibinfo {author}
  {\bibfnamefont {J.-L}\ \bibnamefont {Raggazzoni}}, \bibinfo {author}
  {\bibfnamefont {H}~\bibnamefont {Rakoto}}, \bibinfo {author} {\bibfnamefont
  {M}~\bibnamefont {Rheinstädter}}, \ and\ \bibinfo {author} {\bibfnamefont
  {H.~M}\ \bibnamefont {R{\o}nnow}},\ }\bibfield  {title} {\enquote {\bibinfo
  {title} {{Quantum helimagnetism of the frustrated spin-{$1/2$} chain
  LiCuVO$_4$}},}\ }\href {\doibase 10.1209/epl/i2004-10484-x} {\bibfield
  {journal} {\bibinfo  {journal} {Europhysics Letters ({EPL})}\ }\textbf
  {\bibinfo {volume} {70}},\ \bibinfo {pages} {237--243} (\bibinfo {year}
  {2005})}\BibitemShut {NoStop}%
\bibitem [{\citenamefont {Naito}\ \emph {et~al.}(2007)\citenamefont {Naito},
  \citenamefont {Sato}, \citenamefont {Yasui}, \citenamefont {Kobayashi},
  \citenamefont {Kobayashi},\ and\ \citenamefont {Sato}}]{Naito_LicuVO4_2007}%
  \BibitemOpen
  \bibfield  {author} {\bibinfo {author} {\bibfnamefont {Yutaka}\ \bibnamefont
  {Naito}}, \bibinfo {author} {\bibfnamefont {Kenji}\ \bibnamefont {Sato}},
  \bibinfo {author} {\bibfnamefont {Yukio}\ \bibnamefont {Yasui}}, \bibinfo
  {author} {\bibfnamefont {Yusuke}\ \bibnamefont {Kobayashi}}, \bibinfo
  {author} {\bibfnamefont {Yoshiaki}\ \bibnamefont {Kobayashi}}, \ and\
  \bibinfo {author} {\bibfnamefont {Masatoshi}\ \bibnamefont {Sato}},\
  }\bibfield  {title} {\enquote {\bibinfo {title} {{Ferroelectric Transition
  Induced by the Incommensurate Magnetic Ordering in LiCuVO4}},}\ }\href
  {\doibase 10.1143/JPSJ.76.023708} {\bibfield  {journal} {\bibinfo  {journal}
  {Journal of the Physical Society of Japan}\ }\textbf {\bibinfo {volume}
  {76}},\ \bibinfo {pages} {023708} (\bibinfo {year} {2007})}\BibitemShut
  {NoStop}%
\bibitem [{\citenamefont {Yasui}\ \emph {et~al.}(2011)\citenamefont {Yasui},
  \citenamefont {Sato},\ and\ \citenamefont {Terasaki}}]{yasui_cuo2_chain}%
  \BibitemOpen
  \bibfield  {author} {\bibinfo {author} {\bibfnamefont {Yukio}\ \bibnamefont
  {Yasui}}, \bibinfo {author} {\bibfnamefont {Masatoshi}\ \bibnamefont {Sato}},
  \ and\ \bibinfo {author} {\bibfnamefont {Ichiro}\ \bibnamefont {Terasaki}},\
  }\bibfield  {title} {\enquote {\bibinfo {title} {{Multiferroic Behavior in
  the Quasi-One-Dimensional Frustrated Spin-1/2 System PbCuSO4(OH)2 with CuO2
  Ribbon Chains}},}\ }\href {\doibase 10.1143/JPSJ.80.033707} {\bibfield
  {journal} {\bibinfo  {journal} {Journal of the Physical Society of Japan}\
  }\textbf {\bibinfo {volume} {80}},\ \bibinfo {pages} {033707} (\bibinfo
  {year} {2011})}\BibitemShut {NoStop}%
\bibitem [{\citenamefont {Sch\"apers}\ \emph {et~al.}(2013)\citenamefont
  {Sch\"apers}, \citenamefont {Wolter}, \citenamefont {Drechsler},
  \citenamefont {Nishimoto}, \citenamefont {M\"uller}, \citenamefont
  {Abdel-Hafiez}, \citenamefont {Schottenhamel}, \citenamefont {B\"uchner},
  \citenamefont {Richter}, \citenamefont {Ouladdiaf}, \citenamefont {Uhlarz},
  \citenamefont {Beyer}, \citenamefont {Skourski}, \citenamefont {Wosnitza},
  \citenamefont {Rule}, \citenamefont {Ryll}, \citenamefont {Klemke},
  \citenamefont {Kiefer}, \citenamefont {Reehuis}, \citenamefont {Willenberg},\
  and\ \citenamefont {S\"ullow}}]{schapers_cuo2_chain}%
  \BibitemOpen
  \bibfield  {author} {\bibinfo {author} {\bibfnamefont {M.}~\bibnamefont
  {Sch\"apers}}, \bibinfo {author} {\bibfnamefont {A.~U.~B.}\ \bibnamefont
  {Wolter}}, \bibinfo {author} {\bibfnamefont {S.-L.}\ \bibnamefont
  {Drechsler}}, \bibinfo {author} {\bibfnamefont {S.}~\bibnamefont
  {Nishimoto}}, \bibinfo {author} {\bibfnamefont {K.-H.}\ \bibnamefont
  {M\"uller}}, \bibinfo {author} {\bibfnamefont {M.}~\bibnamefont
  {Abdel-Hafiez}}, \bibinfo {author} {\bibfnamefont {W.}~\bibnamefont
  {Schottenhamel}}, \bibinfo {author} {\bibfnamefont {B.}~\bibnamefont
  {B\"uchner}}, \bibinfo {author} {\bibfnamefont {J.}~\bibnamefont {Richter}},
  \bibinfo {author} {\bibfnamefont {B.}~\bibnamefont {Ouladdiaf}}, \bibinfo
  {author} {\bibfnamefont {M.}~\bibnamefont {Uhlarz}}, \bibinfo {author}
  {\bibfnamefont {R.}~\bibnamefont {Beyer}}, \bibinfo {author} {\bibfnamefont
  {Y.}~\bibnamefont {Skourski}}, \bibinfo {author} {\bibfnamefont
  {J.}~\bibnamefont {Wosnitza}}, \bibinfo {author} {\bibfnamefont {K.~C.}\
  \bibnamefont {Rule}}, \bibinfo {author} {\bibfnamefont {H.}~\bibnamefont
  {Ryll}}, \bibinfo {author} {\bibfnamefont {B.}~\bibnamefont {Klemke}},
  \bibinfo {author} {\bibfnamefont {K.}~\bibnamefont {Kiefer}}, \bibinfo
  {author} {\bibfnamefont {M.}~\bibnamefont {Reehuis}}, \bibinfo {author}
  {\bibfnamefont {B.}~\bibnamefont {Willenberg}}, \ and\ \bibinfo {author}
  {\bibfnamefont {S.}~\bibnamefont {S\"ullow}},\ }\bibfield  {title} {\enquote
  {\bibinfo {title} {{Thermodynamic properties of the anisotropic frustrated
  spin-chain compound linarite PbCuSO${}_{4}$(OH)${}_{2}$}},}\ }\href {\doibase
  10.1103/PhysRevB.88.184410} {\bibfield  {journal} {\bibinfo  {journal} {Phys.
  Rev. B}\ }\textbf {\bibinfo {volume} {88}},\ \bibinfo {pages} {184410}
  (\bibinfo {year} {2013})}\BibitemShut {NoStop}%
\bibitem [{\citenamefont {Hase}\ \emph {et~al.}(2004)\citenamefont {Hase},
  \citenamefont {Kuroe}, \citenamefont {Ozawa}, \citenamefont {Suzuki},
  \citenamefont {Kitazawa}, \citenamefont {Kido},\ and\ \citenamefont
  {Sekine}}]{hase_2004}%
  \BibitemOpen
  \bibfield  {author} {\bibinfo {author} {\bibfnamefont {Masashi}\ \bibnamefont
  {Hase}}, \bibinfo {author} {\bibfnamefont {Haruhiko}\ \bibnamefont {Kuroe}},
  \bibinfo {author} {\bibfnamefont {Kiyoshi}\ \bibnamefont {Ozawa}}, \bibinfo
  {author} {\bibfnamefont {Osamu}\ \bibnamefont {Suzuki}}, \bibinfo {author}
  {\bibfnamefont {Hideaki}\ \bibnamefont {Kitazawa}}, \bibinfo {author}
  {\bibfnamefont {Giyuu}\ \bibnamefont {Kido}}, \ and\ \bibinfo {author}
  {\bibfnamefont {Tomoyuki}\ \bibnamefont {Sekine}},\ }\bibfield  {title}
  {\enquote {\bibinfo {title} {{Magnetic properties of
  ${\mathrm{Rb}}_{2}{\mathrm{Cu}}_{2}{\mathrm{Mo}}_{3}{\mathrm{O}}_{12}$
  including a one-dimensional spin-$1/2$ Heisenberg system with ferromagnetic
  first-nearest-neighbor and antiferromagnetic second-nearest-neighbor exchange
  interactions}},}\ }\href {\doibase 10.1103/PhysRevB.70.104426} {\bibfield
  {journal} {\bibinfo  {journal} {Phys. Rev. B}\ }\textbf {\bibinfo {volume}
  {70}},\ \bibinfo {pages} {104426} (\bibinfo {year} {2004})}\BibitemShut
  {NoStop}%
\bibitem [{\citenamefont {Matsui}\ \emph {et~al.}(2017)\citenamefont {Matsui},
  \citenamefont {Yagi}, \citenamefont {Hoshino}, \citenamefont {Atarashi},
  \citenamefont {Hase}, \citenamefont {Sasaki},\ and\ \citenamefont
  {Goto}}]{matsui_2017}%
  \BibitemOpen
  \bibfield  {author} {\bibinfo {author} {\bibfnamefont {Kazuki}\ \bibnamefont
  {Matsui}}, \bibinfo {author} {\bibfnamefont {Ayato}\ \bibnamefont {Yagi}},
  \bibinfo {author} {\bibfnamefont {Yukihiro}\ \bibnamefont {Hoshino}},
  \bibinfo {author} {\bibfnamefont {Sochiro}\ \bibnamefont {Atarashi}},
  \bibinfo {author} {\bibfnamefont {Masashi}\ \bibnamefont {Hase}}, \bibinfo
  {author} {\bibfnamefont {Takahiko}\ \bibnamefont {Sasaki}}, \ and\ \bibinfo
  {author} {\bibfnamefont {Takayuki}\ \bibnamefont {Goto}},\ }\bibfield
  {title} {\enquote {\bibinfo {title} {{Rb-NMR study of the
  quasi-one-dimensional competing spin-chain compound
  $\mathrm{R}{\mathrm{b}}_{2}\mathrm{C}{\mathrm{u}}_{2}\mathrm{M}{\mathrm{o}}_{3}{\mathrm{O}}_{12}$}},}\
  }\href {\doibase 10.1103/PhysRevB.96.220402} {\bibfield  {journal} {\bibinfo
  {journal} {Phys. Rev. B}\ }\textbf {\bibinfo {volume} {96}},\ \bibinfo
  {pages} {220402} (\bibinfo {year} {2017})}\BibitemShut {NoStop}%
\bibitem [{\citenamefont {Nawa}\ \emph {et~al.}(2015)\citenamefont {Nawa},
  \citenamefont {Yajima}, \citenamefont {Okamoto},\ and\ \citenamefont
  {Hiroi}}]{nawa_1d_J1J2_InorgChem}%
  \BibitemOpen
  \bibfield  {author} {\bibinfo {author} {\bibfnamefont {Kazuhiro}\
  \bibnamefont {Nawa}}, \bibinfo {author} {\bibfnamefont {Takeshi}\
  \bibnamefont {Yajima}}, \bibinfo {author} {\bibfnamefont {Yoshihiko}\
  \bibnamefont {Okamoto}}, \ and\ \bibinfo {author} {\bibfnamefont {Zenji}\
  \bibnamefont {Hiroi}},\ }\bibfield  {title} {\enquote {\bibinfo {title}
  {{Orbital Arrangements and Magnetic Interactions in the Quasi-One-Dimensional
  Cuprates ACuMoO4(OH) (A = Na, K)}},}\ }\href {\doibase
  10.1021/acs.inorgchem.5b00686} {\bibfield  {journal} {\bibinfo  {journal}
  {Inorganic Chemistry}\ }\textbf {\bibinfo {volume} {54}},\ \bibinfo {pages}
  {5566--5570} (\bibinfo {year} {2015})},\ \bibinfo {note} {pMID:
  25988987}\BibitemShut {NoStop}%
\bibitem [{\citenamefont {Goodenough}(1955)}]{goodenough}%
  \BibitemOpen
  \bibfield  {author} {\bibinfo {author} {\bibfnamefont {John~B.}\ \bibnamefont
  {Goodenough}},\ }\bibfield  {title} {\enquote {\bibinfo {title} {{Theory of
  the Role of Covalence in the Perovskite-Type Manganites $[\mathrm{La},
  M(\mathrm{II})]\mathrm{Mn}{\mathrm{O}}_{3}$}},}\ }\href {\doibase
  10.1103/PhysRev.100.564} {\bibfield  {journal} {\bibinfo  {journal} {Phys.
  Rev.}\ }\textbf {\bibinfo {volume} {100}},\ \bibinfo {pages} {564--573}
  (\bibinfo {year} {1955})}\BibitemShut {NoStop}%
\bibitem [{\citenamefont {Kanamori}(1957{\natexlab{a}})}]{kanamori1}%
  \BibitemOpen
  \bibfield  {author} {\bibinfo {author} {\bibfnamefont {Junjiro}\ \bibnamefont
  {Kanamori}},\ }\bibfield  {title} {\enquote {\bibinfo {title} {{Theory of the
  Magnetic Properties of Ferrous and Cobaltous Oxides, I}},}\ }\href {\doibase
  10.1143/PTP.17.177} {\bibfield  {journal} {\bibinfo  {journal} {Progress of
  Theoretical Physics}\ }\textbf {\bibinfo {volume} {17}},\ \bibinfo {pages}
  {177--196} (\bibinfo {year} {1957}{\natexlab{a}})}\BibitemShut {NoStop}%
\bibitem [{\citenamefont {Kanamori}(1957{\natexlab{b}})}]{kanamori2}%
  \BibitemOpen
  \bibfield  {author} {\bibinfo {author} {\bibfnamefont {Junjiro}\ \bibnamefont
  {Kanamori}},\ }\bibfield  {title} {\enquote {\bibinfo {title} {{Theory of the
  Magnetic Properties of Ferrous and Cobaltous Oxides, II}},}\ }\href {\doibase
  10.1143/PTP.17.197} {\bibfield  {journal} {\bibinfo  {journal} {Progress of
  Theoretical Physics}\ }\textbf {\bibinfo {volume} {17}},\ \bibinfo {pages}
  {197--222} (\bibinfo {year} {1957}{\natexlab{b}})}\BibitemShut {NoStop}%
\bibitem [{\citenamefont {Giamarchi}(2004)}]{giamarchi_book}%
  \BibitemOpen
  \bibfield  {author} {\bibinfo {author} {\bibfnamefont {T.}~\bibnamefont
  {Giamarchi}},\ }\href@noop {} {\emph {\bibinfo {title} {Quantum Physics in
  One Dimension}}}\ (\bibinfo  {publisher} {Oxford University Press},\ \bibinfo
  {address} {Oxford},\ \bibinfo {year} {2004})\BibitemShut {NoStop}%
\bibitem [{\citenamefont {Gogolin}\ \emph {et~al.}(2004)\citenamefont
  {Gogolin}, \citenamefont {Nersesyan},\ and\ \citenamefont
  {Tsvelik}}]{gogolin_textbook}%
  \BibitemOpen
  \bibfield  {author} {\bibinfo {author} {\bibfnamefont {Alexander~O}\
  \bibnamefont {Gogolin}}, \bibinfo {author} {\bibfnamefont {Alexander~A}\
  \bibnamefont {Nersesyan}}, \ and\ \bibinfo {author} {\bibfnamefont
  {Alexei~M}\ \bibnamefont {Tsvelik}},\ }\href@noop {} {\emph {\bibinfo {title}
  {Bosonization and strongly correlated systems}}}\ (\bibinfo  {publisher}
  {Cambridge university press},\ \bibinfo {year} {2004})\BibitemShut {NoStop}%
\bibitem [{\citenamefont {Okamoto}\ and\ \citenamefont
  {Nomura}(1992)}]{okamoto_J1J2_critical}%
  \BibitemOpen
  \bibfield  {author} {\bibinfo {author} {\bibfnamefont {Kiyomi}\ \bibnamefont
  {Okamoto}}\ and\ \bibinfo {author} {\bibfnamefont {Kiyohide}\ \bibnamefont
  {Nomura}},\ }\bibfield  {title} {\enquote {\bibinfo {title} {{Fluid-dimer
  critical point in S = 12 antiferromagnetic Heisenberg chain with next nearest
  neighbor interactions}},}\ }\href {\doibase
  https://doi.org/10.1016/0375-9601(92)90823-5} {\bibfield  {journal} {\bibinfo
   {journal} {Physics Letters A}\ }\textbf {\bibinfo {volume} {169}},\ \bibinfo
  {pages} {433--437} (\bibinfo {year} {1992})}\BibitemShut {NoStop}%
\bibitem [{\citenamefont {White}\ and\ \citenamefont
  {Affleck}(1996)}]{white_J1J2_chain}%
  \BibitemOpen
  \bibfield  {author} {\bibinfo {author} {\bibfnamefont {Steven~R.}\
  \bibnamefont {White}}\ and\ \bibinfo {author} {\bibfnamefont {Ian}\
  \bibnamefont {Affleck}},\ }\bibfield  {title} {\enquote {\bibinfo {title}
  {Dimerization and incommensurate spiral spin correlations in the zigzag spin
  chain: Analogies to the kondo lattice},}\ }\href {\doibase
  10.1103/PhysRevB.54.9862} {\bibfield  {journal} {\bibinfo  {journal} {Phys.
  Rev. B}\ }\textbf {\bibinfo {volume} {54}},\ \bibinfo {pages} {9862--9869}
  (\bibinfo {year} {1996})}\BibitemShut {NoStop}%
\bibitem [{\citenamefont {Hikihara}\ and\ \citenamefont
  {Furusaki}(2004)}]{hikihara_coeff_a1b0b1}%
  \BibitemOpen
  \bibfield  {author} {\bibinfo {author} {\bibfnamefont {T.}~\bibnamefont
  {Hikihara}}\ and\ \bibinfo {author} {\bibfnamefont {A.}~\bibnamefont
  {Furusaki}},\ }\bibfield  {title} {\enquote {\bibinfo {title} {Correlation
  amplitudes for the spin-$\frac{1}{2}$ $\mathrm{XXZ}$ chain in a magnetic
  field},}\ }\href {\doibase 10.1103/PhysRevB.69.064427} {\bibfield  {journal}
  {\bibinfo  {journal} {Phys. Rev. B}\ }\textbf {\bibinfo {volume} {69}},\
  \bibinfo {pages} {064427} (\bibinfo {year} {2004})}\BibitemShut {NoStop}%
\bibitem [{\citenamefont {Lukyanov}\ and\ \citenamefont
  {Zamolodchikov}(1997)}]{lukyanov_mass}%
  \BibitemOpen
  \bibfield  {author} {\bibinfo {author} {\bibfnamefont {Sergei}\ \bibnamefont
  {Lukyanov}}\ and\ \bibinfo {author} {\bibfnamefont {Alexander}\ \bibnamefont
  {Zamolodchikov}},\ }\bibfield  {title} {\enquote {\bibinfo {title} {Exact
  expectation values of local fields in the quantum sine-gordon model},}\
  }\href {\doibase https://doi.org/10.1016/S0550-3213(97)00123-5} {\bibfield
  {journal} {\bibinfo  {journal} {Nuclear Physics B}\ }\textbf {\bibinfo
  {volume} {493}},\ \bibinfo {pages} {571 -- 587} (\bibinfo {year}
  {1997})}\BibitemShut {NoStop}%
\bibitem [{\citenamefont {Takayoshi}\ and\ \citenamefont
  {Sato}(2010)}]{takayoshi_coeff}%
  \BibitemOpen
  \bibfield  {author} {\bibinfo {author} {\bibfnamefont {Shintaro}\
  \bibnamefont {Takayoshi}}\ and\ \bibinfo {author} {\bibfnamefont {Masahiro}\
  \bibnamefont {Sato}},\ }\bibfield  {title} {\enquote {\bibinfo {title}
  {{Coefficients of bosonized dimer operators in spin-$\frac{1}{2}$ $XXZ$
  chains and their applications}},}\ }\href {\doibase
  10.1103/PhysRevB.82.214420} {\bibfield  {journal} {\bibinfo  {journal} {Phys.
  Rev. B}\ }\textbf {\bibinfo {volume} {82}},\ \bibinfo {pages} {214420}
  (\bibinfo {year} {2010})}\BibitemShut {NoStop}%
\bibitem [{\citenamefont {Hikihara}\ \emph {et~al.}(2017)\citenamefont
  {Hikihara}, \citenamefont {Furusaki},\ and\ \citenamefont
  {Lukyanov}}]{hikihara_coeff_dimer}%
  \BibitemOpen
  \bibfield  {author} {\bibinfo {author} {\bibfnamefont {Toshiya}\ \bibnamefont
  {Hikihara}}, \bibinfo {author} {\bibfnamefont {Akira}\ \bibnamefont
  {Furusaki}}, \ and\ \bibinfo {author} {\bibfnamefont {Sergei}\ \bibnamefont
  {Lukyanov}},\ }\bibfield  {title} {\enquote {\bibinfo {title} {Dimer
  correlation amplitudes and dimer excitation gap in spin-$\frac{1}{2}$ xxz and
  heisenberg chains},}\ }\href {\doibase 10.1103/PhysRevB.96.134429} {\bibfield
   {journal} {\bibinfo  {journal} {Phys. Rev. B}\ }\textbf {\bibinfo {volume}
  {96}},\ \bibinfo {pages} {134429} (\bibinfo {year} {2017})}\BibitemShut
  {NoStop}%
\bibitem [{\citenamefont {Okamoto}\ and\ \citenamefont
  {Nakamura}(1997)}]{okamoto_J1J2}%
  \BibitemOpen
  \bibfield  {author} {\bibinfo {author} {\bibfnamefont {Kiyomi}\ \bibnamefont
  {Okamoto}}\ and\ \bibinfo {author} {\bibfnamefont {Tota}\ \bibnamefont
  {Nakamura}},\ }\bibfield  {title} {\enquote {\bibinfo {title} {{Critical
  properties of the spin-$\frac{1}{2}$ Heisenberg chain with frustration and
  bond alternation}},}\ }\href {\doibase 10.1088/0305-4470/30/18/012}
  {\bibfield  {journal} {\bibinfo  {journal} {Journal of Physics A:
  Mathematical and General}\ }\textbf {\bibinfo {volume} {30}},\ \bibinfo
  {pages} {6287--6298} (\bibinfo {year} {1997})}\BibitemShut {NoStop}%
\bibitem [{\citenamefont {Lukyanov}(1997)}]{lukyanov_sg}%
  \BibitemOpen
  \bibfield  {author} {\bibinfo {author} {\bibfnamefont {Sergei}\ \bibnamefont
  {Lukyanov}},\ }\bibfield  {title} {\enquote {\bibinfo {title} {Form factors
  of exponential fields in the sine–gordon model},}\ }\href {\doibase
  10.1142/S0217732397002673} {\bibfield  {journal} {\bibinfo  {journal} {Modern
  Physics Letters A}\ }\textbf {\bibinfo {volume} {12}},\ \bibinfo {pages}
  {2543--2550} (\bibinfo {year} {1997})}\BibitemShut {NoStop}%
\bibitem [{\citenamefont {Zamolodchikov}(1995)}]{zamolodchikov_mass}%
  \BibitemOpen
  \bibfield  {author} {\bibinfo {author} {\bibfnamefont {Al.~B.}\ \bibnamefont
  {Zamolodchikov}},\ }\bibfield  {title} {\enquote {\bibinfo {title} {Mass
  scale in the sine--gordon model and its reductions},}\ }\href
  {https://www.worldscientific.com/doi/abs/10.1142/S0217751X9500053X}
  {\bibfield  {journal} {\bibinfo  {journal} {International Journal of Modern
  Physics A}\ }\textbf {\bibinfo {volume} {10}},\ \bibinfo {pages} {1125--1150}
  (\bibinfo {year} {1995})}\BibitemShut {NoStop}%
\bibitem [{\citenamefont {Antonides}\ \emph {et~al.}(1977)\citenamefont
  {Antonides}, \citenamefont {Janse},\ and\ \citenamefont
  {Sawatzky}}]{antonides_3dTM}%
  \BibitemOpen
  \bibfield  {author} {\bibinfo {author} {\bibfnamefont {E.}~\bibnamefont
  {Antonides}}, \bibinfo {author} {\bibfnamefont {E.~C.}\ \bibnamefont
  {Janse}}, \ and\ \bibinfo {author} {\bibfnamefont {G.~A.}\ \bibnamefont
  {Sawatzky}},\ }\bibfield  {title} {\enquote {\bibinfo {title}
  {{$\mathrm{LMM}$ Auger spectra of Cu, Zn, Ga, and Ge. I. Transition
  probabilities, term splittings, and effective Coulomb interaction}},}\ }\href
  {\doibase 10.1103/PhysRevB.15.1669} {\bibfield  {journal} {\bibinfo
  {journal} {Phys. Rev. B}\ }\textbf {\bibinfo {volume} {15}},\ \bibinfo
  {pages} {1669--1679} (\bibinfo {year} {1977})}\BibitemShut {NoStop}%
\bibitem [{\citenamefont {Yin}\ \emph {et~al.}(1977)\citenamefont {Yin},
  \citenamefont {Tsang},\ and\ \citenamefont {Adler}}]{yin_3dTM}%
  \BibitemOpen
  \bibfield  {author} {\bibinfo {author} {\bibfnamefont {Lo~I}\ \bibnamefont
  {Yin}}, \bibinfo {author} {\bibfnamefont {Tung}\ \bibnamefont {Tsang}}, \
  and\ \bibinfo {author} {\bibfnamefont {Isidore}\ \bibnamefont {Adler}},\
  }\bibfield  {title} {\enquote {\bibinfo {title} {{Electron delocalization and
  the characterization of the ${L}_{3}\mathrm{MM}$ Auger spectra of $3d$
  transition metals}},}\ }\href {\doibase 10.1103/PhysRevB.15.2974} {\bibfield
  {journal} {\bibinfo  {journal} {Phys. Rev. B}\ }\textbf {\bibinfo {volume}
  {15}},\ \bibinfo {pages} {2974--2983} (\bibinfo {year} {1977})}\BibitemShut
  {NoStop}%
\bibitem [{\citenamefont {Fujimori}\ \emph {et~al.}(1993)\citenamefont
  {Fujimori}, \citenamefont {Bocquet}, \citenamefont {Saitoh},\ and\
  \citenamefont {Mizokawa}}]{fujimori_3dTM}%
  \BibitemOpen
  \bibfield  {author} {\bibinfo {author} {\bibfnamefont {A.}~\bibnamefont
  {Fujimori}}, \bibinfo {author} {\bibfnamefont {A.E.}\ \bibnamefont
  {Bocquet}}, \bibinfo {author} {\bibfnamefont {T.}~\bibnamefont {Saitoh}}, \
  and\ \bibinfo {author} {\bibfnamefont {T.}~\bibnamefont {Mizokawa}},\
  }\bibfield  {title} {\enquote {\bibinfo {title} {Electronic structure of 3d
  transition metal compounds: systematic chemical trends and multiplet
  effects},}\ }\href {\doibase https://doi.org/10.1016/0368-2048(93)80011-A}
  {\bibfield  {journal} {\bibinfo  {journal} {Journal of Electron Spectroscopy
  and Related Phenomena}\ }\textbf {\bibinfo {volume} {62}},\ \bibinfo {pages}
  {141 -- 152} (\bibinfo {year} {1993})}\BibitemShut {NoStop}%
\bibitem [{\citenamefont {Furuya}(2020)}]{furuya_screw}%
  \BibitemOpen
  \bibfield  {author} {\bibinfo {author} {\bibfnamefont {Shunsuke~C.}\
  \bibnamefont {Furuya}},\ }\bibfield  {title} {\enquote {\bibinfo {title}
  {Field-induced dimer orders in quantum spin chains},}\ }\href {\doibase
  10.1103/PhysRevB.101.134425} {\bibfield  {journal} {\bibinfo  {journal}
  {Phys. Rev. B}\ }\textbf {\bibinfo {volume} {101}},\ \bibinfo {pages}
  {134425} (\bibinfo {year} {2020})}\BibitemShut {NoStop}%
\bibitem [{\citenamefont {Shelton}\ \emph {et~al.}(1996)\citenamefont
  {Shelton}, \citenamefont {Nersesyan},\ and\ \citenamefont
  {Tsvelik}}]{shelton_ladder}%
  \BibitemOpen
  \bibfield  {author} {\bibinfo {author} {\bibfnamefont {D.~G.}\ \bibnamefont
  {Shelton}}, \bibinfo {author} {\bibfnamefont {A.~A.}\ \bibnamefont
  {Nersesyan}}, \ and\ \bibinfo {author} {\bibfnamefont {A.~M.}\ \bibnamefont
  {Tsvelik}},\ }\bibfield  {title} {\enquote {\bibinfo {title}
  {{Antiferromagnetic spin ladders: Crossover between spin S=1/2 and S=1
  chains}},}\ }\href {\doibase 10.1103/PhysRevB.53.8521} {\bibfield  {journal}
  {\bibinfo  {journal} {Phys. Rev. B}\ }\textbf {\bibinfo {volume} {53}},\
  \bibinfo {pages} {8521--8532} (\bibinfo {year} {1996})}\BibitemShut {NoStop}%
\bibitem [{\citenamefont {Nawa}\ \emph {et~al.}(2014)\citenamefont {Nawa},
  \citenamefont {Okamoto}, \citenamefont {Matsuo}, \citenamefont {Kindo},
  \citenamefont {Kitahara}, \citenamefont {Yoshida}, \citenamefont {Ikeda},
  \citenamefont {Hara}, \citenamefont {Sakurai}, \citenamefont {Okubo},
  \citenamefont {Ohta},\ and\ \citenamefont {Hiroi}}]{nawa_1d_J1J2_jpsj}%
  \BibitemOpen
  \bibfield  {author} {\bibinfo {author} {\bibfnamefont {Kazuhiro}\
  \bibnamefont {Nawa}}, \bibinfo {author} {\bibfnamefont {Yoshihiko}\
  \bibnamefont {Okamoto}}, \bibinfo {author} {\bibfnamefont {Akira}\
  \bibnamefont {Matsuo}}, \bibinfo {author} {\bibfnamefont {Koichi}\
  \bibnamefont {Kindo}}, \bibinfo {author} {\bibfnamefont {Yoko}\ \bibnamefont
  {Kitahara}}, \bibinfo {author} {\bibfnamefont {Syota}\ \bibnamefont
  {Yoshida}}, \bibinfo {author} {\bibfnamefont {Shohei}\ \bibnamefont {Ikeda}},
  \bibinfo {author} {\bibfnamefont {Shigeo}\ \bibnamefont {Hara}}, \bibinfo
  {author} {\bibfnamefont {Takahiro}\ \bibnamefont {Sakurai}}, \bibinfo
  {author} {\bibfnamefont {Susumu}\ \bibnamefont {Okubo}}, \bibinfo {author}
  {\bibfnamefont {Hitoshi}\ \bibnamefont {Ohta}}, \ and\ \bibinfo {author}
  {\bibfnamefont {Zenji}\ \bibnamefont {Hiroi}},\ }\bibfield  {title} {\enquote
  {\bibinfo {title} {{NaCuMoO4(OH) as a Candidate Frustrated J1–J2 Chain
  Quantum Magnet}},}\ }\href {\doibase 10.7566/JPSJ.83.103702} {\bibfield
  {journal} {\bibinfo  {journal} {Journal of the Physical Society of Japan}\
  }\textbf {\bibinfo {volume} {83}},\ \bibinfo {pages} {103702} (\bibinfo
  {year} {2014})}\BibitemShut {NoStop}%
\bibitem [{\citenamefont {Nawa}\ \emph {et~al.}(2017)\citenamefont {Nawa},
  \citenamefont {Yoshida}, \citenamefont {Takigawa}, \citenamefont {Okamoto},\
  and\ \citenamefont {Hiroi}}]{nawa_1d_J1J2_prb}%
  \BibitemOpen
  \bibfield  {author} {\bibinfo {author} {\bibfnamefont {Kazuhiro}\
  \bibnamefont {Nawa}}, \bibinfo {author} {\bibfnamefont {Makoto}\ \bibnamefont
  {Yoshida}}, \bibinfo {author} {\bibfnamefont {Masashi}\ \bibnamefont
  {Takigawa}}, \bibinfo {author} {\bibfnamefont {Yoshihiko}\ \bibnamefont
  {Okamoto}}, \ and\ \bibinfo {author} {\bibfnamefont {Zenji}\ \bibnamefont
  {Hiroi}},\ }\bibfield  {title} {\enquote {\bibinfo {title} {{Collinear spin
  density wave order and anisotropic spin fluctuations in the frustrated
  ${J}_{1}\ensuremath{-}{J}_{2}$ chain magnet
  ${\mathrm{NaCuMoO}}_{4}(\mathrm{OH})$}},}\ }\href {\doibase
  10.1103/PhysRevB.96.174433} {\bibfield  {journal} {\bibinfo  {journal} {Phys.
  Rev. B}\ }\textbf {\bibinfo {volume} {96}},\ \bibinfo {pages} {174433}
  (\bibinfo {year} {2017})}\BibitemShut {NoStop}%
\bibitem [{\citenamefont {Zvyagin}\ \emph {et~al.}(2004)\citenamefont
  {Zvyagin}, \citenamefont {Krzystek}, \citenamefont {{van Loosdrecht}},
  \citenamefont {Dhalenne},\ and\ \citenamefont
  {Revcolevschi}}]{zvyagin_cugeo3}%
  \BibitemOpen
  \bibfield  {author} {\bibinfo {author} {\bibfnamefont {S.A.}\ \bibnamefont
  {Zvyagin}}, \bibinfo {author} {\bibfnamefont {J.}~\bibnamefont {Krzystek}},
  \bibinfo {author} {\bibfnamefont {P.H.M.}\ \bibnamefont {{van Loosdrecht}}},
  \bibinfo {author} {\bibfnamefont {G.}~\bibnamefont {Dhalenne}}, \ and\
  \bibinfo {author} {\bibfnamefont {A.}~\bibnamefont {Revcolevschi}},\
  }\bibfield  {title} {\enquote {\bibinfo {title} {{High-field ESR study of the
  dimerized-incommensurate phase transition in the spin-Peierls compound
  CuGeO3}},}\ }\href {\doibase https://doi.org/10.1016/j.physb.2004.01.009}
  {\bibfield  {journal} {\bibinfo  {journal} {Physica B: Condensed Matter}\
  }\textbf {\bibinfo {volume} {346-347}},\ \bibinfo {pages} {1 -- 5} (\bibinfo
  {year} {2004})},\ \bibinfo {note} {proceedings of the 7th International
  Symposium on Research in High Magnetic Fields}\BibitemShut {NoStop}%
\bibitem [{\citenamefont {Glazkov}\ \emph {et~al.}(2010)\citenamefont
  {Glazkov}, \citenamefont {Smirnov}, \citenamefont {Zheludev},\ and\
  \citenamefont {Sales}}]{glazkov_ntenp}%
  \BibitemOpen
  \bibfield  {author} {\bibinfo {author} {\bibfnamefont {V.~N.}\ \bibnamefont
  {Glazkov}}, \bibinfo {author} {\bibfnamefont {A.~I.}\ \bibnamefont
  {Smirnov}}, \bibinfo {author} {\bibfnamefont {A.}~\bibnamefont {Zheludev}}, \
  and\ \bibinfo {author} {\bibfnamefont {B.~C}\ \bibnamefont {Sales}},\
  }\bibfield  {title} {\enquote {\bibinfo {title} {{Modes of magnetic resonance
  of the $S=1$ dimer chain compound NTENP}},}\ }\href {\doibase
  10.1103/PhysRevB.82.184406} {\bibfield  {journal} {\bibinfo  {journal} {Phys.
  Rev. B}\ }\textbf {\bibinfo {volume} {82}},\ \bibinfo {pages} {184406}
  (\bibinfo {year} {2010})}\BibitemShut {NoStop}%
\bibitem [{\citenamefont {Nishi}\ \emph {et~al.}(1994)\citenamefont {Nishi},
  \citenamefont {Fujita},\ and\ \citenamefont {Akimitsu}}]{nishi_cugeo3_ins}%
  \BibitemOpen
  \bibfield  {author} {\bibinfo {author} {\bibfnamefont {M.}~\bibnamefont
  {Nishi}}, \bibinfo {author} {\bibfnamefont {O.}~\bibnamefont {Fujita}}, \
  and\ \bibinfo {author} {\bibfnamefont {J.}~\bibnamefont {Akimitsu}},\
  }\bibfield  {title} {\enquote {\bibinfo {title} {Neutron-scattering study on
  the spin-peierls transition in a quasi-one-dimensional magnet
  ${\mathrm{cugeo}}_{3}$},}\ }\href {\doibase 10.1103/PhysRevB.50.6508}
  {\bibfield  {journal} {\bibinfo  {journal} {Phys. Rev. B}\ }\textbf {\bibinfo
  {volume} {50}},\ \bibinfo {pages} {6508--6510} (\bibinfo {year}
  {1994})}\BibitemShut {NoStop}%
\bibitem [{\citenamefont {Sato}\ \emph {et~al.}(2012)\citenamefont {Sato},
  \citenamefont {Katsura},\ and\ \citenamefont {Nagaosa}}]{sato_raman}%
  \BibitemOpen
  \bibfield  {author} {\bibinfo {author} {\bibfnamefont {Masahiro}\
  \bibnamefont {Sato}}, \bibinfo {author} {\bibfnamefont {Hosho}\ \bibnamefont
  {Katsura}}, \ and\ \bibinfo {author} {\bibfnamefont {Naoto}\ \bibnamefont
  {Nagaosa}},\ }\bibfield  {title} {\enquote {\bibinfo {title} {Theory of raman
  scattering in one-dimensional quantum spin-$\frac{1}{2}$ antiferromagnets},}\
  }\href {\doibase 10.1103/PhysRevLett.108.237401} {\bibfield  {journal}
  {\bibinfo  {journal} {Phys. Rev. Lett.}\ }\textbf {\bibinfo {volume} {108}},\
  \bibinfo {pages} {237401} (\bibinfo {year} {2012})}\BibitemShut {NoStop}%
\bibitem [{\citenamefont {Prosnikov}\ \emph {et~al.}(2018)\citenamefont
  {Prosnikov}, \citenamefont {Smirnov}, \citenamefont {Davydov}, \citenamefont
  {Pisarev}, \citenamefont {Lyubochko},\ and\ \citenamefont
  {Barilo}}]{prosnikov_raman}%
  \BibitemOpen
  \bibfield  {author} {\bibinfo {author} {\bibfnamefont {M.~A.}\ \bibnamefont
  {Prosnikov}}, \bibinfo {author} {\bibfnamefont {A.~N.}\ \bibnamefont
  {Smirnov}}, \bibinfo {author} {\bibfnamefont {V.~Yu.}\ \bibnamefont
  {Davydov}}, \bibinfo {author} {\bibfnamefont {R.~V.}\ \bibnamefont
  {Pisarev}}, \bibinfo {author} {\bibfnamefont {N.~A.}\ \bibnamefont
  {Lyubochko}}, \ and\ \bibinfo {author} {\bibfnamefont {S.~N.}\ \bibnamefont
  {Barilo}},\ }\bibfield  {title} {\enquote {\bibinfo {title} {{Magnetic
  dynamics and spin-phonon coupling in the antiferromagnet
  ${\mathrm{Ni}}_{2}{\mathrm{NbBO}}_{6}$}},}\ }\href {\doibase
  10.1103/PhysRevB.98.104404} {\bibfield  {journal} {\bibinfo  {journal} {Phys.
  Rev. B}\ }\textbf {\bibinfo {volume} {98}},\ \bibinfo {pages} {104404}
  (\bibinfo {year} {2018})}\BibitemShut {NoStop}%
\bibitem [{\citenamefont {Oshikawa}\ and\ \citenamefont
  {Affleck}(1997)}]{oshikawa_staggered_field_prl}%
  \BibitemOpen
  \bibfield  {author} {\bibinfo {author} {\bibfnamefont {Masaki}\ \bibnamefont
  {Oshikawa}}\ and\ \bibinfo {author} {\bibfnamefont {Ian}\ \bibnamefont
  {Affleck}},\ }\bibfield  {title} {\enquote {\bibinfo {title} {{Field-Induced
  Gap in $\mathit{S}=1/2$ Antiferromagnetic Chains}},}\ }\href {\doibase
  10.1103/PhysRevLett.79.2883} {\bibfield  {journal} {\bibinfo  {journal}
  {Phys. Rev. Lett.}\ }\textbf {\bibinfo {volume} {79}},\ \bibinfo {pages}
  {2883--2886} (\bibinfo {year} {1997})}\BibitemShut {NoStop}%
\bibitem [{\citenamefont {Affleck}\ and\ \citenamefont
  {Oshikawa}(1999)}]{oshikawa_staggered_field_prb}%
  \BibitemOpen
  \bibfield  {author} {\bibinfo {author} {\bibfnamefont {Ian}\ \bibnamefont
  {Affleck}}\ and\ \bibinfo {author} {\bibfnamefont {Masaki}\ \bibnamefont
  {Oshikawa}},\ }\bibfield  {title} {\enquote {\bibinfo {title} {{Field-induced
  gap in Cu benzoate and other $S=\frac{1}{2}$ antiferromagnetic chains}},}\
  }\href {\doibase 10.1103/PhysRevB.60.1038} {\bibfield  {journal} {\bibinfo
  {journal} {Phys. Rev. B}\ }\textbf {\bibinfo {volume} {60}},\ \bibinfo
  {pages} {1038--1056} (\bibinfo {year} {1999})}\BibitemShut {NoStop}%
\bibitem [{\citenamefont {Chen}\ \emph
  {et~al.}(2011{\natexlab{a}})\citenamefont {Chen}, \citenamefont {Gu},\ and\
  \citenamefont {Wen}}]{chen_spt_2011a}%
  \BibitemOpen
  \bibfield  {author} {\bibinfo {author} {\bibfnamefont {Xie}\ \bibnamefont
  {Chen}}, \bibinfo {author} {\bibfnamefont {Zheng-Cheng}\ \bibnamefont {Gu}},
  \ and\ \bibinfo {author} {\bibfnamefont {Xiao-Gang}\ \bibnamefont {Wen}},\
  }\bibfield  {title} {\enquote {\bibinfo {title} {Classification of gapped
  symmetric phases in one-dimensional spin systems},}\ }\href {\doibase
  10.1103/PhysRevB.83.035107} {\bibfield  {journal} {\bibinfo  {journal} {Phys.
  Rev. B}\ }\textbf {\bibinfo {volume} {83}},\ \bibinfo {pages} {035107}
  (\bibinfo {year} {2011}{\natexlab{a}})}\BibitemShut {NoStop}%
\bibitem [{\citenamefont {Chen}\ \emph
  {et~al.}(2011{\natexlab{b}})\citenamefont {Chen}, \citenamefont {Gu},\ and\
  \citenamefont {Wen}}]{chen_spt_2011b}%
  \BibitemOpen
  \bibfield  {author} {\bibinfo {author} {\bibfnamefont {Xie}\ \bibnamefont
  {Chen}}, \bibinfo {author} {\bibfnamefont {Zheng-Cheng}\ \bibnamefont {Gu}},
  \ and\ \bibinfo {author} {\bibfnamefont {Xiao-Gang}\ \bibnamefont {Wen}},\
  }\bibfield  {title} {\enquote {\bibinfo {title} {Complete classification of
  one-dimensional gapped quantum phases in interacting spin systems},}\ }\href
  {\doibase 10.1103/PhysRevB.84.235128} {\bibfield  {journal} {\bibinfo
  {journal} {Phys. Rev. B}\ }\textbf {\bibinfo {volume} {84}},\ \bibinfo
  {pages} {235128} (\bibinfo {year} {2011}{\natexlab{b}})}\BibitemShut
  {NoStop}%
\bibitem [{\citenamefont {Nakagawa}\ \emph {et~al.}(2018)\citenamefont
  {Nakagawa}, \citenamefont {Yoshida}, \citenamefont {Peters},\ and\
  \citenamefont {Kawakami}}]{nakagawa_pump}%
  \BibitemOpen
  \bibfield  {author} {\bibinfo {author} {\bibfnamefont {Masaya}\ \bibnamefont
  {Nakagawa}}, \bibinfo {author} {\bibfnamefont {Tsuneya}\ \bibnamefont
  {Yoshida}}, \bibinfo {author} {\bibfnamefont {Robert}\ \bibnamefont
  {Peters}}, \ and\ \bibinfo {author} {\bibfnamefont {Norio}\ \bibnamefont
  {Kawakami}},\ }\bibfield  {title} {\enquote {\bibinfo {title} {Breakdown of
  topological thouless pumping in the strongly interacting regime},}\ }\href
  {\doibase 10.1103/PhysRevB.98.115147} {\bibfield  {journal} {\bibinfo
  {journal} {Phys. Rev. B}\ }\textbf {\bibinfo {volume} {98}},\ \bibinfo
  {pages} {115147} (\bibinfo {year} {2018})}\BibitemShut {NoStop}%
\bibitem [{\citenamefont {Sakai}\ and\ \citenamefont
  {Shiba}(1994)}]{sakai_nep}%
  \BibitemOpen
  \bibfield  {author} {\bibinfo {author} {\bibfnamefont {T{\^o}ru}\
  \bibnamefont {Sakai}}\ and\ \bibinfo {author} {\bibfnamefont {Hiroyuki}\
  \bibnamefont {Shiba}},\ }\bibfield  {title} {\enquote {\bibinfo {title}
  {{Numerical Study of a Model for NENP: One-Dimensional S=1 Antiferromagnet in
  a Staggered Field}},}\ }\href {\doibase 10.1143/JPSJ.63.867} {\bibfield
  {journal} {\bibinfo  {journal} {Journal of the Physical Society of Japan}\
  }\textbf {\bibinfo {volume} {63}},\ \bibinfo {pages} {867--871} (\bibinfo
  {year} {1994})}\BibitemShut {NoStop}%
\bibitem [{\citenamefont {Oshikawa}\ and\ \citenamefont
  {Affleck}(1999)}]{oshikawa_esr_prl}%
  \BibitemOpen
  \bibfield  {author} {\bibinfo {author} {\bibfnamefont {Masaki}\ \bibnamefont
  {Oshikawa}}\ and\ \bibinfo {author} {\bibfnamefont {Ian}\ \bibnamefont
  {Affleck}},\ }\bibfield  {title} {\enquote {\bibinfo {title} {Low-temperature
  electron spin resonance theory for half-integer spin antiferromagnetic
  chains},}\ }\href {\doibase 10.1103/PhysRevLett.82.5136} {\bibfield
  {journal} {\bibinfo  {journal} {Phys. Rev. Lett.}\ }\textbf {\bibinfo
  {volume} {82}},\ \bibinfo {pages} {5136--5139} (\bibinfo {year}
  {1999})}\BibitemShut {NoStop}%
\bibitem [{\citenamefont {Furuya}\ and\ \citenamefont
  {Oshikawa}(2012)}]{furuya_bbs}%
  \BibitemOpen
  \bibfield  {author} {\bibinfo {author} {\bibfnamefont {Shunsuke~C.}\
  \bibnamefont {Furuya}}\ and\ \bibinfo {author} {\bibfnamefont {Masaki}\
  \bibnamefont {Oshikawa}},\ }\bibfield  {title} {\enquote {\bibinfo {title}
  {{Boundary Resonances in $S\mathbf{=}1/2$ Antiferromagnetic Chains Under a
  Staggered Field}},}\ }\href {\doibase 10.1103/PhysRevLett.109.247603}
  {\bibfield  {journal} {\bibinfo  {journal} {Phys. Rev. Lett.}\ }\textbf
  {\bibinfo {volume} {109}},\ \bibinfo {pages} {247603} (\bibinfo {year}
  {2012})}\BibitemShut {NoStop}%
\bibitem [{\citenamefont {Pollmann}\ \emph {et~al.}(2010)\citenamefont
  {Pollmann}, \citenamefont {Turner}, \citenamefont {Berg},\ and\ \citenamefont
  {Oshikawa}}]{pollmann_haldane2010}%
  \BibitemOpen
  \bibfield  {author} {\bibinfo {author} {\bibfnamefont {Frank}\ \bibnamefont
  {Pollmann}}, \bibinfo {author} {\bibfnamefont {Ari~M.}\ \bibnamefont
  {Turner}}, \bibinfo {author} {\bibfnamefont {Erez}\ \bibnamefont {Berg}}, \
  and\ \bibinfo {author} {\bibfnamefont {Masaki}\ \bibnamefont {Oshikawa}},\
  }\bibfield  {title} {\enquote {\bibinfo {title} {Entanglement spectrum of a
  topological phase in one dimension},}\ }\href {\doibase
  10.1103/PhysRevB.81.064439} {\bibfield  {journal} {\bibinfo  {journal} {Phys.
  Rev. B}\ }\textbf {\bibinfo {volume} {81}},\ \bibinfo {pages} {064439}
  (\bibinfo {year} {2010})}\BibitemShut {NoStop}%
\bibitem [{\citenamefont {Pollmann}\ \emph {et~al.}(2012)\citenamefont
  {Pollmann}, \citenamefont {Berg}, \citenamefont {Turner},\ and\ \citenamefont
  {Oshikawa}}]{pollmann_haldane2012}%
  \BibitemOpen
  \bibfield  {author} {\bibinfo {author} {\bibfnamefont {Frank}\ \bibnamefont
  {Pollmann}}, \bibinfo {author} {\bibfnamefont {Erez}\ \bibnamefont {Berg}},
  \bibinfo {author} {\bibfnamefont {Ari~M.}\ \bibnamefont {Turner}}, \ and\
  \bibinfo {author} {\bibfnamefont {Masaki}\ \bibnamefont {Oshikawa}},\
  }\bibfield  {title} {\enquote {\bibinfo {title} {Symmetry protection of
  topological phases in one-dimensional quantum spin systems},}\ }\href
  {\doibase 10.1103/PhysRevB.85.075125} {\bibfield  {journal} {\bibinfo
  {journal} {Phys. Rev. B}\ }\textbf {\bibinfo {volume} {85}},\ \bibinfo
  {pages} {075125} (\bibinfo {year} {2012})}\BibitemShut {NoStop}%
\bibitem [{\citenamefont {Hagiwara}\ \emph {et~al.}(1990)\citenamefont
  {Hagiwara}, \citenamefont {Katsumata}, \citenamefont {Affleck}, \citenamefont
  {Halperin},\ and\ \citenamefont {Renard}}]{hagiwara_haldane_esr}%
  \BibitemOpen
  \bibfield  {author} {\bibinfo {author} {\bibfnamefont {M.}~\bibnamefont
  {Hagiwara}}, \bibinfo {author} {\bibfnamefont {K.}~\bibnamefont {Katsumata}},
  \bibinfo {author} {\bibfnamefont {Ian}\ \bibnamefont {Affleck}}, \bibinfo
  {author} {\bibfnamefont {B.~I.}\ \bibnamefont {Halperin}}, \ and\ \bibinfo
  {author} {\bibfnamefont {J.~P.}\ \bibnamefont {Renard}},\ }\bibfield  {title}
  {\enquote {\bibinfo {title} {{Observation of S=1/2 degrees of freedom in an
  S=1 linear-chain Heisenberg antiferromagnet}},}\ }\href {\doibase
  10.1103/PhysRevLett.65.3181} {\bibfield  {journal} {\bibinfo  {journal}
  {Phys. Rev. Lett.}\ }\textbf {\bibinfo {volume} {65}},\ \bibinfo {pages}
  {3181--3184} (\bibinfo {year} {1990})}\BibitemShut {NoStop}%
\bibitem [{\citenamefont {Yoshida}\ \emph {et~al.}(2005)\citenamefont
  {Yoshida}, \citenamefont {Shiraki}, \citenamefont {Okubo}, \citenamefont
  {Ohta}, \citenamefont {Ito}, \citenamefont {Takagi}, \citenamefont
  {Kaburagi},\ and\ \citenamefont {Ajiro}}]{yoshida_haldane_esr}%
  \BibitemOpen
  \bibfield  {author} {\bibinfo {author} {\bibfnamefont {M.}~\bibnamefont
  {Yoshida}}, \bibinfo {author} {\bibfnamefont {K.}~\bibnamefont {Shiraki}},
  \bibinfo {author} {\bibfnamefont {S.}~\bibnamefont {Okubo}}, \bibinfo
  {author} {\bibfnamefont {H.}~\bibnamefont {Ohta}}, \bibinfo {author}
  {\bibfnamefont {T.}~\bibnamefont {Ito}}, \bibinfo {author} {\bibfnamefont
  {H.}~\bibnamefont {Takagi}}, \bibinfo {author} {\bibfnamefont
  {M.}~\bibnamefont {Kaburagi}}, \ and\ \bibinfo {author} {\bibfnamefont
  {Y.}~\bibnamefont {Ajiro}},\ }\bibfield  {title} {\enquote {\bibinfo {title}
  {{Energy Structure of a Finite Haldane Chain in
  ${\mathrm{Y}}_{2}{\mathrm{BaNi}}_{0.96}{\mathrm{Mg}}_{0.04}{\mathrm{O}}_{5}$
  Studied by High Field Electron Spin Resonance}},}\ }\href {\doibase
  10.1103/PhysRevLett.95.117202} {\bibfield  {journal} {\bibinfo  {journal}
  {Phys. Rev. Lett.}\ }\textbf {\bibinfo {volume} {95}},\ \bibinfo {pages}
  {117202} (\bibinfo {year} {2005})}\BibitemShut {NoStop}%
\bibitem [{\citenamefont {Furuya}\ and\ \citenamefont {Sato}()}]{furuya_dm}%
  \BibitemOpen
  \bibfield  {author} {\bibinfo {author} {\bibfnamefont {Shunsuke~C.}\
  \bibnamefont {Furuya}}\ and\ \bibinfo {author} {\bibfnamefont {Masahiro}\
  \bibnamefont {Sato}},\ }\href@noop {} {}\bibinfo {note}
  {Unpublished.}\BibitemShut {Stop}%
\bibitem [{\citenamefont {Kirilyuk}\ \emph {et~al.}(2010)\citenamefont
  {Kirilyuk}, \citenamefont {Kimel},\ and\ \citenamefont
  {Rasing}}]{kirilyuk_rmp}%
  \BibitemOpen
  \bibfield  {author} {\bibinfo {author} {\bibfnamefont {Andrei}\ \bibnamefont
  {Kirilyuk}}, \bibinfo {author} {\bibfnamefont {Alexey~V.}\ \bibnamefont
  {Kimel}}, \ and\ \bibinfo {author} {\bibfnamefont {Theo}\ \bibnamefont
  {Rasing}},\ }\bibfield  {title} {\enquote {\bibinfo {title} {Ultrafast
  optical manipulation of magnetic order},}\ }\href {\doibase
  10.1103/RevModPhys.82.2731} {\bibfield  {journal} {\bibinfo  {journal} {Rev.
  Mod. Phys.}\ }\textbf {\bibinfo {volume} {82}},\ \bibinfo {pages}
  {2731--2784} (\bibinfo {year} {2010})}\BibitemShut {NoStop}%
\bibitem [{\citenamefont {Oshikawa}\ and\ \citenamefont
  {Affleck}(2002)}]{oshikawa-affleck}%
  \BibitemOpen
  \bibfield  {author} {\bibinfo {author} {\bibfnamefont {Masaki}\ \bibnamefont
  {Oshikawa}}\ and\ \bibinfo {author} {\bibfnamefont {Ian}\ \bibnamefont
  {Affleck}},\ }\bibfield  {title} {\enquote {\bibinfo {title} {Electron spin
  resonance in $s=\frac{1}{2}$ antiferromagnetic chains},}\ }\href {\doibase
  10.1103/PhysRevB.65.134410} {\bibfield  {journal} {\bibinfo  {journal} {Phys.
  Rev. B}\ }\textbf {\bibinfo {volume} {65}},\ \bibinfo {pages} {134410}
  (\bibinfo {year} {2002})}\BibitemShut {NoStop}%
\bibitem [{\citenamefont {Furuya}\ and\ \citenamefont
  {Sato}(2015)}]{furuya_width}%
  \BibitemOpen
  \bibfield  {author} {\bibinfo {author} {\bibfnamefont {Shunsuke~C.}\
  \bibnamefont {Furuya}}\ and\ \bibinfo {author} {\bibfnamefont {Masahiro}\
  \bibnamefont {Sato}},\ }\bibfield  {title} {\enquote {\bibinfo {title}
  {Electron spin resonance in quasi-one-dimensional quantum antiferromagnets:
  Relevance of weak interchain interactions},}\ }\href {\doibase
  10.7566/jpsj.84.033704} {\bibfield  {journal} {\bibinfo  {journal} {Journal
  of the Physical Society of Japan}\ }\textbf {\bibinfo {volume} {84}},\
  \bibinfo {pages} {033704} (\bibinfo {year} {2015})}\BibitemShut {NoStop}%
\bibitem [{\citenamefont {Beaurepaire}\ \emph {et~al.}(1996)\citenamefont
  {Beaurepaire}, \citenamefont {Merle}, \citenamefont {Daunois},\ and\
  \citenamefont {Bigot}}]{beaurepaire_1996}%
  \BibitemOpen
  \bibfield  {author} {\bibinfo {author} {\bibfnamefont {E.}~\bibnamefont
  {Beaurepaire}}, \bibinfo {author} {\bibfnamefont {J.-C.}\ \bibnamefont
  {Merle}}, \bibinfo {author} {\bibfnamefont {A.}~\bibnamefont {Daunois}}, \
  and\ \bibinfo {author} {\bibfnamefont {J.-Y.}\ \bibnamefont {Bigot}},\
  }\bibfield  {title} {\enquote {\bibinfo {title} {Ultrafast spin dynamics in
  ferromagnetic nickel},}\ }\href {\doibase 10.1103/PhysRevLett.76.4250}
  {\bibfield  {journal} {\bibinfo  {journal} {Phys. Rev. Lett.}\ }\textbf
  {\bibinfo {volume} {76}},\ \bibinfo {pages} {4250--4253} (\bibinfo {year}
  {1996})}\BibitemShut {NoStop}%
\bibitem [{\citenamefont {Koopmans}\ \emph {et~al.}(2000)\citenamefont
  {Koopmans}, \citenamefont {van Kampen}, \citenamefont {Kohlhepp},\ and\
  \citenamefont {de~Jonge}}]{koopmans_2000}%
  \BibitemOpen
  \bibfield  {author} {\bibinfo {author} {\bibfnamefont {B.}~\bibnamefont
  {Koopmans}}, \bibinfo {author} {\bibfnamefont {M.}~\bibnamefont {van
  Kampen}}, \bibinfo {author} {\bibfnamefont {J.~T.}\ \bibnamefont {Kohlhepp}},
  \ and\ \bibinfo {author} {\bibfnamefont {W.~J.~M.}\ \bibnamefont
  {de~Jonge}},\ }\bibfield  {title} {\enquote {\bibinfo {title} {Ultrafast
  magneto-optics in nickel: Magnetism or optics?}}\ }\href {\doibase
  10.1103/PhysRevLett.85.844} {\bibfield  {journal} {\bibinfo  {journal} {Phys.
  Rev. Lett.}\ }\textbf {\bibinfo {volume} {85}},\ \bibinfo {pages} {844--847}
  (\bibinfo {year} {2000})}\BibitemShut {NoStop}%
\bibitem [{\citenamefont {Mashkovich}\ \emph {et~al.}(2019)\citenamefont
  {Mashkovich}, \citenamefont {Grishunin}, \citenamefont {Mikhaylovskiy},
  \citenamefont {Zvezdin}, \citenamefont {Pisarev}, \citenamefont {Strugatsky},
  \citenamefont {Christianen}, \citenamefont {Rasing},\ and\ \citenamefont
  {Kimel}}]{mashkovich_2019}%
  \BibitemOpen
  \bibfield  {author} {\bibinfo {author} {\bibfnamefont {E.~A.}\ \bibnamefont
  {Mashkovich}}, \bibinfo {author} {\bibfnamefont {K.~A.}\ \bibnamefont
  {Grishunin}}, \bibinfo {author} {\bibfnamefont {R.~V.}\ \bibnamefont
  {Mikhaylovskiy}}, \bibinfo {author} {\bibfnamefont {A.~K.}\ \bibnamefont
  {Zvezdin}}, \bibinfo {author} {\bibfnamefont {R.~V.}\ \bibnamefont
  {Pisarev}}, \bibinfo {author} {\bibfnamefont {M.~B.}\ \bibnamefont
  {Strugatsky}}, \bibinfo {author} {\bibfnamefont {P.~C.~M.}\ \bibnamefont
  {Christianen}}, \bibinfo {author} {\bibfnamefont {Th.}\ \bibnamefont
  {Rasing}}, \ and\ \bibinfo {author} {\bibfnamefont {A.~V.}\ \bibnamefont
  {Kimel}},\ }\bibfield  {title} {\enquote {\bibinfo {title} {Terahertz
  optomagnetism: Nonlinear thz excitation of ghz spin waves in
  antiferromagnetic ${\mathrm{febo}}_{3}$},}\ }\href {\doibase
  10.1103/PhysRevLett.123.157202} {\bibfield  {journal} {\bibinfo  {journal}
  {Phys. Rev. Lett.}\ }\textbf {\bibinfo {volume} {123}},\ \bibinfo {pages}
  {157202} (\bibinfo {year} {2019})}\BibitemShut {NoStop}%
\bibitem [{\citenamefont {Tzschaschel}\ \emph {et~al.}(2019)\citenamefont
  {Tzschaschel}, \citenamefont {Satoh},\ and\ \citenamefont
  {Fiebig}}]{Tzschaschel_2019}%
  \BibitemOpen
  \bibfield  {author} {\bibinfo {author} {\bibfnamefont {Christian}\
  \bibnamefont {Tzschaschel}}, \bibinfo {author} {\bibfnamefont {Takuya}\
  \bibnamefont {Satoh}}, \ and\ \bibinfo {author} {\bibfnamefont {Manfred}\
  \bibnamefont {Fiebig}},\ }\bibfield  {title} {\enquote {\bibinfo {title}
  {Tracking the ultrafast motion of an antiferromagnetic order parameter},}\
  }\href {\doibase 10.1038/s41467-019-11961-9} {\bibfield  {journal} {\bibinfo
  {journal} {Nature Communications}\ }\textbf {\bibinfo {volume} {10}},\
  \bibinfo {pages} {3995} (\bibinfo {year} {2019})}\BibitemShut {NoStop}%
\bibitem [{\citenamefont {Lenz}\ \emph {et~al.}(2006)\citenamefont {Lenz},
  \citenamefont {Wende}, \citenamefont {Kuch}, \citenamefont {Baberschke},
  \citenamefont {Nagy},\ and\ \citenamefont {J\'anossy}}]{lenz_2006}%
  \BibitemOpen
  \bibfield  {author} {\bibinfo {author} {\bibfnamefont {K.}~\bibnamefont
  {Lenz}}, \bibinfo {author} {\bibfnamefont {H.}~\bibnamefont {Wende}},
  \bibinfo {author} {\bibfnamefont {W.}~\bibnamefont {Kuch}}, \bibinfo {author}
  {\bibfnamefont {K.}~\bibnamefont {Baberschke}}, \bibinfo {author}
  {\bibfnamefont {K.}~\bibnamefont {Nagy}}, \ and\ \bibinfo {author}
  {\bibfnamefont {A.}~\bibnamefont {J\'anossy}},\ }\bibfield  {title} {\enquote
  {\bibinfo {title} {Two-magnon scattering and viscous gilbert damping in
  ultrathin ferromagnets},}\ }\href {\doibase 10.1103/PhysRevB.73.144424}
  {\bibfield  {journal} {\bibinfo  {journal} {Phys. Rev. B}\ }\textbf {\bibinfo
  {volume} {73}},\ \bibinfo {pages} {144424} (\bibinfo {year}
  {2006})}\BibitemShut {NoStop}%
\bibitem [{\citenamefont {Vittoria}\ \emph {et~al.}(2010)\citenamefont
  {Vittoria}, \citenamefont {Yoon},\ and\ \citenamefont
  {Widom}}]{vittoria_2010}%
  \BibitemOpen
  \bibfield  {author} {\bibinfo {author} {\bibfnamefont {C.}~\bibnamefont
  {Vittoria}}, \bibinfo {author} {\bibfnamefont {S.~D.}\ \bibnamefont {Yoon}},
  \ and\ \bibinfo {author} {\bibfnamefont {A.}~\bibnamefont {Widom}},\
  }\bibfield  {title} {\enquote {\bibinfo {title} {Relaxation mechanism for
  ordered magnetic materials},}\ }\href {\doibase 10.1103/PhysRevB.81.014412}
  {\bibfield  {journal} {\bibinfo  {journal} {Phys. Rev. B}\ }\textbf {\bibinfo
  {volume} {81}},\ \bibinfo {pages} {014412} (\bibinfo {year}
  {2010})}\BibitemShut {NoStop}%
\end{thebibliography}%

\end{document}